\documentclass[twocolumn,tighten,iop,times,twocolappendix,trackchanges]{aastex7}
\usepackage{graphicx}
\usepackage{amsmath}
\usepackage{iondefs, farhan-defs}
\usepackage[super]{nth}

\usepackage[T1]{fontenc}
\usepackage[utf8]{inputenc}

\usepackage{bigints}
\usepackage[fulladjust]{marginnote}

\usepackage{appendix}

\usepackage{CJK}

\hypersetup{colorlinks=true, citecolor=blue, filecolor=magenta,urlcolor=blue, linkcolor=blue}

\begin{document}

\begin{CJK*}{UTF8}{gbsn}


\title{\sc Impact of Cosmic Filaments on Galaxy Morphological Evolution and Predictions of Early Cosmic Web Structure for Roman}


\author[0000-0002-0072-0281]{Farhanul Hasan}
\affiliation{Space Telescope Science Institute, 3700 San Martin Drive, Baltimore, MD 21218, USA}
\affiliation{Department of Astronomy, New Mexico State University, Las Cruces, NM 88003, USA}
\email{fhasan@stsci.edu}

\author[0000-0002-4321-3538]{Haowen Zhang  (张昊文)}
\affiliation{Steward Observatory, University of Arizona, 933 N Cherry Ave., Tucson, AZ 85721, USA}
\affiliation{Canadian Institute for Theoretical Astrophysics, University of Toronto, Toronto, ON M5S 3H8, Canada}
\email{hwzhang0595@arizona.edu}

\author[0000-0002-2499-9205]{Viraj Pandya}
\altaffiliation{NASA Hubble Fellow}
\affiliation{Columbia Astrophysics Laboratory, Columbia University, 550 West 120th Street, New York, NY 10027, USA}
\email{viraj.pandya@gmail.com}

\author[0000-0002-9946-4731]{Marc Rafelski}
\affiliation{Space Telescope Science Institute, 3700 San Martin Drive, Baltimore, MD 21218, USA}
\affiliation{Department of Physics and Astronomy, Johns Hopkins University, 3400 North Charles Street, Baltimore, MD 21218, USA}
\email{mrafelski@stsci.edu}

\author[0000-0002-1979-2197]{Joseph N. Burchett}
\affiliation{Department of Astronomy, New Mexico State University, Las Cruces, NM 88003, USA}
\email{jnb@nmsu.edu}

\author[0009-0005-7451-0614]{Douglas Hellinger}
\affiliation{Santa Cruz Institute for Particle Physics (SCIPP), University of California, Santa Cruz, CA 95064, USA }
\email{dhelling@ucsc.edu}

\author[0000-0001-5294-8002]{Kalina~V.~Nedkova}
\affiliation{Department of Physics and Astronomy, Johns Hopkins University, 3400 North Charles Street, Baltimore, MD 21218, USA}
\affiliation{Space Telescope Science Institute, 3700 San Martin Drive, Baltimore, MD 21218, USA}
\email{knedkova@stsci.edu}


\author[0009-0007-8470-5946]{Ilias Goovaerts}
\affiliation{Space Telescope Science Institute, 3700 San Martin Drive, Baltimore, MD 21218, USA}
\email{igoovaerts@stsci.edu}

\author[0000-0001-8057-5880]{Nir Mandelker}
\affiliation{Centre for Astrophysics and Planetary Science, Racah Institute of Physics, The Hebrew University, Jerusalem 91904, Israel}
\email{nir.mandelker@mail.huji.ac.il}

\author[0000-0002-6766-5942]{Daisuke Nagai}
\affiliation{Department of Physics, Yale University, New Haven, CT 06520, USA}
\email{daisuke.nagai@yale.edu}

\author[0000-0003-0028-4130]{Grecco A. Oyarz\'un}
\affiliation{Department of Physics and Astronomy, Johns Hopkins University, 3400 North Charles Street, Baltimore, MD 21218, USA}
\affiliation{Space Telescope Science Institute, 3700 San Martin Drive, Baltimore, MD 21218, USA}
\email{goyarzun@stsci.edu}

\author[0000-0001-5091-5098]{Joel R. Primack}
\affiliation{Department of Physics, University of California, Santa Cruz, CA 95064, USA}
\email{joel@ucsc.edu}

\author[0000-0002-2462-2619]{Joanna Woo}
\affiliation{Department of Physics, Simon Fraser University, 8888 University Drive, Burnaby, BC, V5A 1S6, Canada}
\email{j_woo@sfu.ca}


\correspondingauthor{Farhanul Hasan} 
\email{fhasan@stsci.edu}

\shortauthors{Hasan et al.}

\begin{abstract}

We leverage the IllustrisTNG cosmological simulations to test how the large-scale cosmic web shapes galaxy morphology and to forecast the early cosmic web structure that the Nancy Grace Roman Space Telescope will reveal. In the hydrodynamic TNG50 and $N$-body TNG50-Dark runs, we reconstruct the cosmic web at redshifts $z=$0, 0.5, 1, 2, 3, and 4 with the Monte Carlo Physarum Machine density estimator and the DisPerSE structure identification framework. We confirm that dark matter halos  start out predominantly prolate (elongated), and their shapes are aligned with their nearest filaments; prolate galaxies retain strong shape-alignment with their outer halos to later times. At $z\geq1$, the fraction of prolate (spheroidal) halos increases (decreases) toward lower stellar mass, higher redshift, and lower filament density.
At $z<1$, more spheroidal (oblate) stellar structures preferentially reside  in higher-density (lower-density) filaments.
We also find that higher-density filaments favor extended rotationally-supported disks, whereas lower-density filaments more often host smaller dispersion-supported systems. 
Then, generating mock galaxy samples from TNG100 and TNG50, we predict the early cosmic web accessible to Roman.
We find that the spectroscopic emission-line depth planned for the High-Latitude Wide-Area Survey (HLWAS) yields a highly incomplete galaxy sample that does not accurately trace the $z=1$ cosmic web. A survey $\geq$2.5$\times$ deeper over a few square degrees would enable a proper reconstruction and reveal qualitatively correct filament-galaxy morphology relationships. Nevertheless, the planned HLWAS Deep field should still identify most galaxy overdensities; targeted deeper spectroscopy of these regions would efficiently and adequately map the early filamentary structure.


\end{abstract}


\section{Introduction}
\label{sec:intro}

The anisotropic collapse of matter under the force of gravity results in the formation of the web-like patterns of the large-scale structure known as the cosmic web, consisting of filaments, sheets, nodes, and voids. During this process, tidal torques of the cosmic web impart angular momentum, or spin, to galaxies as they form \citep[e.g.,][]{Peebles69,Zeldovich70,White84,Codis15}. This has major implications for the development of galaxy structure, including the formation of disks \citep[e.g.,][]{Pichon11,Stewart13,Danovich15} and elongated or prolate galaxies \citep[e.g.,][]{ceverino15,tomassetti16,Pandya19}.
Crucially, this effect of the \textit{anisotropic} configuration of the cosmic web depends on geometry and is distinct from the long-established morphology-density relation, where the morphology of a galaxy depends on some \textit{isotropic} measure of the environment, such as local density \citep[e.g.,][]{Dressler80,PG84,Kauffmann04}.

A growing body of evidence suggests that the shapes of galaxies and/or their parent dark matter (DM) halos retain some signatures of the anisotropic accretion of matter from the cosmic web out to later times. 
Most cosmological simulations predict that the major (minor) axes of galaxies/halos are preferentially aligned (misaligned) with the direction of their nearest filaments -- so-called ``shape-filament alignments'' \citep[e.g.,][]{AC07b,Hahn07,FR14,Codis18,GV19}.
Such shape alignments are also generally found to vary with factors such as redshift, stellar/halo mass, and even the density/thickness of the filaments themselves \citep[e.g.,][]{Borzyszkowski17,GV18,GV21,WK18,Lopez21,Zhang23}.
While several observational studies have lent support to these phenomena \citep[e.g.,][]{Tempel13,Zhang13,Chen17}, some others have  found no statistically significant alignment signals \citep[e.g.,][]{Pahwa16,Krolewski19}.
We expect stronger alignments with filaments in the outer regions of DM halos which are sensitive to more recent mass accretion \citep[e.g.,][]{Wang11}, while the inner regions bear complexities beyond the linear transfer of tidal forces from the large-scale structure \citep[e.g.,][]{Codis12,Codis15,Dubois14,Laigle15}.

The transfer of mass and angular momentum from the halo scale to the galaxy scale involves non-linear and baryonic processes acting on the inflowing gas. This gas will be subject to torques and other forces from the circumgalactic medium (CGM) and interstellar medium (ISM) of galaxies, leading to a loss of angular momentum of accreting gas \citep[e.g.,][]{Danovich15,Cadiou22}.
Once infalling gas cools and forms stars, angular momentum is imparted to the stars, and the salient morphological characteristics of galaxies are established, such as spheroidal (bulge-dominated), oblate (disk-dominated), or prolate (elongated) stellar shapes and spin, which determine the degree of rotational vs.~dispersion support. Thus, despite the complexities of baryonic physics, we expect the cosmic web to be causally linked to the emergence of galaxy morphology.

In the classical $\Lambda$CDM picture of galaxy formation, disk galaxies emerge naturally from the gravitational collapse of baryons in the early Universe \citep[see, e.g., ][]{FW80,Blumenthal84,MMS98},  but observations of high-redshift galaxy morphologies have revealed a richer and more complicated story. 
Many lines of (decades-long) evidence from the Hubble Space Telescope (HST) and other observatories hint at a large population of galaxies with elongated and/or irregular morphologies \citep[e.g.,][]{Cowie95,Elmegreen05,Ravindranath06,Law12,Yuma12}.
By comparing the projected axis ratio distributions of $z\!<\!0.1$ galaxies in the Sloan Digital Sky Survey \citep[SDSS;][]{York00} and $z\!\sim\!1.5-2$ galaxies observed by HST, \citet{vanderwel14} concluded that a significant fraction of high-redshift galaxies must have preferentially elongated, rather than oblate, stellar structures.
\citet{zhang19} jointly modeled galaxy axis ratios and sizes and extended the sample to redshifts $z\!=\!0.5-2.5$ to robustly demonstrate that the \textit{majority} of $z\!\geq\!2$ galaxies with stellar masses $9.5\!<\!{\logms}\!<\!10$ and $z\!\geq\!1$ galaxies with $9\!<\!{\logms}\!<\!9.5$ are elongated.

Recently, using higher resolution and deeper James Webb Space Telescope (JWST) imaging, \citet{Pandya24} confirmed this phenomenon of low-mass, high-redshift galaxies being more likely prolate than disky. They showed that the fraction of prolate galaxies with $9<{\logms}<9.5$ drops from $\sim$80\% to $\sim$25\% from $z=8$ to $z=1$, while the fraction of oblate galaxies of this mass rises from $\sim$20\% to $\sim$60\% between $z=8$ and 1. High-mass (${\logms}>10$) galaxies tend to be a mix of oblates and spheroids at $z<3$ and mostly oblate at $z\geq3$. However, establishing the link between the emergence of these structures and their location within the cosmic web remains elusive in observational data.

Some zoom-in hydrodynamical simulations predict a crucial role of the cosmic web in the formation of high-redshift prolate dwarf galaxies and their subsequent evolution to oblate and spheroidal shapes.
\citet{ceverino15} and \citet{tomassetti16} found that the {\sc VELA} simulations \citep{Ceverino14} produce ${\logms}\lesssim9.5$ prolate galaxies inside preferentially prolate halos at $z\approx2-4$, which they argue is a consequence of asymmetric accretion of DM and gas from narrow filaments and mergers. Feedback from early star formation prevents the overabundance of baryons at the cores of these halos, resulting in DM-dominated cores, where DM exerts torques on the stars and causes strong alignment between the major axes of the halo and stars. 
\citet{Pandya19} proposed that intrinsic alignments of these early elongated galaxies could be used as a novel cosmological probe of high-redshift cosmic web filaments, and \citet{Pandya25} recently found preliminary evidence for such alignments in sub-regions of one JWST ``blank'' deep field. 
At later times, the buildup of stellar mass causes baryons to dominate the cores and lose angular momentum, causing a so-called ``wet compaction'' event wherein gas falls to the center of the galaxy and forms compact stellar disks \citep[e.g.,][]{DB14,Zolotov15,Tacchella16}. Combining these phenomena, \citet{Lapiner23} outlined a morphological evolutionary sequence of galaxies from prolate to oblate/spheroidal, and dispersion-dominated to rotation-dominated kinematics.

Despite the compelling picture painted by these results, the {\sc VELA} simulations evolve individual halos and, as such, cannot explicitly characterize the cosmic web filaments and therefore directly constrain their effect on galactic morphological evolution.
To date, no hydrodynamical simulations with a fully cosmological volume have probed the effect of the mass and angular momentum from the cosmic web on the morphological evolution of a statistically large sample of galaxies \citep[see][for the relationship between accreting gas and galactic gas at $z=0$]{Woo25}.

Accordingly, the first goal of this paper is to better understand how the accretion of matter from the cosmic web affects the angular momentum acquisition and morphological evolution of galaxies using state-of-the-art hydrodynamical cosmological simulations. To this end, we investigate galaxies in the context of their cosmic web environment in the high-resolution TNG50 simulation \citep[][]{Nelson19,Pillepich19} of the IllustrisTNG \citep[][]{Nelson18} simulation suite. 
We first characterize the cosmic web filaments in TNG50 using a novel filament identification framework first presented in \citet[hereafter \citetalias{Hasan24}]{Hasan24}. 
Their approach enables highly complete filament identification, including both high- and low-density filaments, as well as a simple characterization of a filament's 1D line density.
At redshifts, $z=4$, 3, 2, 1, 0.5, and 0, we measure the 3D shapes of both the stellar and DM components of galaxies down to stellar mass ${\logms}=8$. We study (1) the shape alignment with the nearest filaments, (2) the effect of the density of the nearest filament on 3D shapes, and (3) the connection between filament density and kinematic morphological properties, such as stellar spin.

The second goal of this paper is to predict the structure of the early cosmic web as we enter the golden age of wide-field galaxy surveys with groundbreaking facilities such as the Dark Energy Spectroscopic Instrument \citep[DESI;][]{DESIEDR24,DESIDR25}, Euclid \citep[][]{Euclid25_overview,Euclid25_laigle}, Subaru Prime Focus Spectrograph \citep[PFS;][]{Takada14,Greene22}, SPHEREx \citep[][]{Dore16,Feder24}, and the Nancy Grace Roman Space Telescope \citep[Roman;][]{Wang:2022_romanHLSS}.
These observatories will provide the wide-area coverage and sensitivity required to map the cosmic web -- and for the first time, at higher redshifts ($z\!\geq\!1$). 
This is a major step towards establishing how well we can predict the influence of the cosmic web on the formation of galaxy morphology.
Our analyses will uniquely set the scene for us to understand the role of the large-scale environment in assembling galactic structure during this peak epoch of galaxy formation
\citep[e.g.,][]{MD14,FS20}.

This paper is organized as follows. In \S~\ref{sec:data}, we describe the simulation data, our measurements of galaxies, and cosmic web reconstruction methods. In \S~\ref{sec:results}, we present our findings on the dependence of the cosmic web on the evolution of galaxy morphology. In \S~\ref{sec:pred}, we present predictions of early cosmic web structure for galaxy surveys with Roman. We discuss our results in \S~\ref{sec:discuss} and conclude the paper in \S~\ref{sec:conc}. Throughout this paper, we adopt the {\it Planck 2016} cosmology \citep{Planck15}, with $H_{0}=67.74$ {\kmsmpc}, $\Omega_{\mathrm{M},0} = 0.3089$, and $\Omega_{\Lambda,0} = 0.6911$. Unless otherwise stated, all distances and lengths are in comoving units.


\section{Simulation Data and Analysis}
\label{sec:data}


\subsection{TNG Simulation Data}
\label{sec:tngdata}

We analyze the outputs from the IllustrisTNG magneto-hydrodynamical cosmological simulations \citep{Marinacci18,Nelson18,TNGDR19,Pillepich18,Springel18}, which use the {\sc AREPO} moving-mesh hydrodynamic code \citep{Springel10} to simulate the evolution of gas, stars, dark matter, and black holes from redshifts $z\!=\!127$ to $z\!=\!0$, assuming a {\it Planck 2016} cosmology \citep{Planck15}. 
Since we investigate the morphologies of galaxies, we primarily use the highest resolution run of the TNG50 simulation \citep{Nelson19,Pillepich19}, TNG50-1, which has a box size of approximately 51.7 comoving Mpc per side, a minimum baryonic particle mass of $\sim\!8.4 \!\times\! 10^{4}~{\Msun}$, DM particle mass of $\sim\!4.6 \!\times\! 10^{5}~{\Msun}$, a $z=0$ gravitational force softening length in DM and stars of 288 pc, and a minimum gas softening length of 72 physical pc.

We analyze 6 snapshots -- at $z=4$, 3, 2, 1, 0.5, 0 -- of the TNG50-1 simulation (hereafter TNG50) from the online data repository,
as presented first in \citet{TNGDR19}. In each snapshot, the ``Group'' catalogs were constructed using the friends-of-friends (FoF) substructure identification algorithm, and the ``subhalo'' catalogs were constructed using the {\sc Subfind} algorithm \citep{Springel01,Dolag09}. The ``subhalo'' catalogs contain gravitationally bound objects within a FoF group, while the ``Group'' catalogs identify halos, i.e., FoF groups. 
We extract data from both of these types of catalogs. Following the convention of TNG, we refer to groups as halos and subhalos as galaxies.
We also make use of gas, DM, and star particle data in all snapshots.
Finally, to constrain the effects of baryonic feedback physics on the morphological properties of galaxies, we also analyze the $N$-body version of TNG50, i.e., TNG50-Dark, for all 6 snapshots, making use of the subhalos, halos, and DM particles.

We limit our analyses to galaxies with stellar mass ${\logms}\geq8$ at each redshift, which ensures that all galaxies contain over 1000 star particles each. This allows us to confidently resolve the morphologies of galaxies with ${\logms}\geq8$, at spatial scales $\lesssim$100~pc. We recognize that TNG100, which has $\sim$8 times the volume of TNG50, would yield a larger statistical sample of galaxies and cosmic web structures. However, its particle mass resolution is $\sim16$ times lower, and its gravitational softening length, hence spatial resolution, is $\approx$2-3 times lower than that of TNG50. Thus, we could only resolve the morphologies of much more massive galaxies (${\logms}\gtrsim9.5$) in TNG100 with the same confidence that we resolve ${\logms}\geq8$ galaxies in TNG50.

\subsection{Measurements of galaxy morphological properties}
\label{sec:measure}

We first measure the intrinsic shapes of stars and DM in TNG50 galaxies using particle data. For these measurements, we closely followed the methods of \cite{Pillepich19}.
First, we obtain the comoving center of mass for each galaxy and measure the 3D radius containing half of the stellar mass of each galaxy, hereafter referred to as the stellar half-mass radius, {\rstar}. Our measurements do not take into account observational considerations, such as 2D projection effects, surface brightness dimming, and instrumental sensitivity; therefore, these 3D radii are not meant to be ``mock'' measurements that replicate observational choices for a truly apples-to-apples comparison with real data.

We similarly measure the DM half-mass radius, {\rdm}, which contains half of the DM mass of a galaxy (or subhalo). 
This choice is made so that we can compare DM properties at the same relative scale as the stellar component -- 2{\rdm} vs. 2{\rstar}. 
While {\rdm} is usually smaller than the halo virial radius {\Rvir}, it represents an appreciable fraction of {\Rvir} for typical halo concentrations \citep[e.g.,][]{Klypin16}.
For our galaxy sample, {\rdm} is typically $\sim$0.2--0.5{\Rvir} at $z\!\leq\!2$ and $\sim$0.3--0.6{\Rvir} at $z\!=\!3$ and 4. 
Therefore, we define 2{\rdm} as the ``outer'' halo radius, where baryonic effects are minimal (many tens to $\sim$100~kpc). Likewise, 0.1{\rdm} is defined as the ``inner'' halo radius, which is the scale of the central galaxy ($\sim$kpc).

Next, we calculate the 3D shapes of stars and DM in galaxies, which we briefly describe below. This is identical to \cite{Pillepich19}, whose measurements are not publicly available in the online catalogs. 
First, we define $A$, $B$, and $C$, as the lengths of the major, middle, and minor axes, respectively.
Next, we initialize 3D elliptical shells to have axis ratios $B/A\!=C/A\!=1$, at a distance of 2{\rstar} from the galaxy center and with a thickness of 0.4{\rstar}. If the initial shell contains fewer than 3 particles, we simply do not calculate the shape and label the galaxy with a flag.
We then iteratively diagonalize the mass tensor and update the elliptical shells until the direction and magnitude of the eigenvectors of the mass tensor converge. The length of the major axis is fixed at 2{\rstar}, but the direction is allowed to vary. 
Note that the length of the major axis and the elliptical shell above is set to 0.1{\rdm}, 2{\rdm}, or 2{\rstar}, depending on the scale at which the shape is measured.

We measure and analyze the 3D stellar and DM shapes of TNG50 galaxies at 0.1{\rdm} and 2{\rdm},
and likewise for the DM shapes of subhalos in TNG50-Dark at 0.1{\rdm} and 2{\rdm}. 
These choices of radii enable consistent comparisons between different physical scales and between TNG50 and TNG50-Dark as there are no stars in the latter.
However, we also analyze stellar shapes at 2{\rstar} instead of 0.1{\rdm} in TNG50, and find that the results for the inner halo do not change appreciably between these two radii.

We characterize the shapes of stellar and DM components of galaxies using several different metrics commonly used in the literature.
We follow the classification method originally devised in \citet[][see their Fig. 2]{vanderwel14}, where the intrinsic middle-to-major axis ratio $B/A$ and minor-to-major axis ratio $C/A$ are used to categorize the shape into spheroidal, oblate (disky), and prolate (elongated). Spheroidal shapes have similar lengths $A$, $B$, and $C$, whereas oblates have much larger $B/A$ than $C/A$ (i.e., strongly flattened along one dimension), Prolates have a much larger $A$ than $B$ or $C$, with the smaller axes being closer to each other than oblates (i.e., flattened along two dimensions). 
We refer the reader to the details of the classifications and visual depictions presented in \cite{zhang19} and \cite{Pandya24}.

We further leverage publicly available catalogs for additional characterizations of TNG50 galaxy morphological properties, based on measurements of \cite{Genel15} and \cite{Genel18}. 
We obtain kinematics-based morphological quantities, including the specific angular momentum (sAM) of stars in galaxies, which is a measure of stellar spin. We also make use of the mass fraction of stars in disk-like and bulge-like orbits. 
By aligning a galaxy such that its spin vector points upward along the $z$-axis, the circularity parameter $\epsilon$ is defined for each particle as the ratio of the sAM along the $z$-axis and maximum sAM possible at the specific binding energy of the particle \citep[see][]{Vb14}. 
Stars in disk-like orbits, i.e., with a high degree of rotation about the center, have $\epsilon$ close to 1, while those in more random orbits comprising the bulge have very small positive to highly negative $\epsilon$. 
Following common definitions in the literature \citep[e.g.,][]{Abadi03}, we define the mass fraction of stars (1) in disk-like orbits as the fraction with $\epsilon>0.7$ minus those with $\epsilon<-0.7$, and (2) in bulge-like orbits as the fraction with $\epsilon<0$.
For (1), we remove the $\epsilon<-0.7$ fraction to remove any contribution of bulge stars to the disk -- assuming the bulge is symmetric around $\epsilon=0$.
Not doing so does not change any of our results appreciably. We use all the measurements above for all stars, but limiting them to a 10{\rstar} radius \citep[as also available from the catalog curated by][]{Genel15} yields identical results.

We find that our shape and size measurements are consistent with the existing literature.
The majority of ${\logms}=8-9$ galaxies are spheroidal at all times. At $z\geq2$, spheroidal shapes dominate the ${\logms}9-10$ mass range, but at lower redshifts, oblate shapes make up the majority. For ${{\logms}>10}$, oblates are most abundant at $z<3$. 
Prolate shapes are always subdominant at $z\leq4$ at ${{\logms}\geq8}$, making up at most $\approx$30\%.
These results confirm the predominant morphological transformation of galaxies in TNG from mostly spheroids to increasing disk-dominance with increasing time and mass assembly \citep{Pillepich19,Tacchella19,Du20,Varma22}.

While qualitatively reproducing many observed morphological properties of galaxies, TNG50 still falls short in many aspects.
The oblate fraction approximately matches the most recent observational constraints with JWST, but too few prolates and somewhat too many spheroids with ${\logms}\sim9-10$ are formed in the simulation \citep[e.g.,][]{Kartaltepe23, Pandya24}. The tension in the former is very strong, as prolates are underproduced by a factor of $\sim$2 at $z=3-4$ and up to $\sim4$ at $z=1-2$. The simulated intermediate-mass spheroidal fraction is twice as large as observed at $z>1$, but closer to agreement at lower redshifts.
We discuss possible sources of these tensions in \S~\ref{sec:disccomb}.

\subsection{Cosmic Web Reconstruction}
\label{sec:cw}

Our methods for reconstructing the cosmic web and identifying filaments from the locations of galaxies are based on the method outlined in \citetalias{Hasan24}. Here, we briefly describe the method and the specific choices made for this work.

\subsubsection{{\disperse} and MCPM}

To identify filamentary structure, we leverage the widely-used topologically-motivated Discrete Persistent Structure Extraction ({\disperse}) algorithm \citep{disperse1,disperse2}. {\disperse} takes point sets, typically galaxies or halos, as input tracers. In the standard workflow, it then estimates a cosmic matter density field by applying the Delaunay Tessellation Field Estimator \citep[DTFE;][]{SV00,VS09} technique and segmenting space into tetrahedrons, whose vertices are the locations of the tracers. Following this, the gradients of the density field are computed, and critical points are identified where the gradient vanishes. Filaments are sets of individual line segments (filament segments) that connect maxima (nodes) to saddle points.
Structures are characterized by their ``persistence,'' which quantifies their robustness with respect to shot noise in the input data. In practice, a persistence threshold is used, which is akin to a signal-to-noise threshold for identified structures.

Using the TNG100 simulation, \citetalias{Hasan24} showed that the use of DTFE in the density estimation step of {\disperse} leads to an incomplete characterization of the cosmic web. They found, in agreement with past studies \citep[e.g.,][]{BW20}, that DTFE, by only estimating density at the locations of tracers, does not accurately estimate the ``true'' matter density, as traced by DM, especially in low-density regions. Instead, they substituted the DTFE density field with the MCPM density field, based on the algorithm first introduced in \cite{Burchett20}. MCPM mimics the growth patterns of the unicellular \textit{Physarum polycephalum} or \textit{slime mold} organism by employing virtual ``agents'' that navigate through a domain to connect ``food sources,'' which, in this context, are galaxies/halos weighted by mass \citep{Elek21,Elek22}. It generates an equilibrium density field as a probabilistic solution of the agents' trajectories across many different realizations of this process, which serves as a robust proxy of the cosmic matter density field. 
The MCPM model was thoroughly calibrated by \citetalias{Hasan24}, such that its outputs most closely map to the cosmic matter density \citep[see also][for more details]{Elek22}. Furthermore, the MCPM density in real data qualitatively agrees with several other large-scale environmental metrics, such as galaxy overdensity \citep[][]{Wilde23,Oyarzun24}.

MCPM accurately traces the true cosmic matter density across a large range of densities \citep{Burchett20,Elek22}, and its use in the {\disperse} pipeline, as opposed to DTFE, results in a more complete characterization of cosmic web structure \citepalias{Hasan24}. From the same inputs (galaxies with ${\logms}\geq8$), MCPM drastically outperforms DTFE in tracing DM density and is able to identify filaments of a wide range of persistence values. This results in the MCPM-based reconstruction identifying \textit{low-density} filamentary structures that the DTFE-based method misses, similar to ``thin'' filaments or ``tendrils,'' which were also reported in previous studies \citep[e.g.,][]{Alpaslan14b,CroneOdekon:2018aa}. These structures have dramatically different effects than conventional high-density filaments on galactic phenomena such as star formation activity and gas supply \citepalias[see][]{Hasan24}.

We characterize the 1D filament line density, {\sigmafil}, as the sum of the MCPM overdensity per unit length along a filament segment, as defined in \citetalias{Hasan24}. 
This quantity is directly measurable in observational data, as it uses the MCPM density field. Recently, {\sigmafil} was also found to tightly correlate with gas density in TNG100 filaments \citep{Woo25}.
In the following sections, we often quote the filament line density of the nearest filament segment to a galaxy, but we note that galaxies are often not at the center of a filament.

\begin{figure*}[htbp]
\vspace{-10pt}
\centering
\gridline{\fig{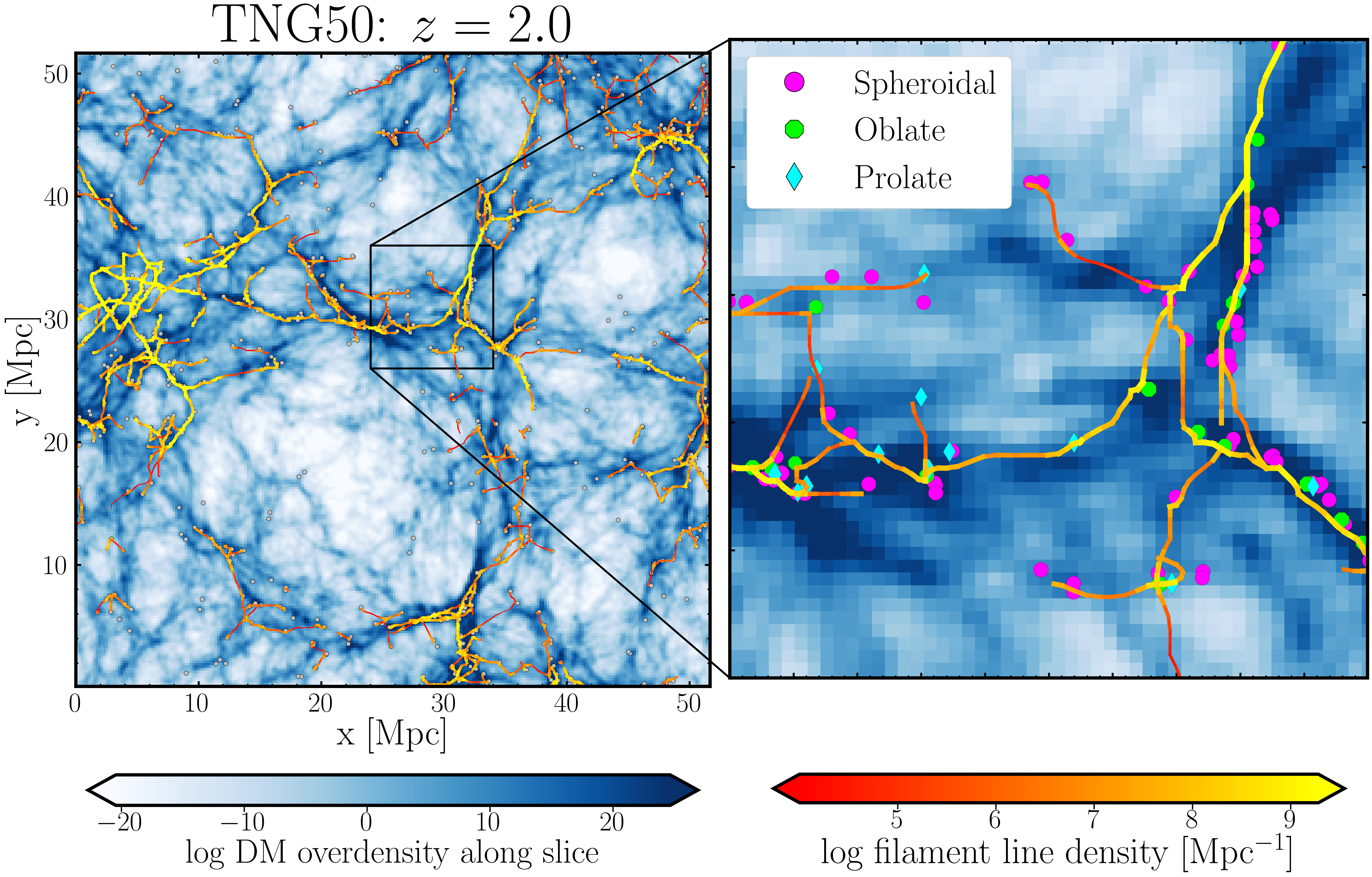}{0.8\textwidth}{}}
\vspace{-25pt}
\caption{
2D projection  of a 10 Mpc slice of the cosmic web in TNG50 at $z=2$ (in the x-y plane), with the full box on the left and a zoomed-in $10\times10$ Mpc region on the right.
The filaments (curves whose color and thickness are proportional to the 1D line density, as defined in \citetalias{Hasan24}) and galaxies  (markers) are overlaid on the integrated DM overdensity (in the $z$-direction). In the zoomed-in panel on the right, prolate, oblate, and spheroidal galaxies are represented by cyan diamonds, lime green octagons, and magenta circles, respectively (the galaxy markers are not to scale).
MCPM-based filaments can accurately trace the underlying DM distribution and reconstruct from the spatial distribution of galaxies alone.
}
\label{fig:vis}
\end{figure*}

\subsubsection{The Cosmic Web in TNG50}

We apply the techniques described above to reconstruct and characterize the cosmic web in each of our six chosen snapshots in TNG50. The inputs to our pipeline are the locations and masses of all galaxies with stellar mass ${\logms}\geq8$. For a given redshift, we keep most of the MCPM density field estimation parameters the same as those used by \citetalias[]{Hasan24} (see their Table 1). We make a small change in the \textit{sense distance} parameter, which is related to the volume of the domain and the data resolution, as these change between TNG100 and TNG50. We also smoothed the density field on $256^3$ grids, which correspond to $\approx$0.2~Mpc in length for the TNG50 box (for TNG100, the same grid resolution is about $\approx$0.4~Mpc)
We verify that all of these choices maximize the correlation of the MCPM density field with the DM density field. \citetalias[]{Hasan24} provides further detail on how the quality of the MCPM reconstruction only varies mildly with different input parameters. 

We verify that the persistence cut of 25 that we use in {\disperse} \citepalias[see][]{Hasan24} maximizes the match between the locations of the identified cosmic web nodes and the peaks of the DM density field (centers of the most massive halos), as we expect a strong correspondence between the two \citep[see, e.g.,][for more discussion regarding matching the cosmic web skeleton to the underlying density field]{GE24}.
Note that since we use a continuous density field (from MCPM), we adopt a persistence cut of 25. A more traditional approach uses discretely sampled density fields (from DTFE), for which the persistence threshold is defined in terms of the number of sigmas \citep[e.g.,][]{GE22,Malavasi22}.
However, the results of this paper are not sensitive to small changes in our cosmic web persistence parameter. Finally, we apply a single degree of smoothing to the final filamentary skeleton in order to smooth out any sharp or unphysical edges of the filament spine.

Figure~\ref{fig:vis} shows an example of a 2D projection of the $z=2$ cosmic web in TNG50, where the filaments are curves colored based on the filament line density, galaxies are small gray circles, and the DM overdensity is the blue color-map in the background. The right panel zooms in on a selected $10~\mathrm{Mpc}\times10~\mathrm{Mpc}$ portion, wherein prolate, oblate, and spheroidal galaxies are indicated. The filaments mostly trace the DM overdensities, with low-density filaments (red) located in low-overdensity regions (lighter blue).

Finally, to study the effects of baryonic feedback physics, we also reconstruct the cosmic web in the same snapshots of TNG50-Dark via the same techniques but using DM subhalos instead of galaxies as tracers. Subhalos are the equivalent units of galaxies for the DM-only simulation, but it is not trivial to convert a galaxy stellar mass threshold to a subhalo DM mass threshold. We choose subhalos with mass $M_{\mathrm{sh}}\geq10^{10}$~{\Msun} as tracers of the cosmic web in TNG50-Dark. This cut is chosen to approximately map from stellar mass ${\logms}\geq8$ to a minimum halo mass using stellar-to-halo-mass relations \citep[e.g.,][]{RP17,Behroozi19}. With this cut, we obtain a density of tracers in TNG50-Dark that is within a few times that in TNG50. However, changing this cut somewhat does not affect our scientific conclusions. For the TNG50-Dark reconstructions, we use the same MCPM and {\disperse} parameters as in TNG50.


\section{Cosmic web-dependence of galaxy morphology}
\label{sec:results}

\begin{figure*}[htbp]
\vspace{-10pt}
\centering
\gridline{
\fig{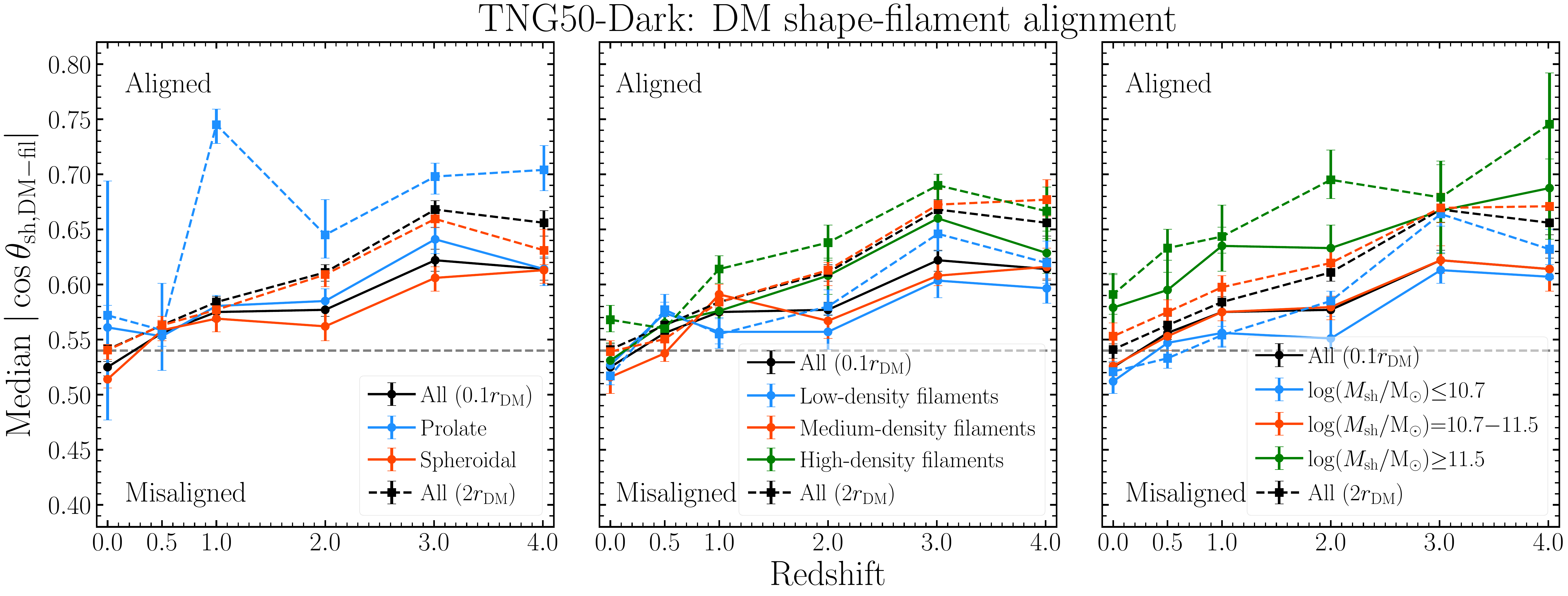}{0.95\textwidth}{}
}
\vspace{-25pt}
\gridline{
\fig{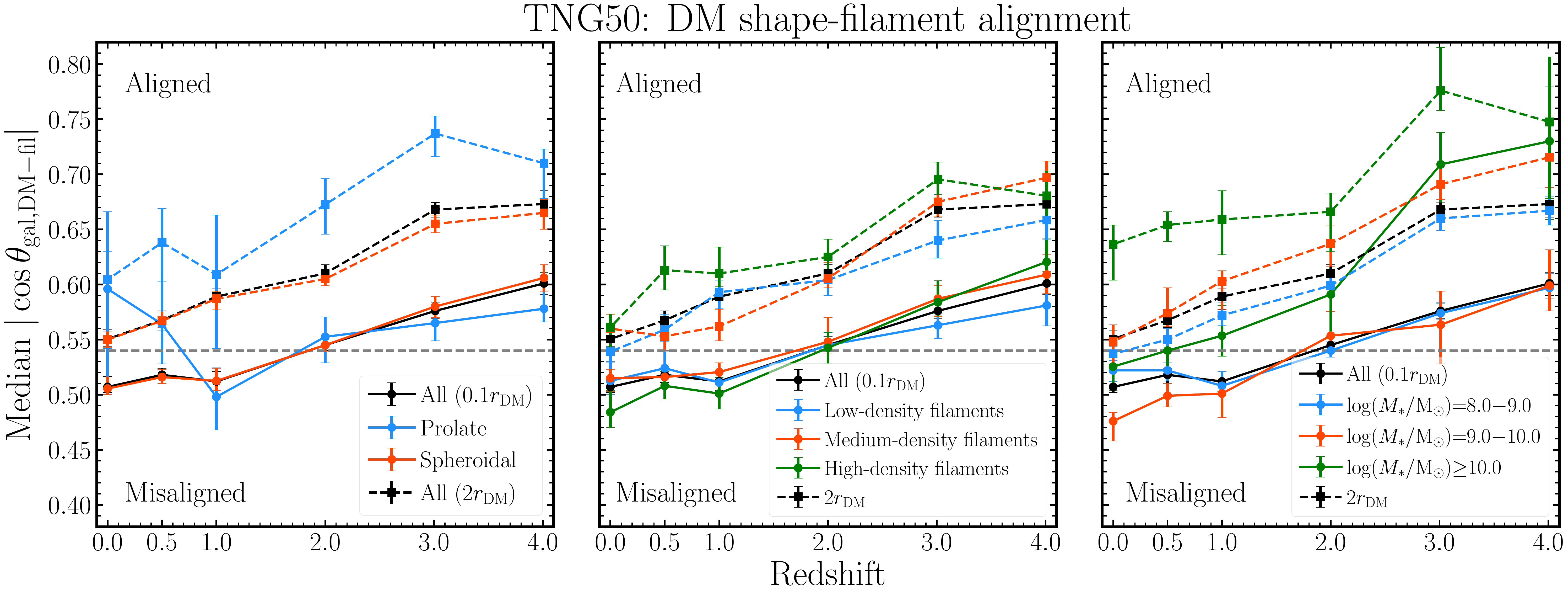}{0.95\textwidth}{}
}
\vspace{-25pt}
\gridline{
\fig{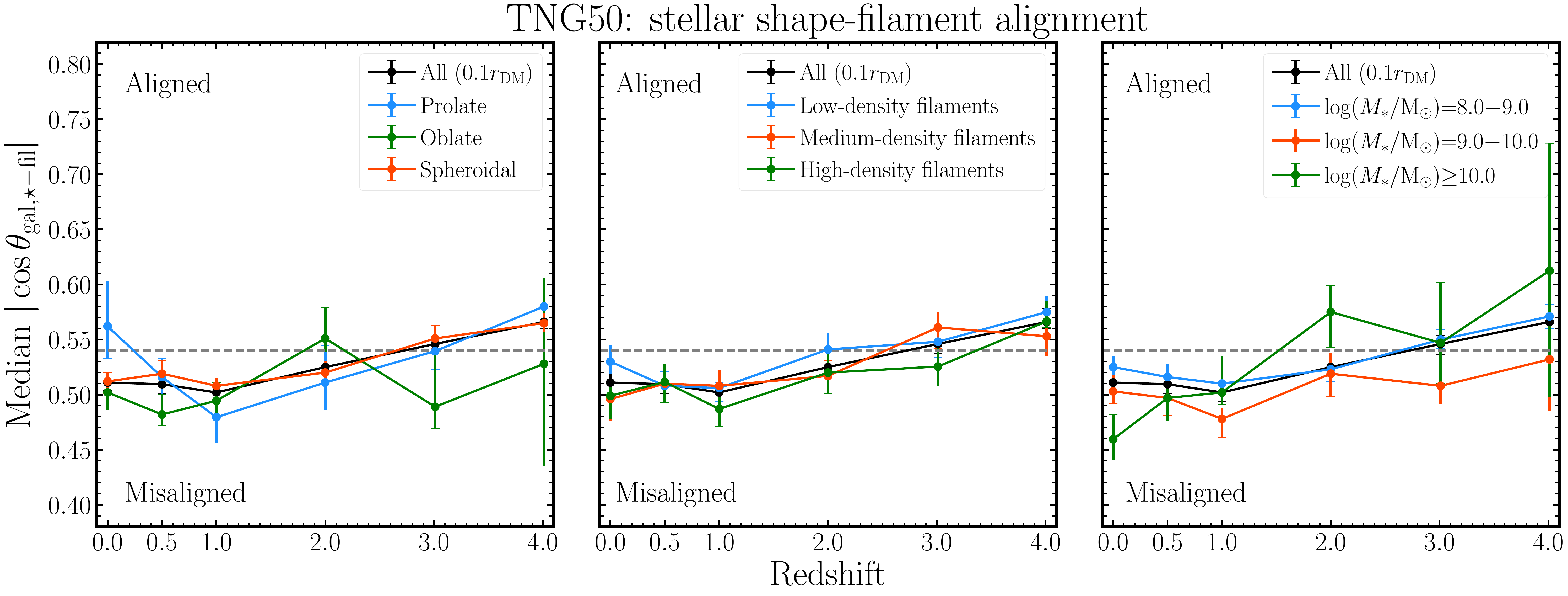}{0.95\textwidth}{}
}
\vspace{-25pt}
\caption{
Redshift-evolution of the median angle of alignment (and $\pm1\sigma$ bootstrapped errors) between the major axis of galaxies/subhalos and the direction of the nearest filament segment, in the DM-only TNG50-Dark (top row) and the hydrodynamic TNG50 (middle and bottom rows) simulations. Dashed lines and square markers represent the alignment at 2{\rdm} (outer halos), while solid lines and circles represent alignment at 0.1{\rdm} (inner halo). 
The left, middle, and right columns subdivide subhalos/galaxies by 3D shape, line density of nearest filament segment, and mass, respectively. The dashed horizontal line represents no preferential shape alignment.
DM shape-filament alignments are down to at least $z\sim2$ in TNG50 and $z\sim0.5$ in TNG50-Dark, but stellar shape-filament alignment disappears at $z<4$ and tends toward misalignment at $z\leq2$.
At the outer halo scale, prolate, massive, and halos near high-density filaments show the strongest shape alignments.
The inner halo shapes are less aligned in the baryonic than in the DM-only simulation. 
}
\label{fig:medangle}
\end{figure*}

In this section, we study the connection between the cosmic web environment of TNG50 galaxies and their 3D shapes and kinematic morphological properties. 
We restrict all analyses below to \textit{central} (non-satellite) galaxies, as satellite galaxies are subjected to various halo and/or local environmental effects \citep[e.g.,][]{Peng12,Wetzel13,Oyarzun23} that complicate interpretations of their morphological evolution. 
Furthermore, we only include galaxies that are within ${\dfil}\leq2$~Mpc of a filament spine to ensure that we probe a width within which filaments are expected to exert some effect \citep[e.g.,][]{BJ25,Yang25}. Only a small fraction of galaxies are excluded by this last requirement, as most galaxies live within this distance from a filament spine \citepalias[see also][]{Hasan24}.
Our final results are not affected by small changes in the above cuts, e.g., changing from ${\dfil}\leq2$~Mpc to ${\dfil}\leq1$~Mpc.


\subsection{Shape Alignments}
\label{sec:align}

We first investigate shape alignments with respect to filaments to understand the geometry of matter accretion from the cosmic web to galaxies. 
In particular, we probe the alignment between the longest axes of galaxies/halos and the axes of the nearest filament segment. 
We measure the angle between the direction of the major axis of a given component of a galaxy and that of the nearest filament segment. This angle is computed for the direction vectors of DM and stars measured at both 0.1{\rdm} and 2{\rdm}, corresponding to inner and outer halo scales, respectively.
We calculate the cosine of this angle -- {\thetadmfil} for DM shape-filament alignment, and {\thetastarfil} for stellar shape-filament alignment -- and report its absolute value. This varies from 0 -- the major axis completely misaligned or perpendicular to the filament -- to 1 -- the major axis completely aligned or parallel to the filament.
In the analysis below, we remove all structures whose two longest axes are equal, i.e., $A=B$, as these do not have a unique major axis. Only a few stellar or DM structures meet this criterion.

For ease of interpretation, we begin with the alignment of DM and filaments in subhalos of the $N$-body TNG50-Dark simulation to study the halo shape alignment as DM accretes from the cosmic web (with no baryonic effects present). 
Since we only include central subhalos hosted by halos, and they are slightly smaller DM structures tracing larger structures (their parent halos), we refer to subhalos and halos interchangeably in the following.

In the top row of Figure~\ref{fig:medangle}, we present the redshift-evolution of the DM shape alignment angle $|{\thetadmfil}|$ at both the 0.1{\rdm} (solid lines and circles) and 2{\rdm} (dashed lines and squares) scales in TNG50-Dark. The markers and error-bars represent the median $|{\thetadmfil}|$ and its $\pm1\sigma$ bootstrapped uncertainties. In each panel, black represents alignment for the entire population, while other colors divide the sample by 3D shape (left), the nearest filament line density (middle), and DM subhalo mass $M_{\mathrm{sh}}$ (right). The dashed horizontal line indicates the null result of no statistical alignment signal.
Only prolate and spheroidal DM shapes are shown in the left panel, since dissipationless DM almost never settles into a disk configuration \citep[e.g.,][]{Blumenthal84}. For the middle panel, we divide the filament line densities into low-, medium-, and high-density quantiles at each redshift (each hosting $1/3$ of the halos within ${\dfil}\leq2$~Mpc).  In the right panel, we separate the sample by low-, intermediate-, and high-mass subhalos. Note, however, that it is not trivial to compare $M_{\mathrm{sh}}$ (based on DM) to {\mstar} (based on stars) and we do not attempt to do so here. Rather, we assess the qualitative effect of increasing mass on the DM shape-filament alignment.

The DM shape-filament alignment decreases with decreasing redshift (at later times), and the alignment is generally higher at outer halo scales. On average, the DM major axis is fairly aligned with filaments at $z=4$, but this alignment virtually disappears by $z=1$ at both the inner halo and outer halo scales. At the outer halo scale, prolate halos exhibit a much stronger shape alignment than spheroidal halos down to $z=0.5$.
At lower redshifts, the errors on the prolate shape alignment dominate, likely due to low number statistics. 
At the inner halo-scale, there is a much smaller difference between the shape alignment of prolate and spheroidal halos.

Halos in higher-density filaments typically have stronger shape alignment than those in lower-density filaments, until $z=1$. While this trend is observed at both the inner and outer halo scales, there is a somewhat larger segregation between the shape alignment in higher and lower-density filaments at the inner halo. 
This filament density-dependence disappears at $z\leq1$. 
We also find stronger shape alignment with increasing subhalo mass, such that high-mass ($\log (M_{\mathrm{sh}}/{\Msun})\geq11.5$) subhalos retain significant shape alignment down to $z=0$, at both inner and outer halo scales. Lower mass subhalos lose their preferential shape alignment at $z\lesssim1$.

We now turn our attention to the fully hydrodynamic TNG50 simulation and repeat our analysis of both star and DM shape filament alignments. 
In the middle and bottom rows of Figure~\ref{fig:medangle}, we present the evolution of the median DM shape-filament alignment and stellar shape-filament alignment, respectively, in TNG50.
The solid lines and circles (in both the middle and bottom rows) represent the shape alignment at 0.1{\rdm} while the dashed lines and squares (in the middle row) represent this alignment at 2{\rdm}.
Similar to above, black represents the total population, while the colors subdivide galaxies by DM/stellar 3D shape (left column), nearest filament density (middle), and stellar mass (right).

\begin{figure*}[htbp]
\vspace{-10pt}
\centering
\gridline{
\fig{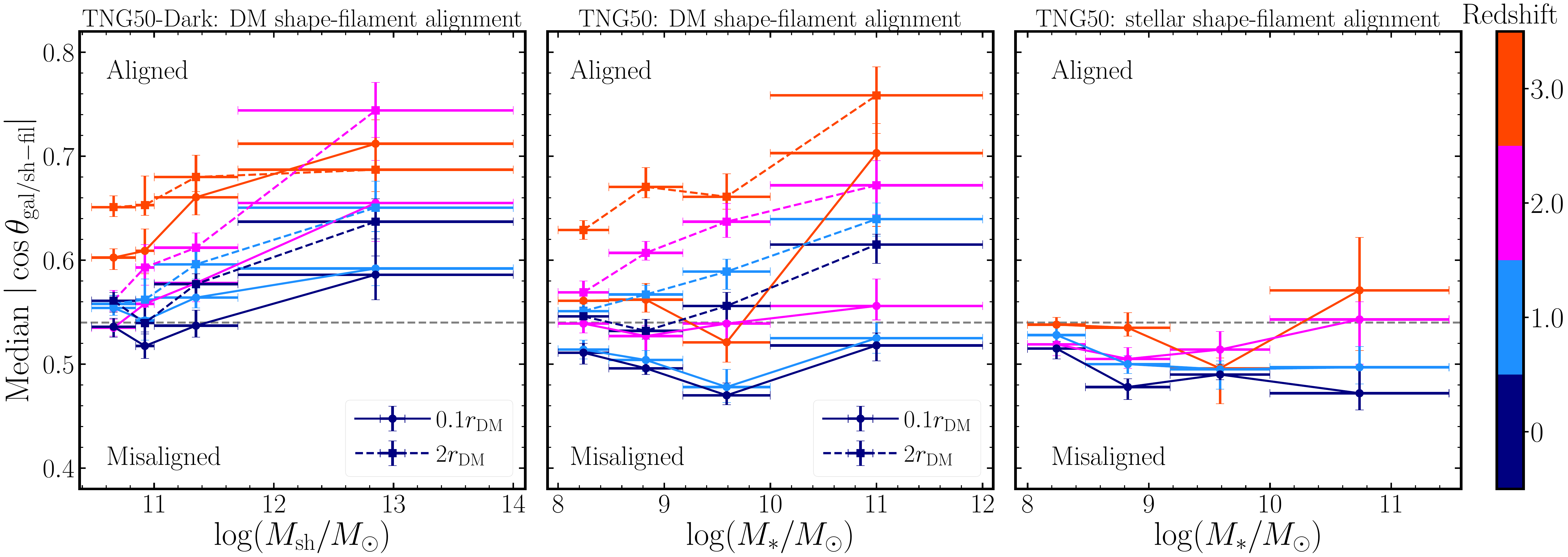}{0.9\textwidth}{}
}
\vspace{-25pt}
\caption{
Average shape-filament alignment as a function of mass, at a few different redshifts (indicated by the color-bar on the right).
The left panel shows the alignment between DM subhalo shape and the nearest filament at inner (solid) and outer (dashed) halo scales as a function of DM subhalo mass in TNG50-Dark.
The middle and right panels present the DM shape-filament alignment and stellar shape-filament alignment, respectively, as functions of stellar mass in TNG50. 
DM halo shapes are generally more aligned with nearby filaments with increasing mass, while stars show almost no alignment with rising mass.
}
\label{fig:medangle_mass}
\end{figure*}

The DM shape-filament alignment decreases with time and is much stronger at the outer halo scale. 
At the outer halo scale, the average shape alignment of the overall population is quantitatively very similar between TNG50 and TNG50-Dark. In contrast, there is a weaker shape alignment at the inner halo scale in TNG50 than in TNG50-Dark. In fact, this difference is much greater at lower $z$; inner halo shapes become slightly preferentially misaligned with nearby filaments at $z<2$ in TNG50, but at $z\sim0$ in TNG50-Dark.  
Within the same galaxy subpopulation, TNG50 generally shows a larger gap between the average alignment at inner and outer halo scales than TNG50-Dark.
Similar to its DM-only counterpart, TNG50 shows, at almost all redshifts, the strongest outer DM halo shape alignment in prolate halos, high-mass halos, and halos in high-density filaments.
However, these trends are not seen for inner halos.

There is virtually no statistically significant stellar shape alignment with filaments. Only at $z<4$ is a slight preferential alignment seen. Similar to inner halos, the stellar shape tends to be misaligned with nearby filaments at low redshifts ($z\lesssim2$). Within the errors, there are no distinctions in the average alignment signal with 3D shape, stellar mass, and nearest filament density.

Next, we investigate the mass-dependence of shape alignments in TNG50 and TNG50-Dark. 
Our result that high-mass galaxies show strong halo shape-filament alignment is consistent with a sizable body of work which reports that the mass of a galaxy/halo significantly affects its shape alignment \citep[e.g.,][]{GV18,GV21,Zhang23}. We present the average shape-filament alignment as a function of mass at redshifts $z=0$, 1, 2, and 3 in Figure~\ref{fig:medangle_mass}.
The left panel shows the DM shape-filament alignment in TNG50-Dark as a function of subhalo mass, while the middle (right) panel shows the DM shape-filament (stellar shape-filament) alignment in TNG50 as a function of stellar mass. Each of the colors represents a different redshift.

In TNG50, the DM shape alignment steeply rises with increasing mass in outer halos, and the correlation is stronger at earlier times. In the inner halo, the shape alignments of DM and stars are weakly dependent on mass, with a rising alignment signal with increasing mass only seen at $z>1$.
In contrast, TNG50-Dark predicts a significant rise in both the inner and outer halo alignment with increasing subhalo mass at all times. 
The difference in inner and outer halo shape alignments is always higher in TNG50 than in TNG50-Dark.

\begin{figure}[htbp]
\centering
\gridline{
\fig{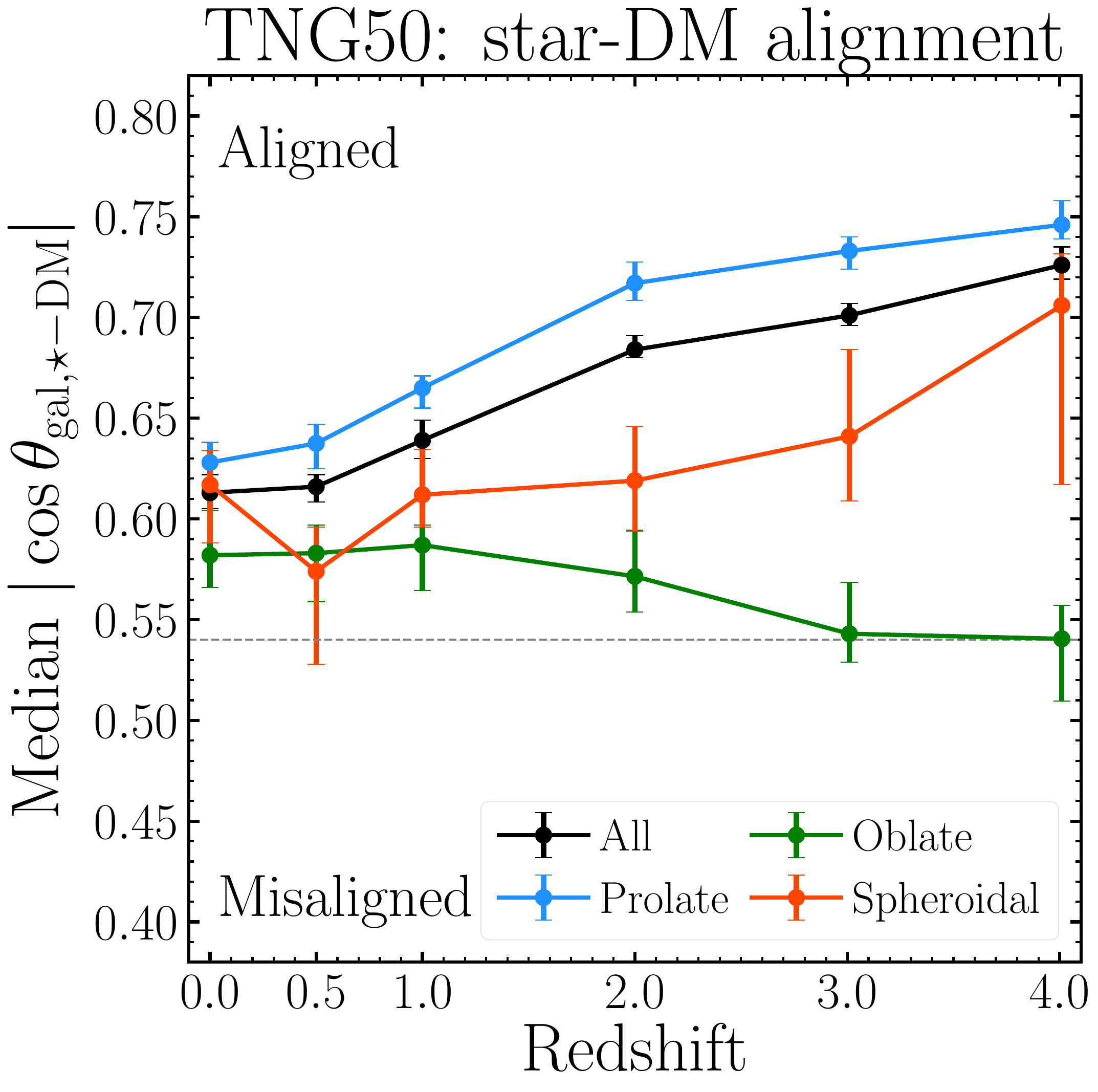}{0.38\textwidth}{}
}
\vspace{-25pt}
\caption{
Evolution of the median alignment angle between stellar (at 0.1{\rdm}) and DM (at 2{\rdm}) shapes shown for all stellar shapes (black), as well as separated by shape. 
}
\label{fig:medangle_starvdm}
\end{figure}

Finally, we measure the alignment between the major axes of stars and outer DM halos to better understand 
the geometrical relationship between matter accreted from the cosmic web and that closer to galaxy centers.
In Figure~\ref{fig:medangle_starvdm}, we present the evolution of the median angle between the major axes of DM at 2{\rdm} and stars at 0.1{\rdm}. All galaxies are shown in black, and those with prolate, oblate, and spheroidal stellar shapes are shown in blue, green, and orange, respectively.

\begin{figure*}[htbp]
\vspace{-10pt}
\centering
\gridline{
\fig{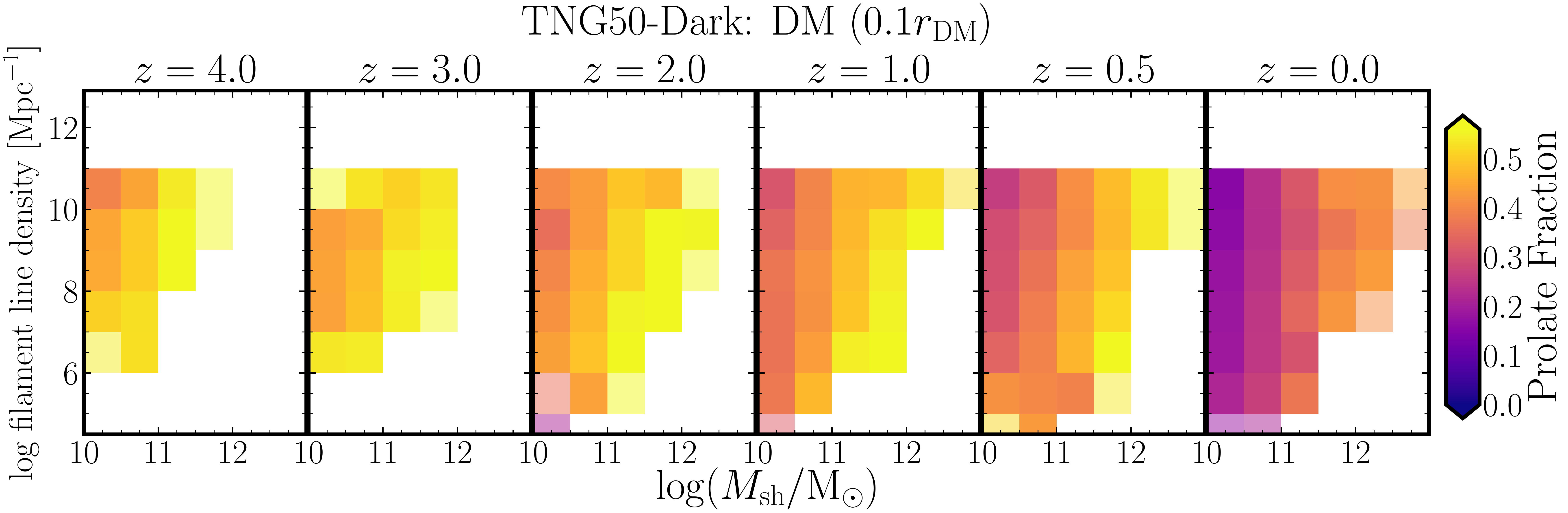}{0.75\textwidth}{}
}
\vspace{-25pt}
\gridline{
\fig{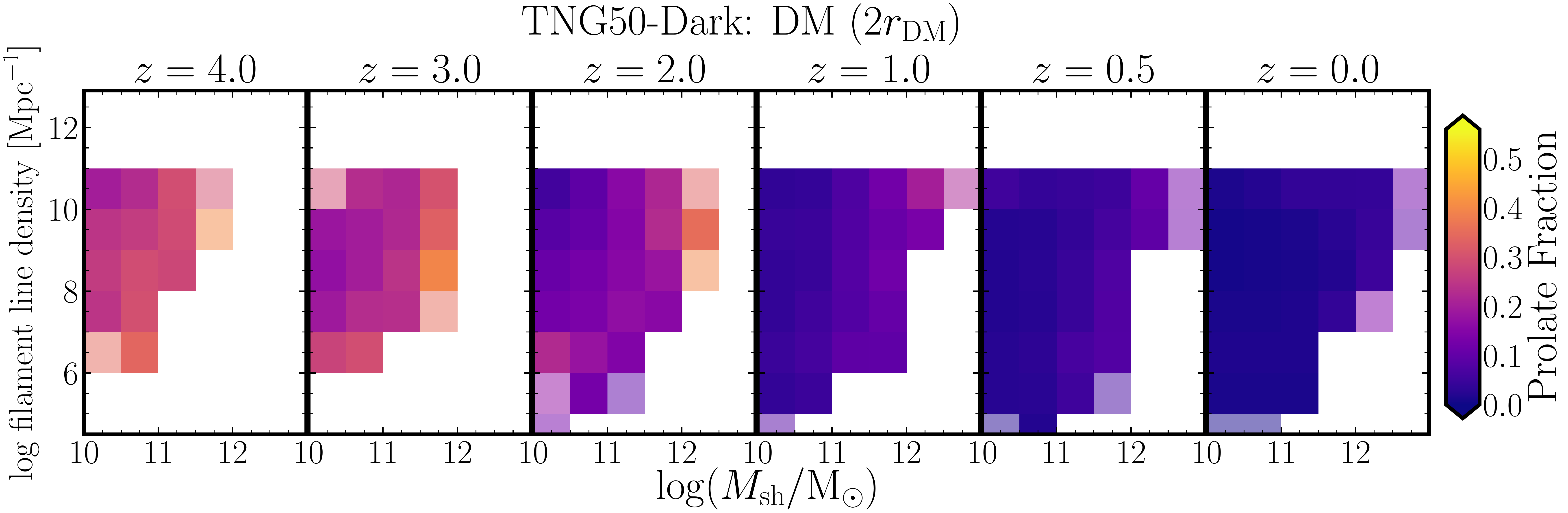}{0.75\textwidth}{}
}
\vspace{-25pt}
\caption{
Fraction of prolate subhalos (color-bar) as a function of filament line density ($y$-axis) and subhalo DM mass ($x$-axis) at different redshifts in TNG50-Dark. The top and bottom rows represent the shape measured within 0.1{\rdm} (inner halo) and 2{\rdm} (outer halo), respectively. In this DM-only simulation, higher-mass halos (especially inner halos) are more prolate.
}
\label{fig:pfrac-dark}
\end{figure*}

On average, stellar shapes are well-aligned with outer DM halo shapes, but this alignment decreases considerably with time.
At almost all redshifts, prolate stellar structures exhibit the strongest shape alignment with outer DM halos, while oblates show the weakest shape alignment.
The strong alignment of prolate stellar bodies with outer halos agrees with the findings of zoom-in hydrodynamical simulations \citep{ceverino15,tomassetti16}. We discuss this further in \S~\ref{sec:physical}.

\subsection{Influence of filaments on 3D shapes}
\label{sec:3dshape}

We now study how both the filament line density and mass affect salient morphological characteristics of galaxies and halos at different redshifts.

\subsubsection{Prolate fraction vs. mass and filament density}
\label{sec:pfrac}

\begin{figure*}[htbp]
\vspace{-10pt}
\centering
\gridline{
\fig{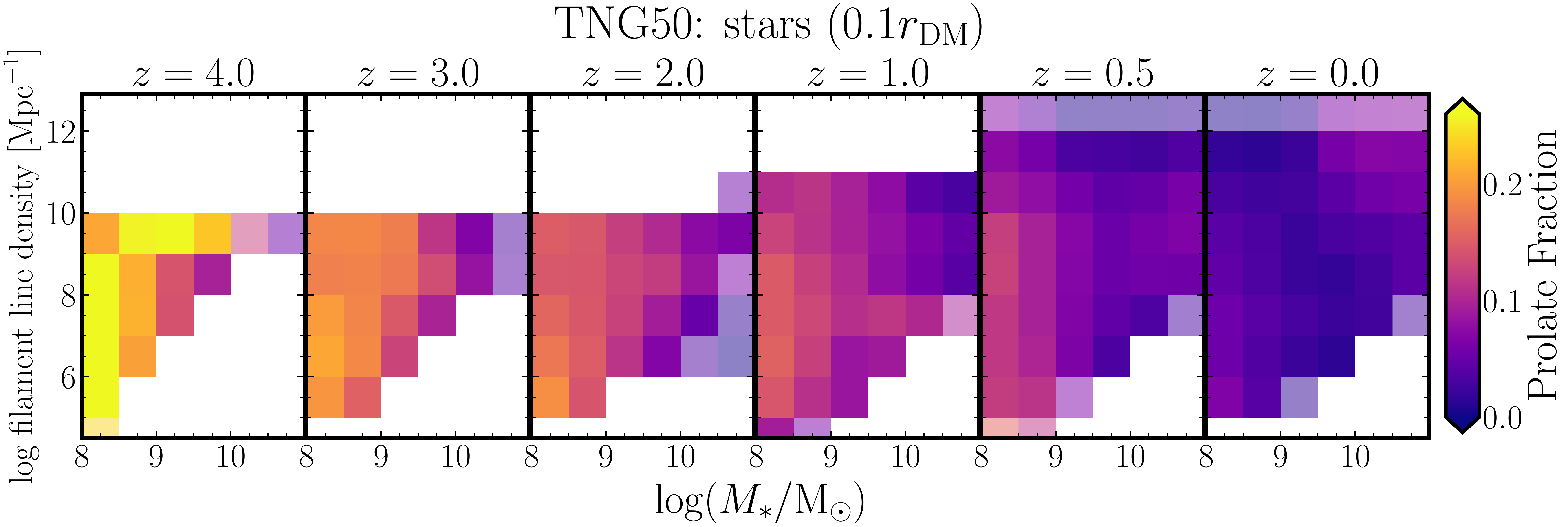}{0.8\textwidth}{}
}
\vspace{-25pt}
\gridline{
\fig{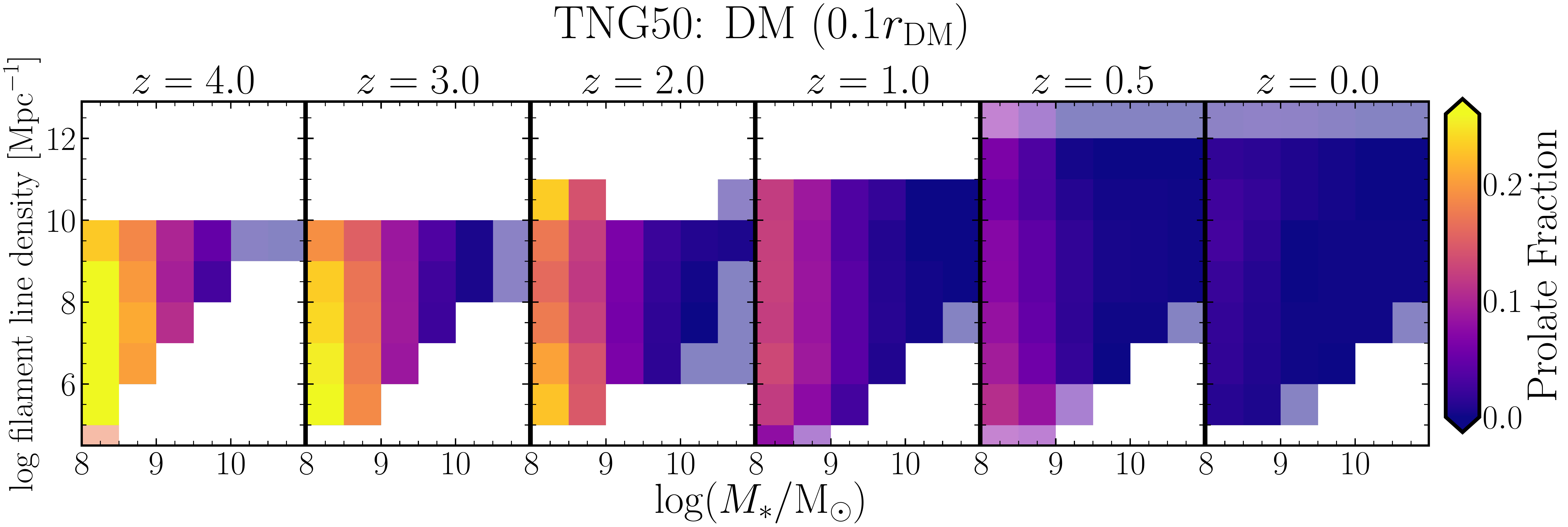}{0.8\textwidth}{}
}
\vspace{-25pt}
\gridline{
\fig{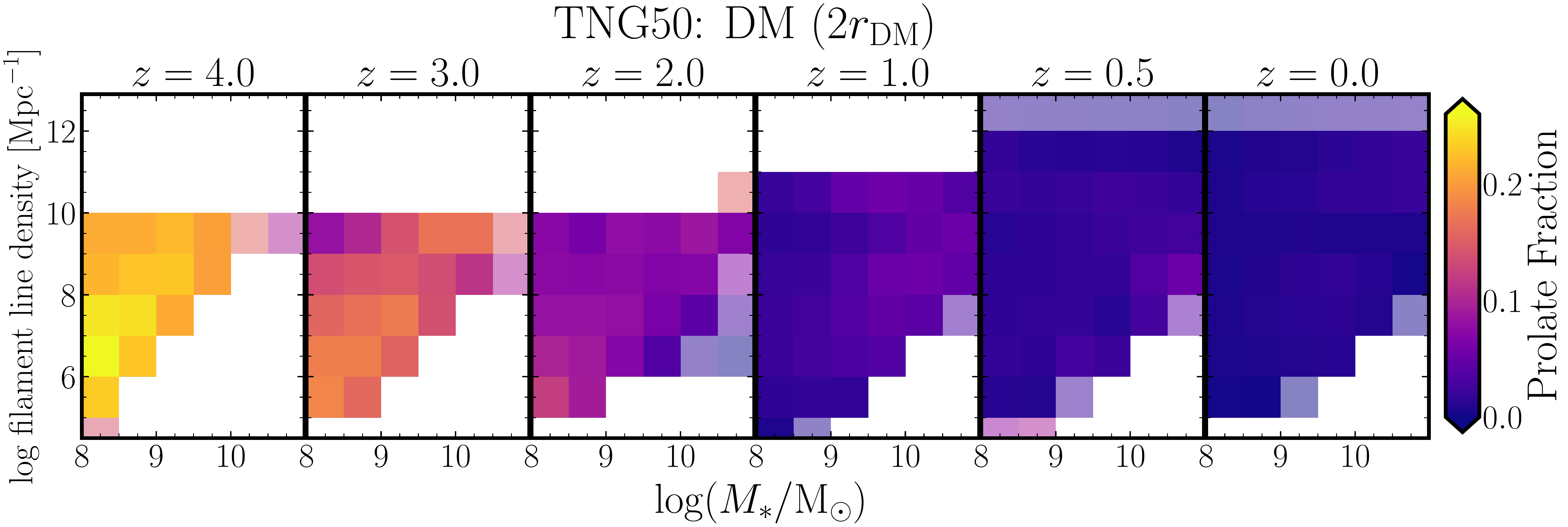}{0.8\textwidth}{}
}
\vspace{-25pt}
\caption{
The fraction of galaxies with prolate stellar structures (top row), inner halos (middle row), and outer halos (bottom row) as a function of filament density, stellar mass, and redshift in TNG50.
Lower-mass galaxies and (primarily inner) halos are more prolate, but at fixed mass, early low-mass structures are more likely to be prolate at lower filament density.
}
\label{fig:pfrac}
\end{figure*}

Given the expectation that prolate galaxy formation is likely influenced by cosmic web accretion, we first compute the fraction of prolate galaxies as a function of both stellar mass and the density of the nearest filament at six different redshift snapshots.
As in the previous section, we begin with the DM-only TNG50-Dark simulation to interpret the effects of gravity alone. In Figure~\ref{fig:pfrac-dark}, we present the prolate fraction (prolate over total population: color-bar) of subhalos in a given bin of subhalo mass $M_{\mathrm{sh}}$ ($x$-axis) and nearest filament line density ($y$-axis) at redshifts $z=4$, 3, 2, 1, 0.5, and 0 (each a different column). The top (bottom) row shows the prolate fraction for DM measured within 0.1{\rdm} (2{\rdm}).
To give a sense of the number of objects in each 2D bin, the color is opaque if the number in that bin is greater than 20 and semi-transparent if not.

In TNG50-Dark, the fraction of prolate inner halos increases with increasing mass at all redshifts, and the same is true for outer halos at $z\gtrsim1$. 
Inner halos have much higher (in some cases by a factor $>$2) prolate fractions at fixed mass than outer halos at any given redshift, which implies substantial elongation of halo cores relative to their outskirts at all times. 
Most outer halos are spheroidal at $z<1$, regardless of mass. 
We also notice a slight drop in the prolate fraction with increasing filament density at fixed mass. This is more prominent for inner halos and higher redshifts ($z\geq1$).

We now repeat the experiment with TNG50. In Figure~\ref{fig:pfrac}, we present the stellar mass and filament density-dependence of the prolate fraction of stars at 0.1{\rdm} (top row), DM at 0.1{\rdm} (middle row), and DM at 2{\rdm} (bottom row) at six different redshifts.
We note a stark and fundamental difference between the prolate fractions in TNG50 and TNG50-Dark: in the former, prolate fractions decrease with increasing stellar mass. 
Prolate stellar and DM shapes are mostly seen at high redshifts ($z\geq1$) and low masses (${\logms}<9$).
With the inclusion of baryons, galaxies and halos become less prolate with increased mass assembly -- exactly opposite to the effect of mass assembly due to gravity alone.

We again observe a small secondary dependence of the prolate fraction on the density of the nearest filament. 
At fixed stellar mass, there is a small (up to $\approx$10\%) decrease in the prolate fraction of stars and DM halos with increasing filament line density, especially for ${\logms}\lesssim10$ galaxies at $z\geq1$. This implies that low-density filaments may help form some low-mass prolate galaxies at early times. This trend is stronger for inner halos and stars than for outer halos.

\begin{figure*}[htbp]
\vspace{-10pt}
\centering
\gridline{
\fig{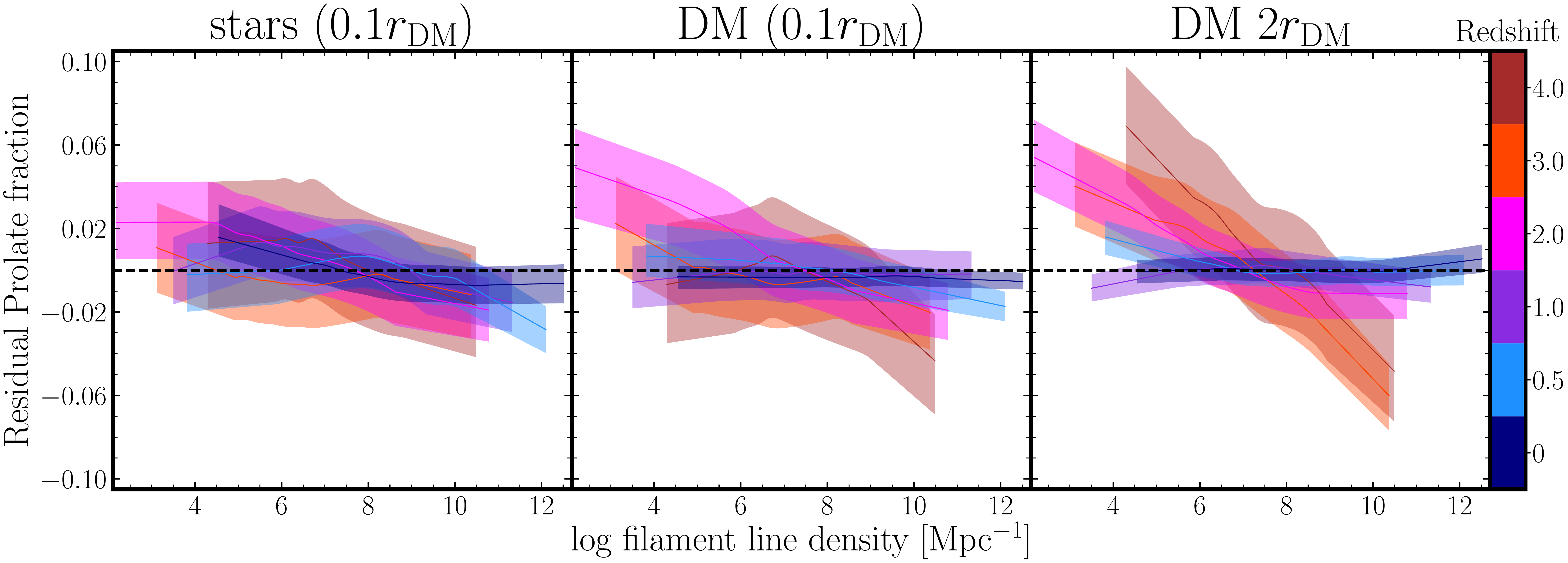}{0.99\textwidth}{}
}
\vspace{-25pt}
\gridline{
\fig{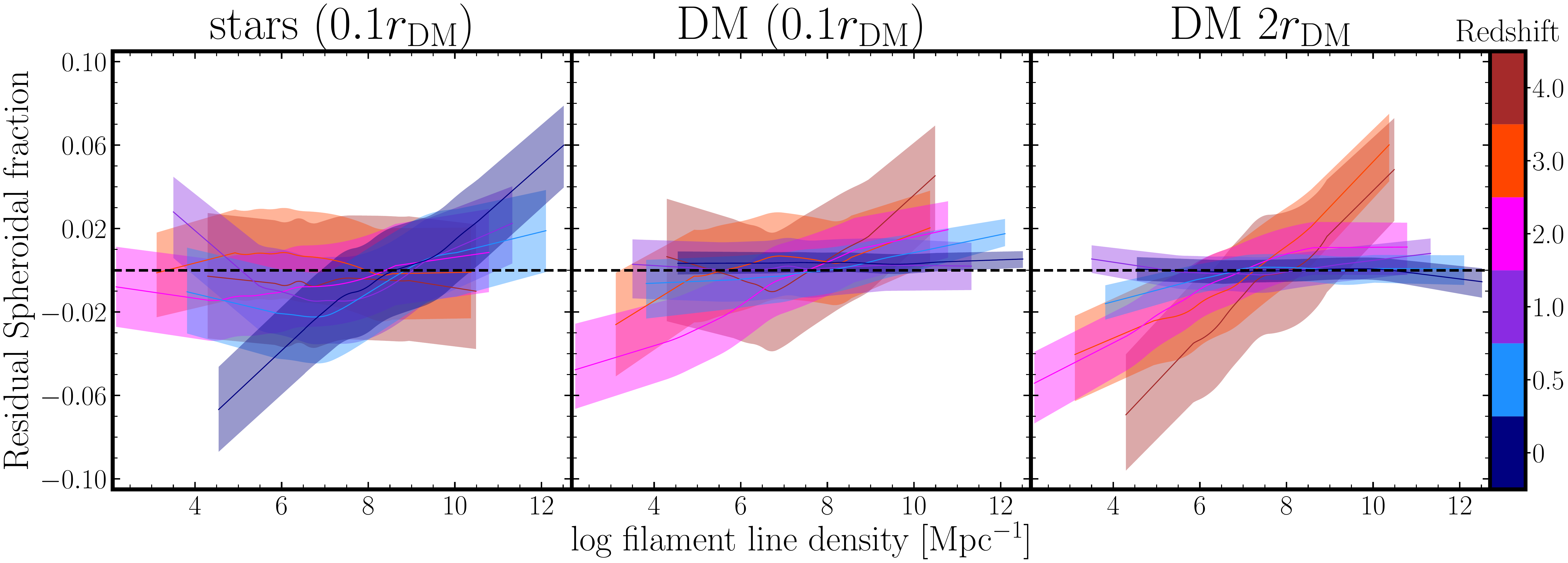}{0.715\textwidth}{}
\raisebox{2pt}{\fig{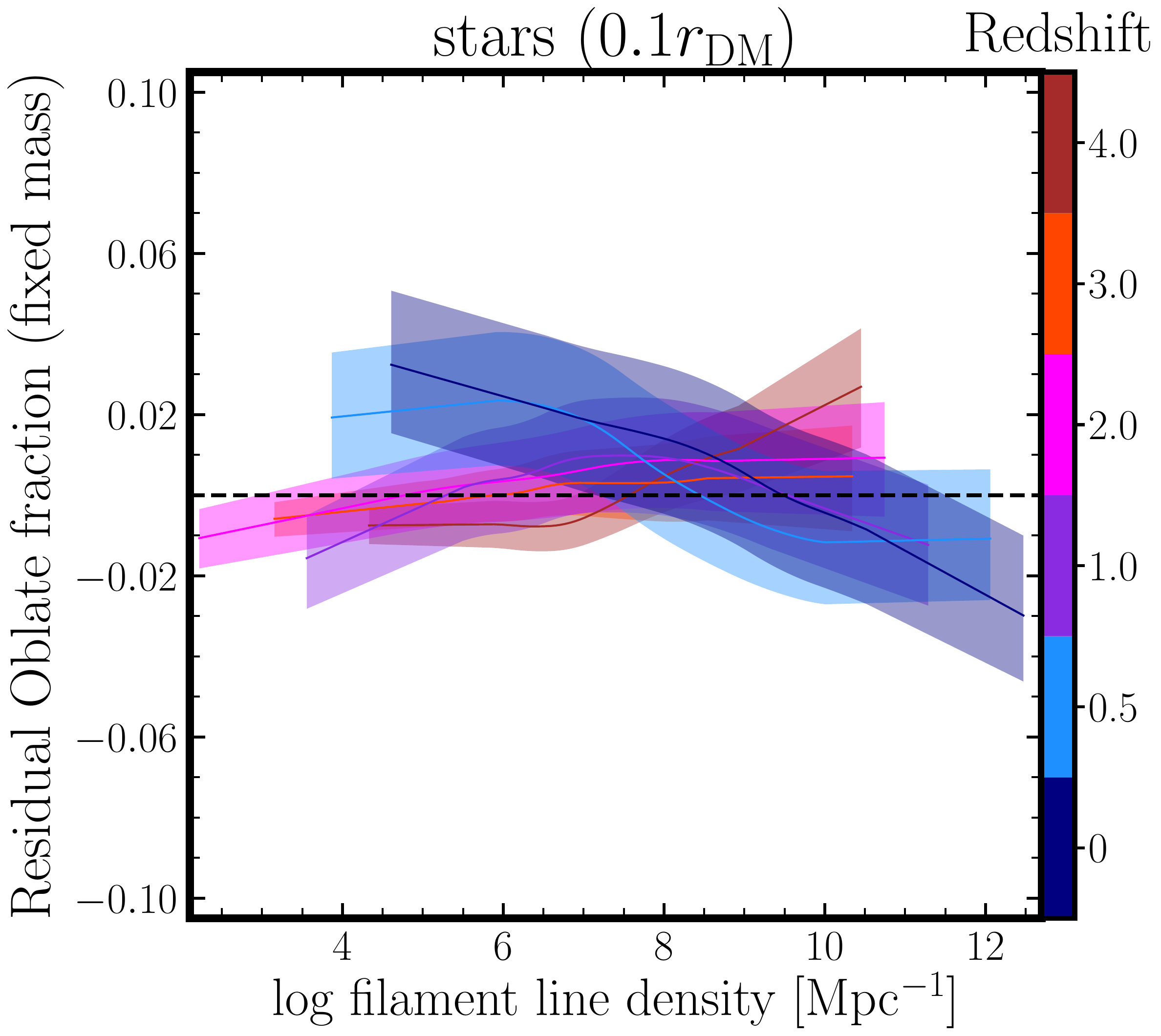}{0.275\textwidth}{}}
}
\vspace{-25pt}
\caption{
Residual morphological fractions at fixed stellar mass, shown as functions of the nearest filament density. Each color is a different redshift, with the median and $\pm1\sigma$ uncertainties being represented by solid lines and lighter shaded region, respectively.
The top row presents the residual prolate fractions for stars, inner DM halo, and outer DM halo in the different columns. 
The bottom row presents the residual spheroidal fractions for stars, inner DM halo, and outer DM halo in the left three columns, and the residual oblate fraction of stars in the right column (DM halos are rarely oblate so we omit those panels for the residual oblate fraction). 
For stars, the residual prolate fraction is higher in low-density filaments at early times, and the residual spheroidal (oblate) fraction is higher in high-density (low-density) filaments at later times.
For halos, the residual prolate fraction decreases and the spheroidal fraction increases with filament density at $z>1$. 
}
\label{fig:resfrac}
\end{figure*}

Interestingly, we find that there are somewhat higher prolate fractions in stars than DM at low redshifts ($z\leq1$), which is unexpected. This includes a small population of high-mass (${\logms}>10$) prolates at $z<1$, which are not seen in the real Universe \citep[e.g.,][]{vanderwel14,zhang19,Pandya24} or expected from zoom-in simulations, which naturally produce high prolate fractions at early times \citep[e.g.,][]{ceverino15,tomassetti16}. One possibility for these galaxies is that they are irregularly shaped (likely due to mergers), i.e., their stellar shapes cannot be approximated as triaxial ellipsoids. Thus, our shape classification criteria (see \S~\ref{sec:measure}) cannot properly classify them, and they are instead identified as prolates. Visual investigation of the stars and gas in a few of these candidates shows that this may indeed be the case. 
Another possibility is that they were formed by dry mergers, i.e., mergers of gas-poor galaxies which do not trigger starbursts (unlike wet mergers) and produce slowly rotating structures. Indeed, dry mergers were found to produce a population of massive prolates at $z=0$ in the Illustris simulation by \citet{Li18}.
Detailed studies of the origin of such galaxies with irregular shapes may prove valuable, but this is beyond the scope of this paper.

\subsubsection{Residual dependence of shapes on filament density}
\label{sec:resfrac}

It is clear that stellar mass plays a dominant role in setting morphological structures, but a subtle dependence of filament density at fixed mass is also observed.
Here, we aim to remove the mass dependence and measure any remaining residual dependence of the 3D shape on the density of the nearest filament.
Therefore, we extract the external effect of cosmic web filaments on these morphological properties \textit{beyond} the internal effects parameterized by mass. While there are many possible approaches to this, we consider the residuals of a given parameter with respect to the stellar mass \citep[somewhat similar to the approach of][]{CroneOdekon:2018aa}.

Briefly, our method is as follows: we use the stellar prolate fraction as an example.
First, for a given redshift, we estimate the median stellar prolate fraction as a function of stellar mass, which we denote as the expected stellar prolate fraction. 
Then, for the population at large, we compute the deviation from this expected stellar prolate fraction in a few bins of nearest filament line density\footnote{Our filament density bins are chosen such that they have roughly equal numbers of galaxies.}, which we denote as the ``residual'' stellar prolate fraction at fixed mass. 
This allows us to take into account and separate the effect arising from the mass alone and determine the residual effect of the filament line density on the prolate fraction. 
Finally, we apply the Savitzky-Golay filter \citep{SG64} to smooth the binned residual prolate fraction-filament density relation and its $\pm1\sigma$ bootstrapped uncertainties (with 1000 resamples). The smoothing helps us remove random stochastic variations in the data and obtain a simple qualitative description of the residual dependence of the filament density on the prolate fraction.

\begin{figure*}[htbp]
\centering
\gridline{
\fig{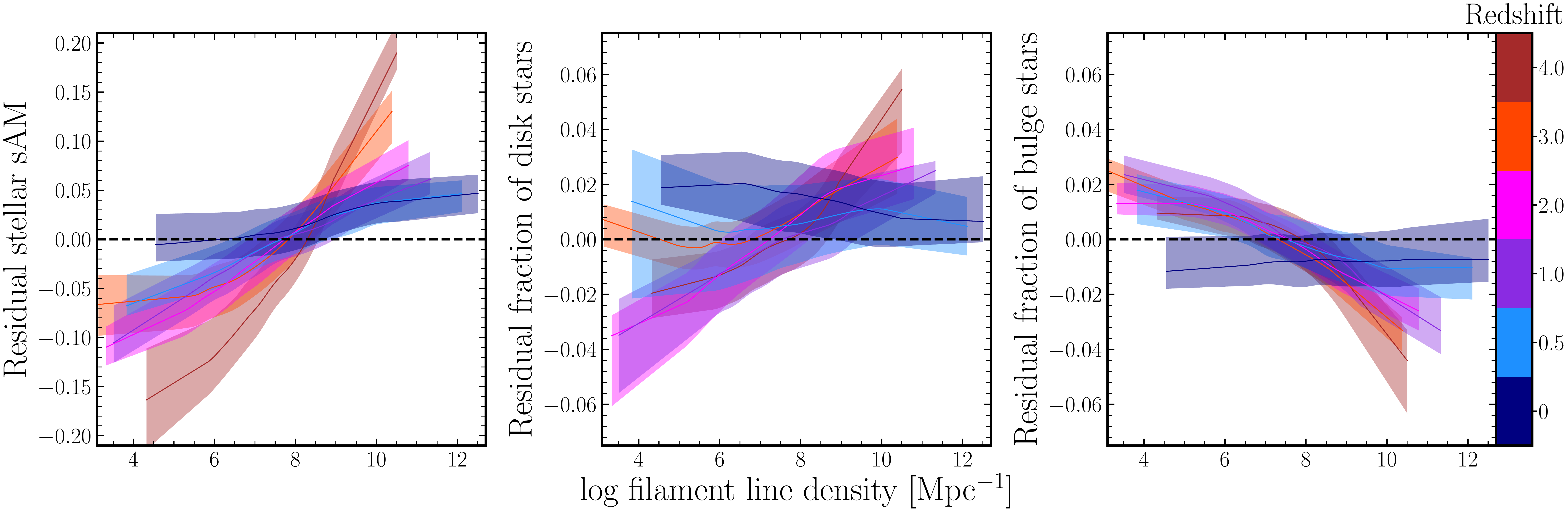}{0.98\textwidth}{}
}
\vspace{-25pt}
\caption{
The residual effect of filament density on stellar kinematic morphological properties at fixed mass.
The three panels show the residual stellar sAM (left), fraction of stars with disk orbits (middle), and fraction of stars with bulge orbits (right). The much larger vertical range for the left panel reflects the much stronger residual dependence of stellar sAM on the filament density. The bulge and disk fractions have a smaller residual dependence on the filament density.
}
\label{fig:reskin}
\end{figure*}

We follow this procedure for the prolate, spheroidal, and oblate fractions of stars at 0.1{\rdm}, and the prolate and spheroidal fractions of DM at 0.1{\rdm} and 2{\rdm}.
Figure~\ref{fig:resfrac} presents the residual prolate (top row) and spheroidal and oblate (bottom row) fractions as functions of the nearest filament density at different redshifts in TNG50. The solid lines and shaded regions represent the smoothed medians and $\pm1\sigma$ bootstrapped uncertainties therein.

As we have already qualitatively inferred, there is a residual dependence, at fixed mass, of the prolate fraction of stars and DM halos on filament density.
At $z>1$, the residual prolate fraction of both inner and outer DM halos increases with decreasing filament density, while the residual stellar prolate fraction rises less steeply in lower density filaments. 
Quantitatively, the residual prolate fraction decreases by up to $\sim5$\% for stars, $\sim10$\% for inner halos, and $\sim15$\% for outer halos, from the lowest to the highest density filaments. 
However, the filament density-dependence of the prolate fraction vanishes at $z\leq1$.

Both the spheroidal and oblate fractions show a residual filament density dependence at fixed mass.
For stars, the residual spheroidal fraction rises strongly with increasing filament density at $z<0$, by $\approx$15\% from the lowest to the highest density filaments.
The variation in this fraction is up to $\sim10\%$ for inner halos, and $\sim15\%$ for outer halos.
In contrast, the residual oblate fraction of stars is higher in lower density filaments at $z\leq0.5$ (by up to $\sim$10\%).

Finally, we briefly comment on the total morphological fractions in TNG50 at a given redshift.
Across all redshifts, fewer spheroidal and more oblate stellar structures are formed with increasing redshift. The majority of low-mass (${\logms}8-9$) galaxies at almost all redshifts are spheroidal, but spheroidal shapes are increasingly rare in ${\logms}\gtrsim10$ galaxies ($\lesssim$40\%) at $z\leq2$. The majority of ${\logms}\gtrsim10$ galaxies at $z\leq2$ are instead oblate. Conversely, hardly any ${\logms}=8-9$ oblate galaxies exist at $z\geq2$, and only a handful of such galaxies exist at lower redshifts. Prolate shapes are always subdominant at $z\leq4$ at any masses above ${\logms}=8$. These results, identical to those first presented by \citet{Pillepich19}, confirm the predominant morphological transformation of galaxies in TNG50 from mostly spheroids to increasing disk-dominance with increasing time and mass assembly,

\subsection{Influence of filaments on kinematic morphologies}

\label{sec:kin}

\begin{figure*}[htbp]
\vspace{-10pt}
\centering
\gridline{
\fig{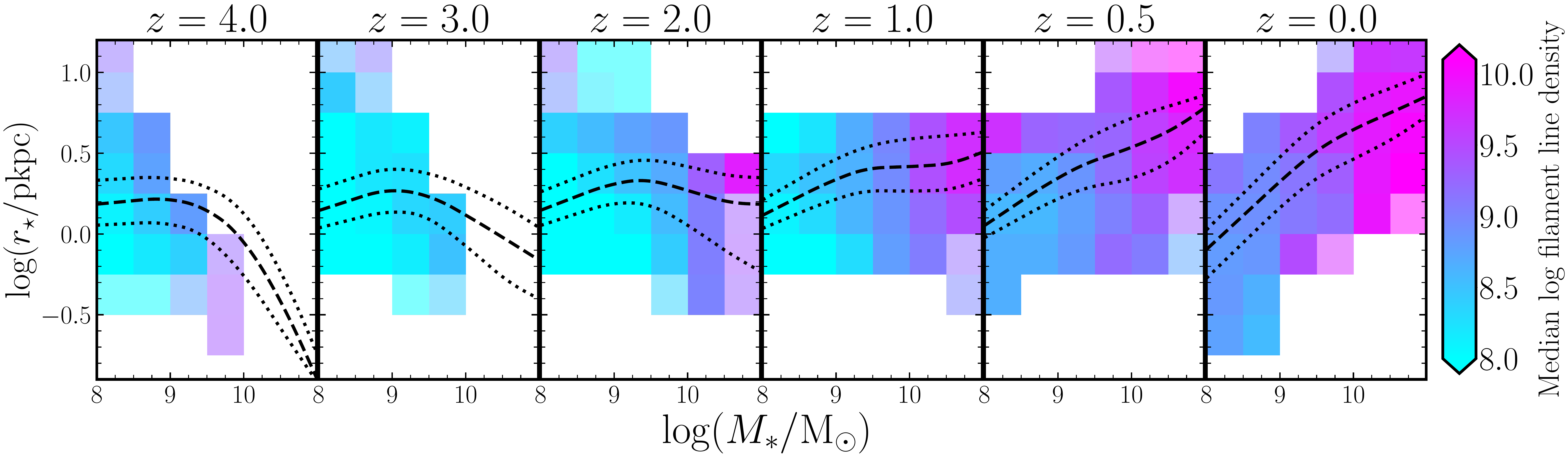}{0.9\textwidth}{}
}
\vspace{-25pt}
\gridline{
\fig{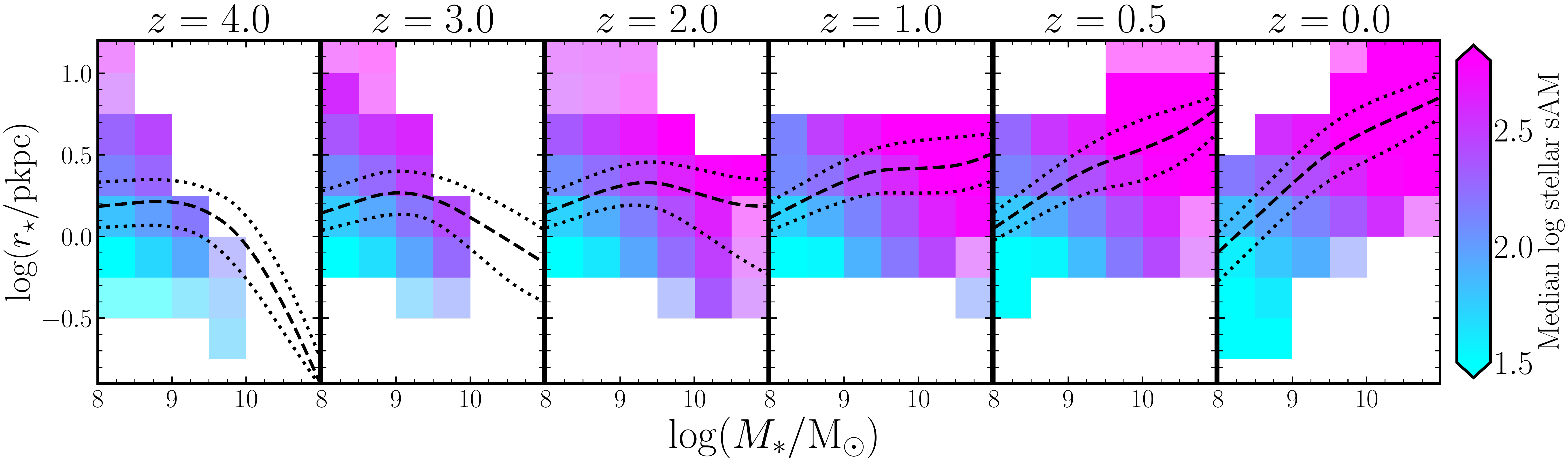}{0.9\textwidth}{}
}
\vspace{-25pt} 
\caption{
Galaxies in stellar mass-size space at different redshifts, colored by the median nearest filament density (top row) and the median stellar sAM (bottom row). The dashed and dotted curves correspond to the median relationship and \nth{25} and \nth{75} percentiles, respectively. More extended galaxies typically have higher spin and at $z\geq0.5$ live in higher-density filaments.
}
\label{fig:masssize}
\end{figure*}

Beyond the 3D shape, galaxy morphologies can also be characterized by various kinematic properties of their stellar populations. Here, we investigate the specific angular momentum (sAM) of stars and the fraction of stars in disk-like and bulge-like orbits in TNG50, as defined by \citet{Genel15}. We study how filaments may influence these quantities and stellar mass.
Similar to the previous section, we quantify the residual dependence of these kinematic morphological parameters on the nearest filament density after removing the stellar mass dependence. 
Here, we compute, for each galaxy, the deviation of a given parameter (e.g., stellar sAM) from the median value given its stellar mass.  
Afterwards, we compute the median of this deviation as a function of the nearest filament density, as before.

In Figure~\ref{fig:reskin}, we present the residual stellar sAM (left), the fraction of disk stars (middle), and the fraction of bulge stars (right) as functions of the filament line density at six different redshifts (each represented by a different color). 
The residual effect of filament density is far stronger on stellar sAM than on the other quantities at most redshifts, for which the vertical scale on the left panel is $\sim$3 times larger in dynamic range than the other panels. 
The residual stellar sAM increases significantly with increasing filament density at all times, but this dependence weakens over time.  
The variation in the residual sAM from the lowest to the highest-density filaments goes from $\sim$40\% at $z=4$ to $\sim$7\% at $z=0$.

The neighboring filament density has a smaller residual effect on the fraction of bulge and disk stars. The residual disk fraction increases with filament density by up to $\sim$10\% at $z>1$, but this relationship flattens at $z<1$.
The residual bulge fraction drops with increasing filament density by $\lesssim$7\% at $z\geq0.5$, and flattens out at later times. 
We note that the definition of ``bulge-like'' orbits only excludes stars in the disk, but not stars in more irregular or boxy orbits typical of prolate structures. Hence, it is more appropriate to think of the bulge fraction as the fraction of \textit{velocity dispersion-dominated} stars. 
Similarly, the disk fraction can be interpreted loosely as the fraction of rotation-dominated stars.

\subsection{Filaments and galaxy sizes and spins}
\label{sec:size}

Here, we investigate the sizes of galaxies to further complete our physical picture of morphological evolution as it is affected by the cosmic web. We determine the stellar mass-size relations at each redshift and investigate the average stellar spins and filamentary environments of galaxies in this parameter space. 
In the top row of Figure~\ref{fig:masssize}, we present the median density of the nearest filament color-coded as a function of stellar mass within twice the half-mass radius and half-mass radius, in six different redshifts. In the bottom row, we color each {\ms}-{\rstar} bin instead by the median stellar sAM. The median and \nth{25} and \nth{75} percentiles of the relationships are also indicated in each panel. As in Figures~\ref{fig:pfrac-dark} and \ref{fig:pfrac}, we reduce the transparency in bins with fewer than 20 galaxies.

A strong dependence of stellar size on stellar sAM is found, as is a noticeable dependence on the density of the nearest filament, particularly at high redshifts ($z\gtrsim1$). 
The stellar spin is, on average, much higher for galaxies larger than the median size for a given mass (i.e., extended galaxies) than for galaxies below the median relation (i.e., compact galaxies). 
This is a consequence of  galaxies retaining the original angular momentum they acquired from their parent halos, resulting in a tight relation between sAM and size \citep[e.g.,][]{Genel15}. 
This helps form extended stellar disks with ordered rotation, while galaxies that lose most of their halo angular momentum instead form more compact spheroidal structures. This is true at all times. We also find a $z\geq1$ trend of extended galaxies residing in somewhat higher density filaments than more compact galaxies.  
We further discuss our results and their physical implications in \S~\ref{sec:discuss}.


\section{Predictions of the Cosmic Web Observable by Roman}
\label{sec:pred}

The predictions presented in this paper on the fundamental influences of the cosmic web on galaxy formation can be tested with new and upcoming observatories. 
Most of our results highlight a more prominent dependence of filaments on galaxy morphological properties at earlier times than at later times, which necessitates higher redshift observations to test our most salient predictions.

\begin{figure*}[htbp]
\vspace{-10pt}
\centering
\gridline{\fig{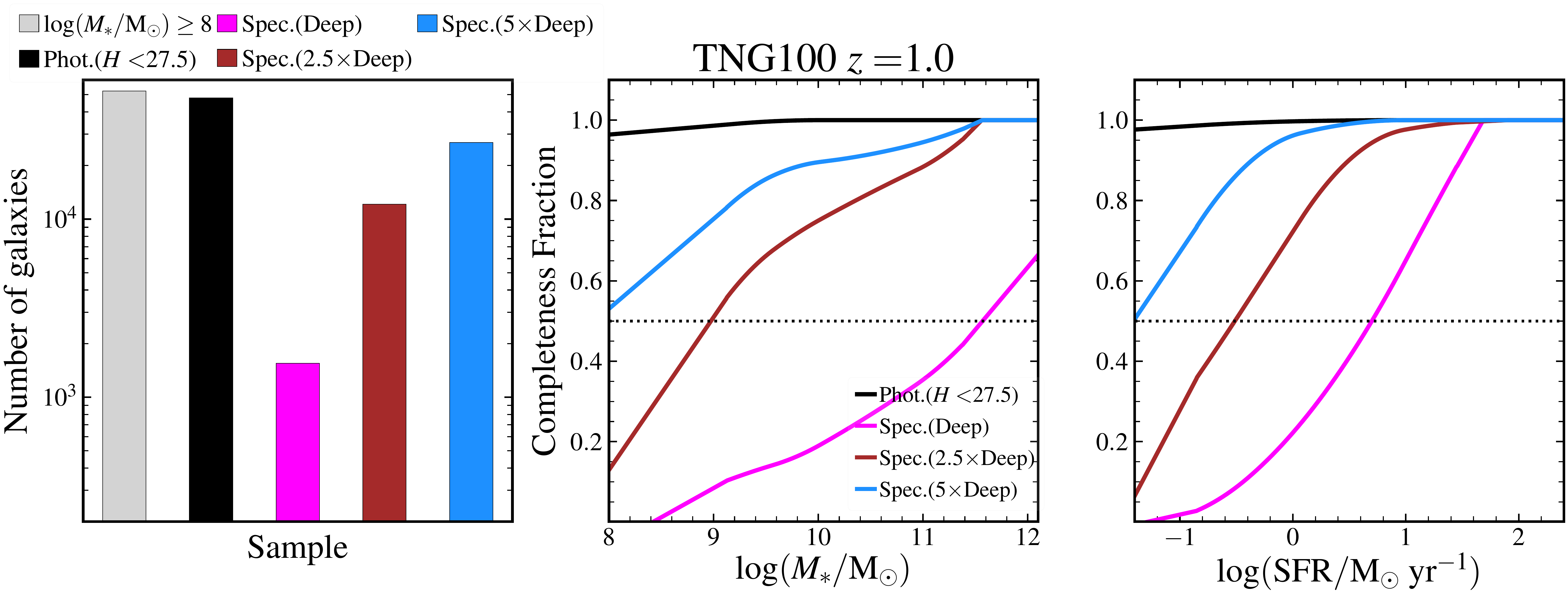}{0.9\textwidth}{}
}
\vspace{-25pt}
\gridline{\fig{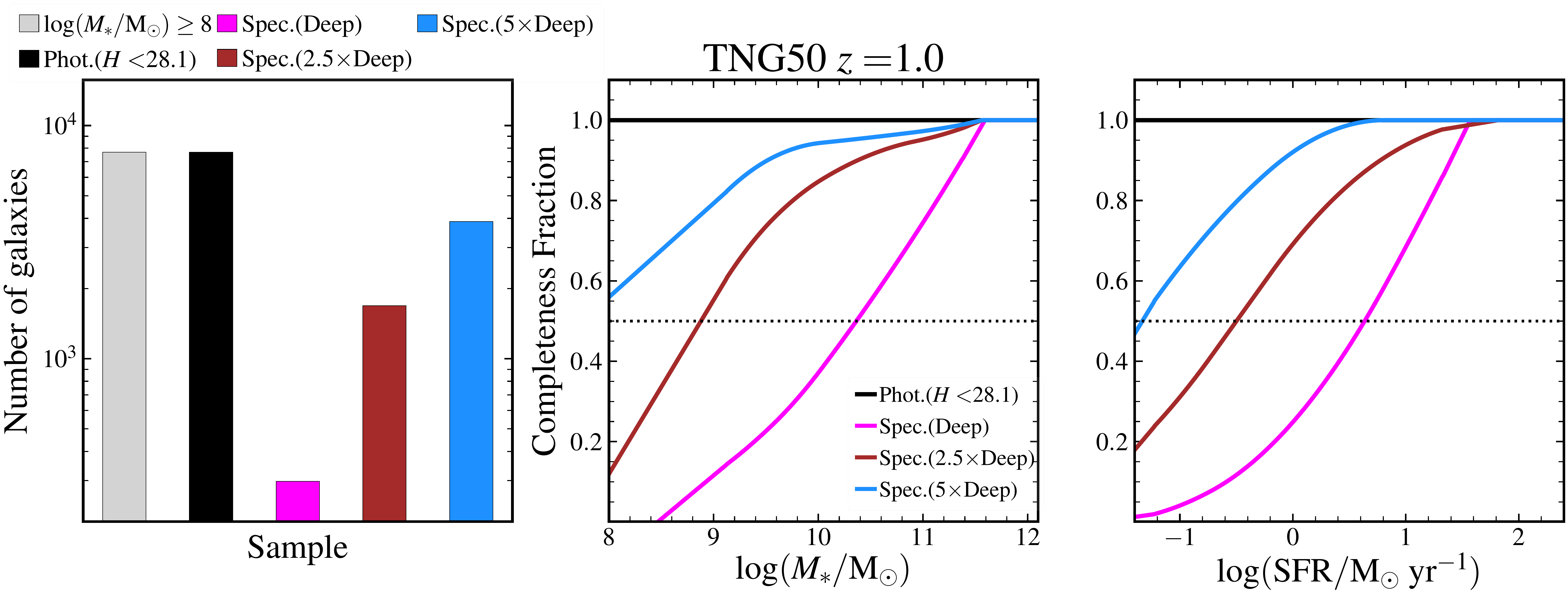}{0.9\textwidth}{}
}
\vspace{-25pt}
\caption{
Properties of different galaxy samples at $z=1$ in TNG100 (top row; comparable to the size of HLWAS-Deep) and TNG50 (bottom row; comparable to HLWAS-Ultra deep). 
The samples are as follows: gray = full sample of ${\logms}\geq8$ galaxies, black = photometric sample with mock $H$-band magnitude $m_H<27.5$ (top) and $m_H<28.1$ (bottom), magenta, brown, blue =  samples satisfying both the photometric cut and a minimum SFR cut from spectroscopic emission line flux limits corresponding to 1$\times$, 2.5$\times$, and 5$\times$ the depth of HLWAS-Deep, respectively.
\textit{Left}: Total number of galaxies in a given sample; 
\textit{Middle} and \textit{Right}: Completeness fraction as a function of mass and SFR. Very deep spectroscopy is needed to obtain fairly complete galaxy samples with which to trace the cosmic web.
}
\label{fig:mock_gal}
\end{figure*}

Unfortunately, a critical gap exists in current reconstructions of the cosmic web at earlier times; at $z>0.5$, current observations are very limited due to several reasons. Most existing wide-field galaxy surveys are conducted from the ground, for which there are high UV and IR backgrounds. This makes it very difficult to probe very faint magnitudes that are necessary to sample higher $z$ galaxies to reasonable completeness levels. The ubiquitous and well-known optical emission lines used to identify and accurately characterize the redshifts of galaxies such as {\Ha}, {\Hb}, and {\OIIIf}, are shifted to NIR bands at redshifts significantly greater than 0.5. For example, due to its r-band magnitude limit $m_r<17.77$~mag, SDSS rapidly loses completeness at ${\logms}\gtrsim10$ at $z\gtrsim0.1$. Thus, only the rarer massive galaxies are accessible for cosmic web reconstruction at even intermediate redshifts.
Deeper surveys over smaller areas have shown some potential for $z>0.5$ reconstruction; for instance, the VIPERS survey \citep{Guzzo14} at $z\sim0.7$ by \citet{Malavasi17}. But even deep ground-based surveys have often been limited to fairly bright galaxies (in addition to smaller volumes).

With the launch of space-based wide-field observatories, the landscape for high-$z$ cosmic web reconstructions is changing dramatically. 
In particular, the early data from Euclid was recently released \citep{Euclid25_overview}, and Roman, NASA's next great observatory, is scheduled for launch in less than a year.
Both of these observatories will provide the high-resolution imaging and spectroscopy required to (1) trace the early cosmic web with large galaxy samples and (2) constrain how the morphological properties of galaxies are shaped by the cosmic web.
However, we focus on Roman in this paper for a few reasons \citep[see also][]{Schlieder24,EuclidNISP,ROTAC25}. Roman (1) is designed to be more sensitive than Euclid in both imaging and spectroscopy of emission-line sources; (2) has a higher grism spectral resolution than Euclid; (3) has a slightly higher imaging spatial resolution than that of Euclid, and (4) will conduct core community surveys whose data will be immediately available to the public.

In this section, we predict the cosmic web structure that can be traced by the deepest wide-field galaxy surveys that Roman will conduct.
We undertake this task by generating ``mock'' galaxy samples from the deep observations planned with Roman's High-Latitude Wide-Area Survey \citep[HLWAS;][]{Wang:2022_romanHLSS}. The goal is to understand both what some of the planned observational depths will allow us to learn, and the effect of depth in accurately reconstructing the high-$z$ cosmic web (focusing on $z=1$). These will lead us to predict how accurately we can constrain the galaxy morphology-cosmic web connection from real data.

\subsection{Modeling the cosmic web with ``mock'' surveys}
\label{sec:mocksamp}

We begin by selecting samples of galaxies based on the design of Roman's HLWAS and the capabilities of its Wide-field instrument (WFI). Whereas the ``Wide'' and ``Medium'' tiers of the HLWAS, with survey areas $\sim$2700 and $\sim$2415~{\sqdeg} respectively, will enable transformative cosmological analyses of the nature of DM, dark energy, etc., we focus our attention on the much smaller and deeper ``Deep'' and ``Ultra-deep'' tiers, over $\sim$19 and $\sim$5~{\sqdeg} areas, respectively (the Ultra-deep field is contained within the Deep field). The Roman Time Allocation Committee report recommends 
8-band imaging and grism spectroscopy for the ``Deep'' tier, and additional deeper 3-band imaging for the ``Ultra-deep'' tier \citep{ROTAC25}.
However, the ``Ultra-deep'' field will not have deeper spectroscopy than the ``Deep'' field.
The much larger ``medium'' and ``wide'' tiers would only find the most massive galaxies that are vigorously star-forming, for reasons described below.

To map the high-$z$ cosmic web with high accuracy and completeness, one needs highly complete samples of galaxies that are representative of the complete galaxy population as tracers of the cosmic web. 
This is difficult due to the biases introduced by various selection methods. Photometric selection of galaxies leads to a higher likelihood of observing brighter and, therefore, more massive galaxies. Spectroscopic selection, typically using emission line features, biases samples towards higher (recent) star formation activity.

\begin{deluxetable*}{cccccc}
\tablecaption{Mock galaxy samples (in Figure~\ref{fig:mock_gal}) and their characteristics\label{tab:sample}}
\tablehead{
\colhead{Length (Mpc)} &
\colhead{Area ({\sqdeg})} &
\colhead{Sample definition} &
\colhead{No. of galaxies} &
\colhead{$\sim$50\% {\logms}\tablenotemark{a}} &
\colhead{$\sim$50\% {\logsfr}\tablenotemark{a}}
}
\startdata
 &  & ${\logms}\geq8$ & 52496 & $<8.0$  & $-3.0$ \\
 &  & $H<27.5$~mag & 47879 & $<8.0$ & $-2.0$ \\
110.7 & 14.0 & $\tablenotemark{b} ~f_{\Ha}\geq5.80\times10^{-17}$ & 1552  & 11.5 & 0.7 \\
 &  & $~\tablenotemark{b}f_{\Ha}\geq2.32\times10^{-17}$ & 12131 & 9.1  & $-0.5$ \\
 & & $\tablenotemark{b}~f_{\Ha}\geq1.16\times10^{-17}$ & 26954 & 8.4  & $-1.4$ \\
\hline 
  &   & ${\logms}\geq8$ & 7696 & $<8.0$  & $-3.0$ \\
  &   & $H<28.1$~mag   & 7693 & $<8.0$ & $-3.0$ \\
51.7  & 3.1  & $\tablenotemark{b}~f_{\Ha}\geq5.80\times10^{-17}$ & 298   & 10.3 & 0.6 \\
  &   & $\tablenotemark{b}~f_{\Ha}\geq2.32\times10^{-17}$ & 1687  & 8.9  & $-0.5$ \\
  &   & $\tablenotemark{b}~f_{\Ha}\geq2.32\times10^{-17}$ & 3872  & 8.3  & $-1.3$ \\ 
\enddata
\tablenotetext{a}{\scriptsize 
Approximate 50\% completeness limit ($<$ indicates upper limit).}
\vspace{-5pt}
\tablenotetext{b}{\scriptsize In addition to the H-band magnitude cut above. Fluxes are in units of {\ergscm}.}
\vspace{-20pt}
\end{deluxetable*}

\begin{figure*}[htbp]
\vspace{-10pt}
\centering
\gridline{\fig{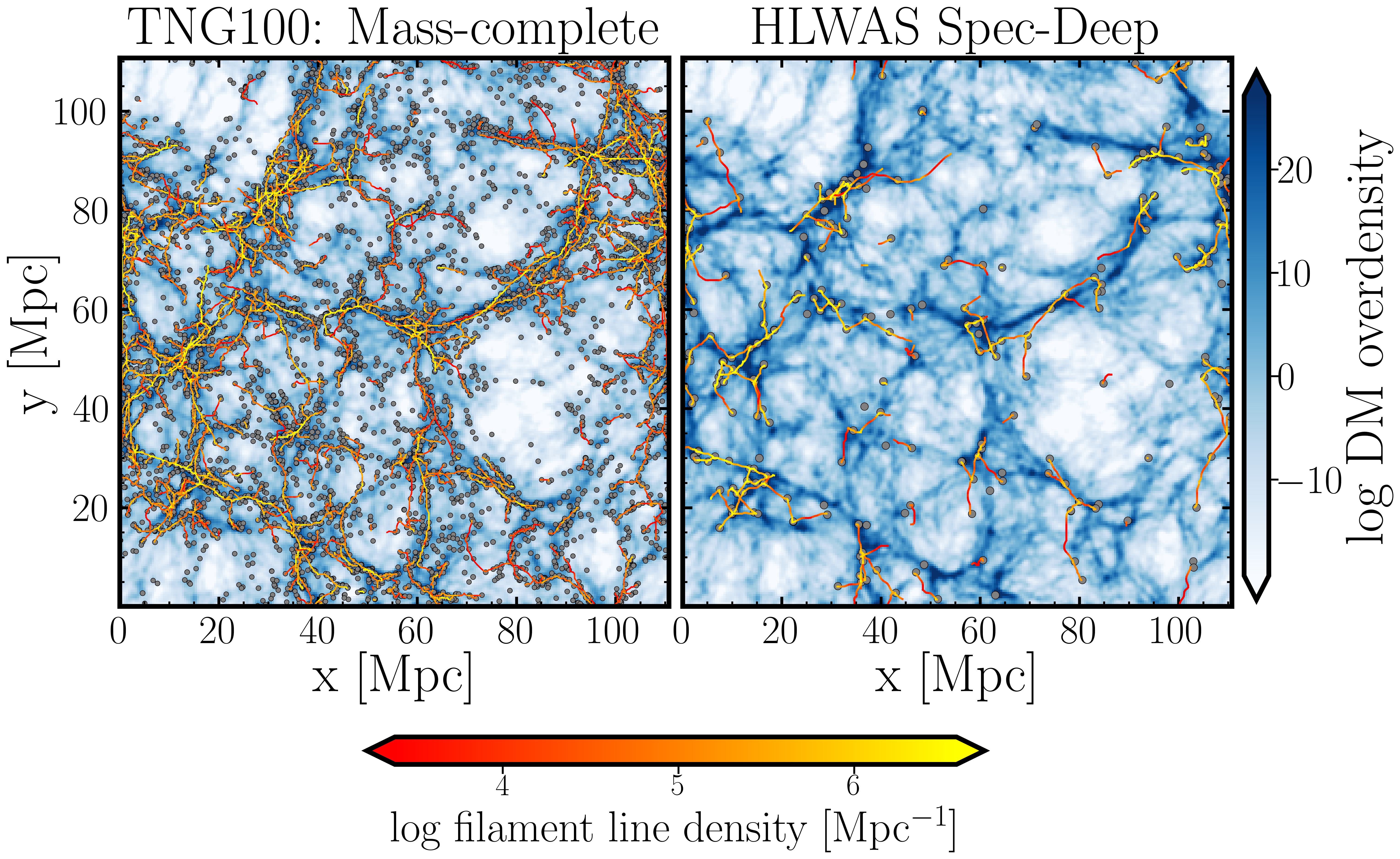}{0.8\textwidth}{}
}
\vspace{-25pt}
\gridline{\fig{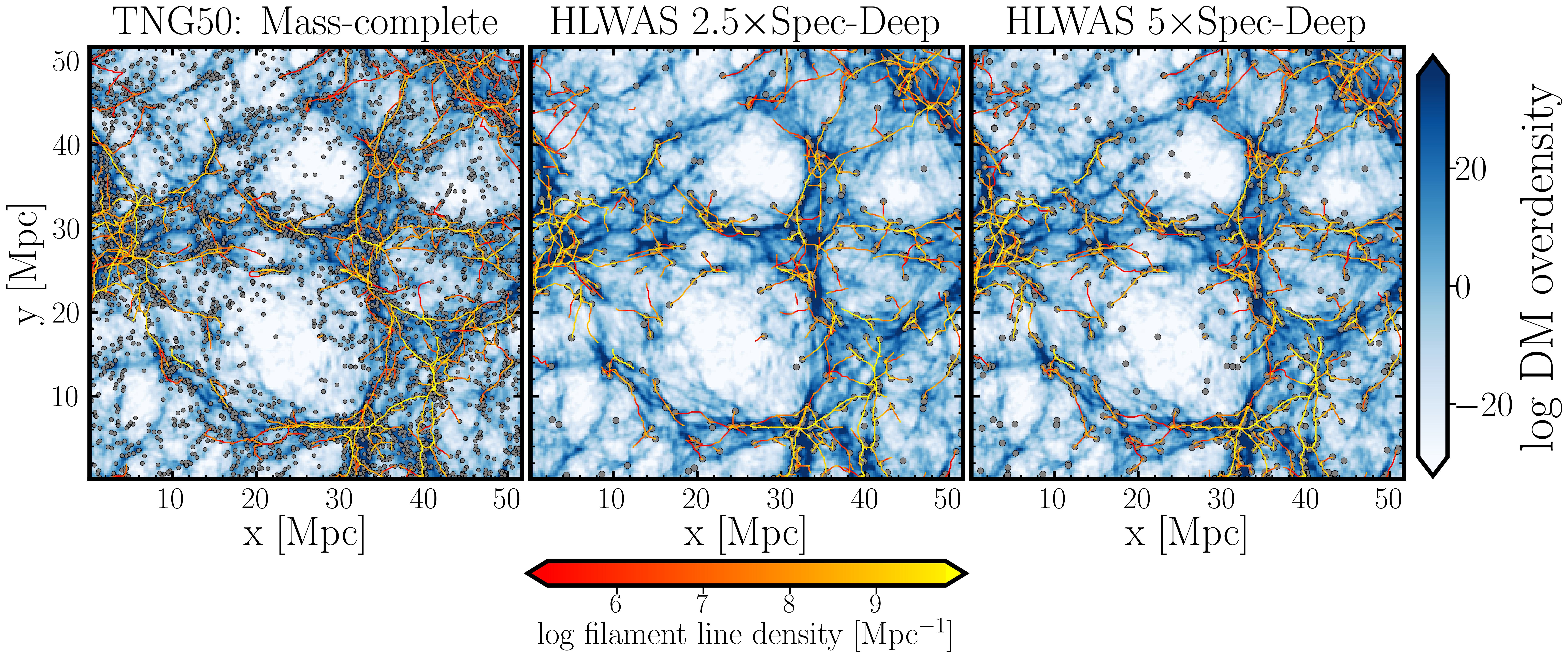}{0.99\textwidth}{}
}
\vspace{-25pt}
\caption{
2D projections, in 20 Mpc thick slices, of the cosmic web identified at $z=1$ from different galaxy samples.
\textit{Top row}: The mass-complete TNG100 sample (left) vs. a mock sample with HLWAS-Deep photometric and spectroscopic depths (right);
\textit{Bottom row}: The mass-complete TNG50 sample (left) vs. mock samples with both HLWAS-Ultra deep photometric depth and $2.5\times$HLWAS-Deep spectroscopic depth (middle) and $5\times$HLWAS-Deep (right). 
Filaments, galaxies, and DM overdensity are denoted similar to Fig.~\ref{fig:vis}. The planned spectroscopic depths of the HLWAS will only allow partial cosmic web reconstruction; deeper spectroscopy is essential.
}
\label{fig:vis_mock}
\end{figure*}

We generate ``mock'' HLWAS Deep and Ultra-deep galaxy samples using TNG100 and TNG50.
\citet{ROTAC25} recommended a limiting 5$\sigma$ $H$-band AB magnitude $H\!< \!27.5$~mag ($H\!<\!28.1$~mag) for the HLWAS Deep (Ultra-deep) survey, and a 5$\sigma$ spectroscopic line sensitivity limit of $f=5.8\!\times\!10^{-17}$~{\ergscm} for the HLWAS Deep survey. The HLWAS Ultra-deep tier does not have deeper spectroscopy planned than HLWAS Deep. 
We assume that galaxies will be identified (1) in photometry in the $H$-band, and (2) in spectroscopy with the commonly-observed {\Hawave} emission line. The {\Ha} line is typically the strongest and one of the most easily observable emission lines \citep[e.g.,][]{KE12}, and it would be accessible by the wavelengths observable by Roman at $z=1$. The criterion for (2) implies that spectroscopic redshifts are available for all galaxies to precisely pinpoint their 3D location.

We describe our methodology to estimate mock Roman magnitudes and spectroscopic line flux limits in detail in Appendix~\ref{app:mock}, but briefly describe them here. We transform mock photometric magnitudes from existing magnitudes in the TNG catalogs \citep{TNGDR19} to the Roman $H$-band, or the F158W filter. For TNG100, we convert the mock $J$-band magnitudes to $H$-band using $J-H$ colors of galaxies observed at $z\sim1$ with HST \citep{Skelton14}. For TNG50, we transform mock $z$-band magnitudes to $H$-band using stellar population synthetic modeling \citep{Conroy09}. 
We also transform the observable spectroscopic {\Ha} flux limits to intrinsic SFR limits by scaling {\Ha} flux to SFR \citep{KE12}. For this, we apply a dust attenuation correction using an empirically-derived dust law \citep{Reddy15} and the observed dust content of $z\sim1$ galaxies \citep{Nedkova24}.
Finally, we do not consider the effects of line resolution from the Roman G150 grism -- whose resolution is $\sim$600 at the observed wavelength of {\Ha} at $z=1$ -- and therefore assume the ``best-case'' scenario for deriving a redshift for any given line.

In the rest of this section, we define {\ms} to be the stellar mass within 2{\rstar}, as this is more directly comparable to observations. We also adopt the average SFR over 10 Myr, as nebular emission lines are sensitive to star formation on these shorter timescales.
We construct eight different $z=1$ mock galaxy samples, whose characteristics are shown in Figure~\ref{fig:mock_gal}.

The top row presents predictions from TNG100, which has a similar survey area to HLWAS Deep with its $\sim$111 Mpc per side box, and the $\sim$51 Mpc-sized TNG50, whose survey area is comparable to HLWAS Ultra-deep, is shown on the bottom row. Aside from the ``full'' mass-complete sample containing all galaxies with ${\logms}\geq8$ (gray), we also construct (i)~a ``mock'' photometrically-selected sample corresponding to the appropriate $H$-band magnitude limit from \citet{ROTAC25} (black) and (ii)~samples that satisfy both the photometric cut and an additional spectroscopic cut. Specifically, these correspond to the planned {\Ha} emission line flux limits at the planned depth (magenta), $2.5\times$ the planned depth (brown), and $5\times$ the planned depth (blue) of HLWAS-Deep. Note that a factor of increase in depth corresponds to a decrease in the line flux limit by the inverse of that factor.
The left column of Figure~\ref{fig:mock_gal} shows the number of galaxies in the given sample, while the middle and right columns show the completeness fractions in {\logms} and SFR, respectively, for the mock samples relative to the mass-complete sample.
Horizontal dashed lines in the middle and right panels denote the 50\% completeness limits.

Figure~\ref{fig:mock_gal} shows that the selection of the sample has a significant effect on the inferred physical properties of the population. The photometric magnitude cuts select almost all galaxies for both TNG100 and TNG50; however, the spectroscopic cut dramatically reduces the number and completeness of galaxies. At the spectroscopic line flux depth of HLWAS-Deep, less than 5\% of the ${\logms}\!\geq\!8$ galaxies are detected in either TNG100 or TNG50. 
The completeness fractions of both {\ms} and SFR are very low -- below 50\% at ${\logms}\!<\!10$ and ${{\logsfr}\!<\!0.5}$ -- for this planned depth.

Increasing the spectroscopic line flux depth dramatically improves the completeness levels. The 50\% completeness limit in {\logms} is pushed to $\approx$1.5--2.5 ($\approx$2--3) dex lower with 2.5$\times$ (5$\times$) the spectroscopic depth of HLWAS-Deep. For {\logsfr}, this limit is pushed to $\approx$1 ($\approx$2) dex lower with 2.5$\times$ (5$\times$) HLWAS-Deep. 
By design, spectroscopic selection via emission lines biases samples towards high {\ms} and SFR, but increasing the line flux depth can help identify the much more common lower mass and less star-forming galaxies. Thus, deeper grism spectroscopic surveys lead to more complete samples that can help accurately trace the cosmic matter density and large-scale structure. 
A summary of the properties of the different galaxy samples, including the number of galaxies and completeness limits, is given in Table~\ref{tab:sample}.

We select certain galaxy subsamples from above as tracers of the cosmic web by applying our reconstruction approach outlined in \S~\ref{sec:cw}. Assuming we have well-characterized spectroscopic redshifts (spec-$z$) with negligible errors, as well as accurate stellar masses from spectral energy distribution (SED)-fitting to multi-band photometry, our mock galaxy samples would have robust 3D positions and stellar masses. We input these into the MCPM+{\disperse} pipeline for different realizations of the cosmic web at $z=1$. 
Each realization is run with the same parameters, including the {\disperse} persistence cut and the density field smoothing scale.

\subsection{Results of the mock cosmic web}
\label{sec:mockres}

\begin{figure*}[htbp]
\vspace{-10pt}
\centering
\gridline{
\fig{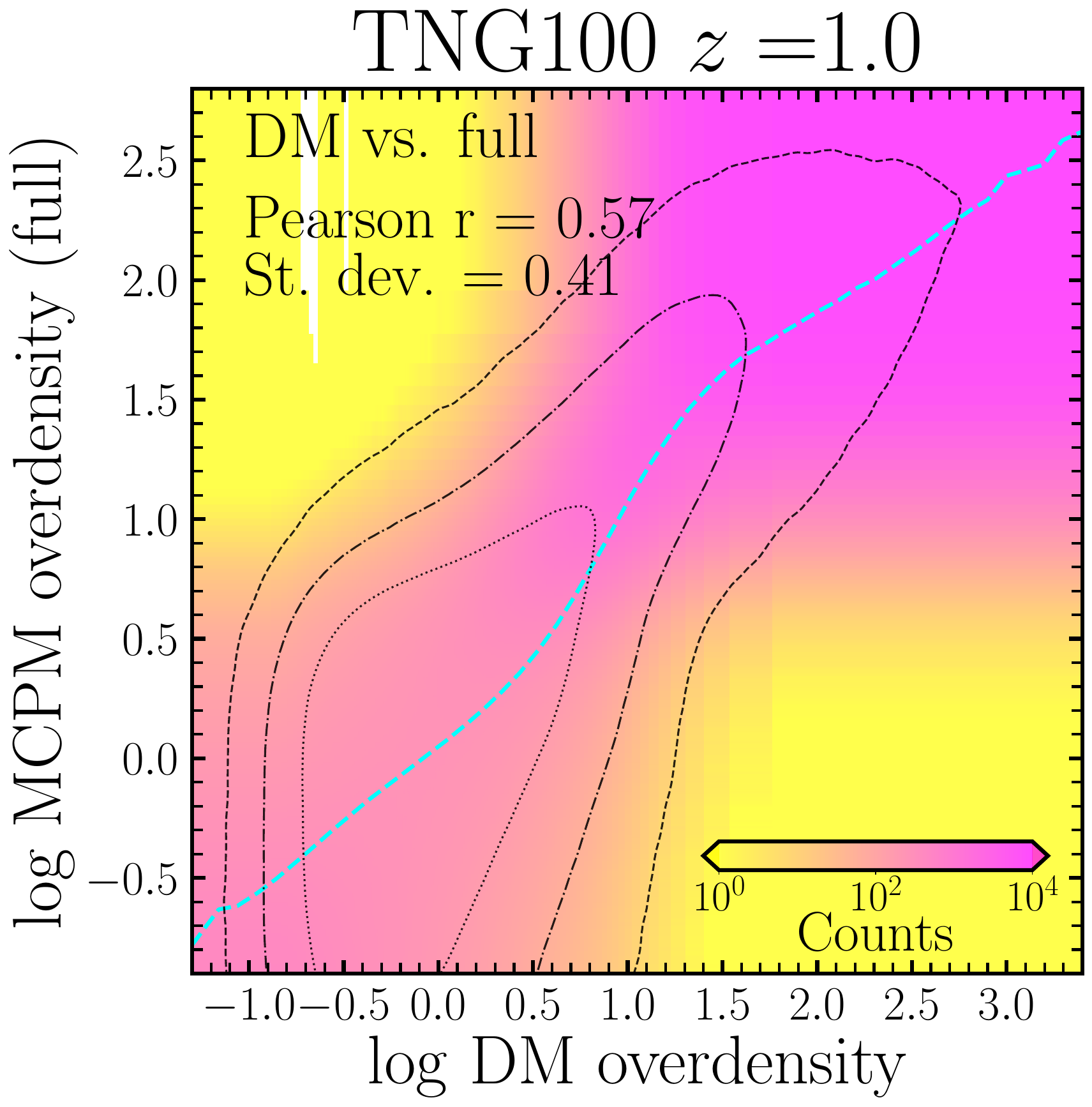}{0.3\textwidth}{}
\fig{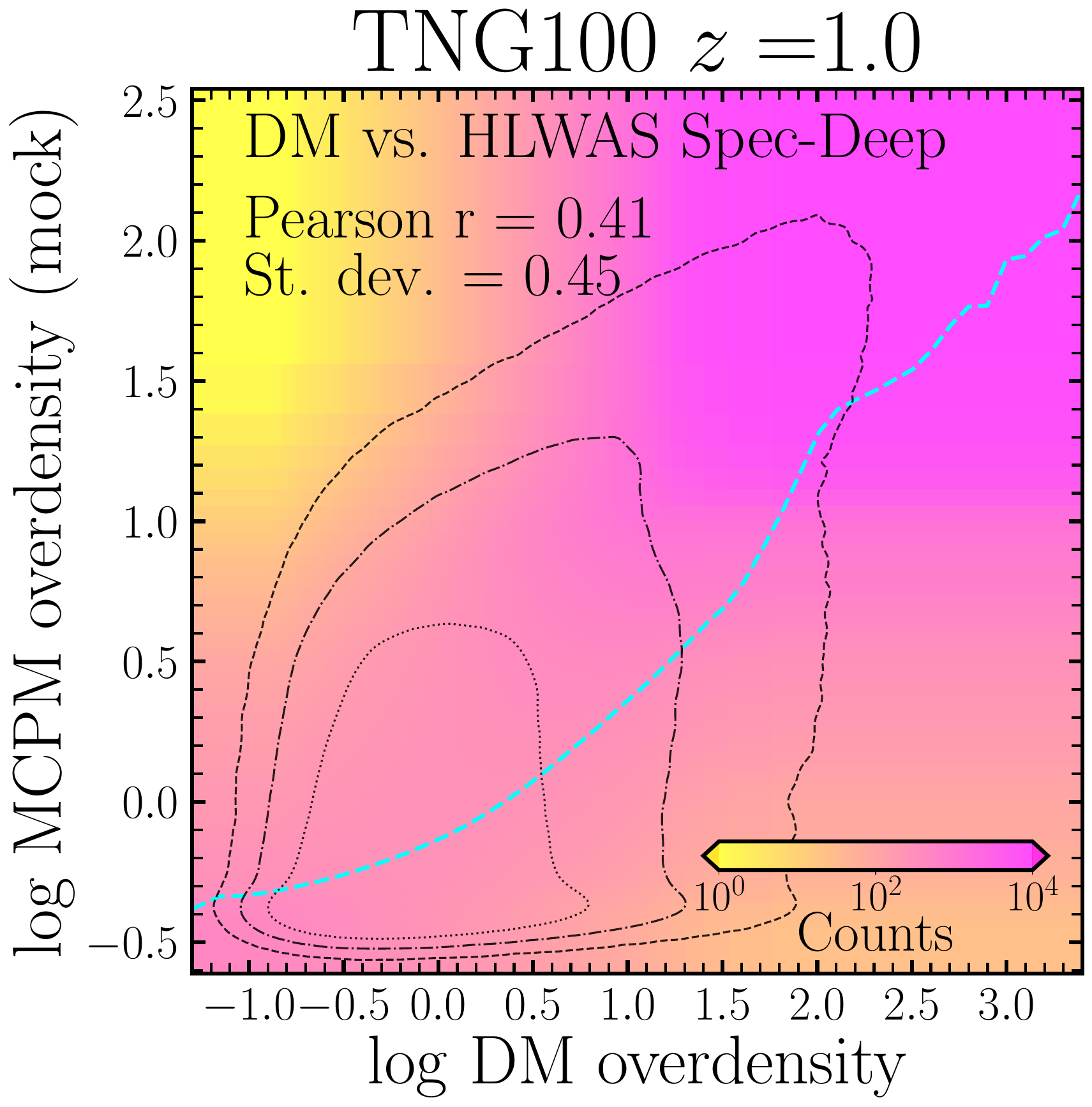}{0.3\textwidth}{}
}
\vspace{-25pt}
\gridline{
\fig{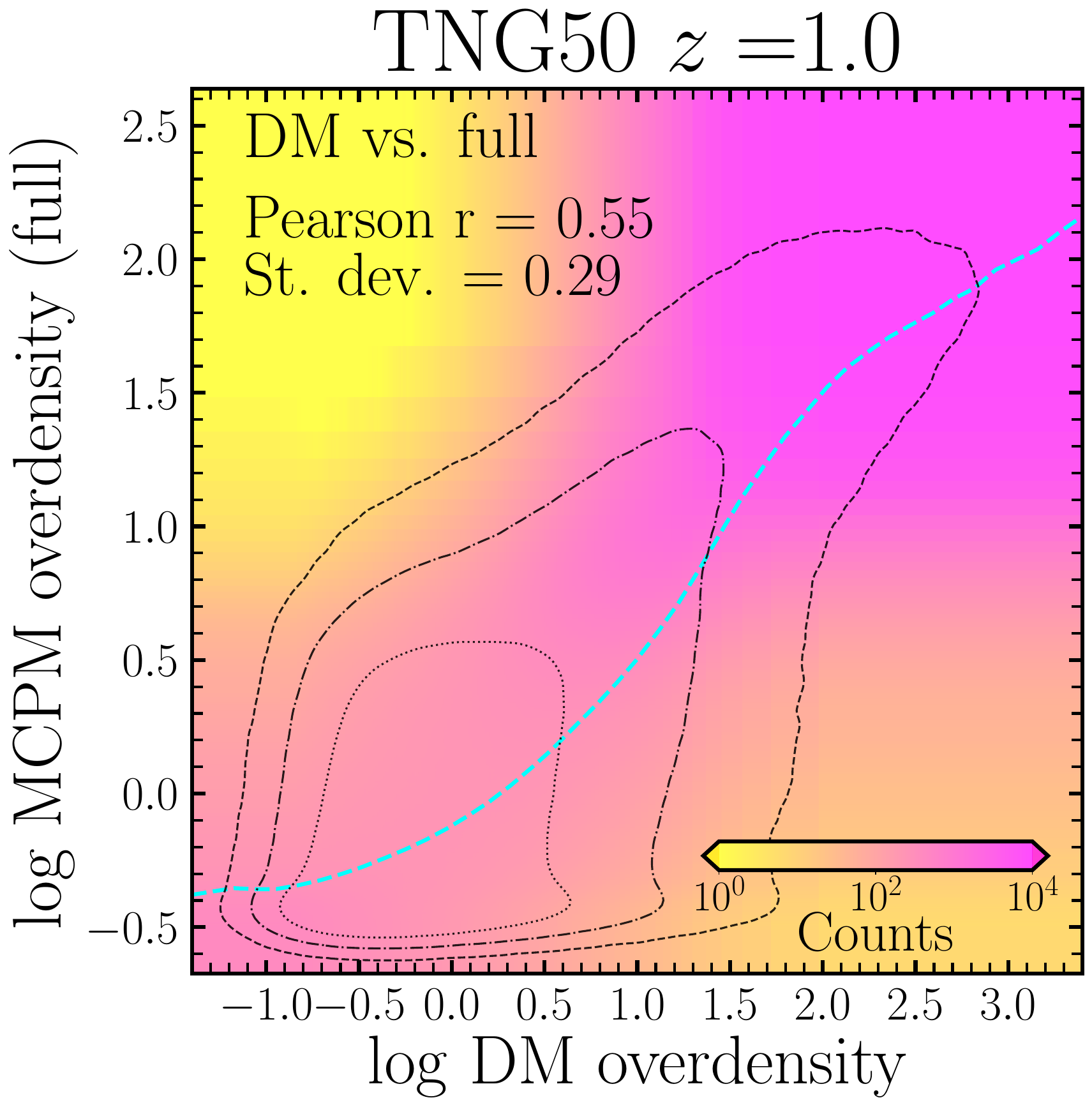}{0.3\textwidth}{}
\fig{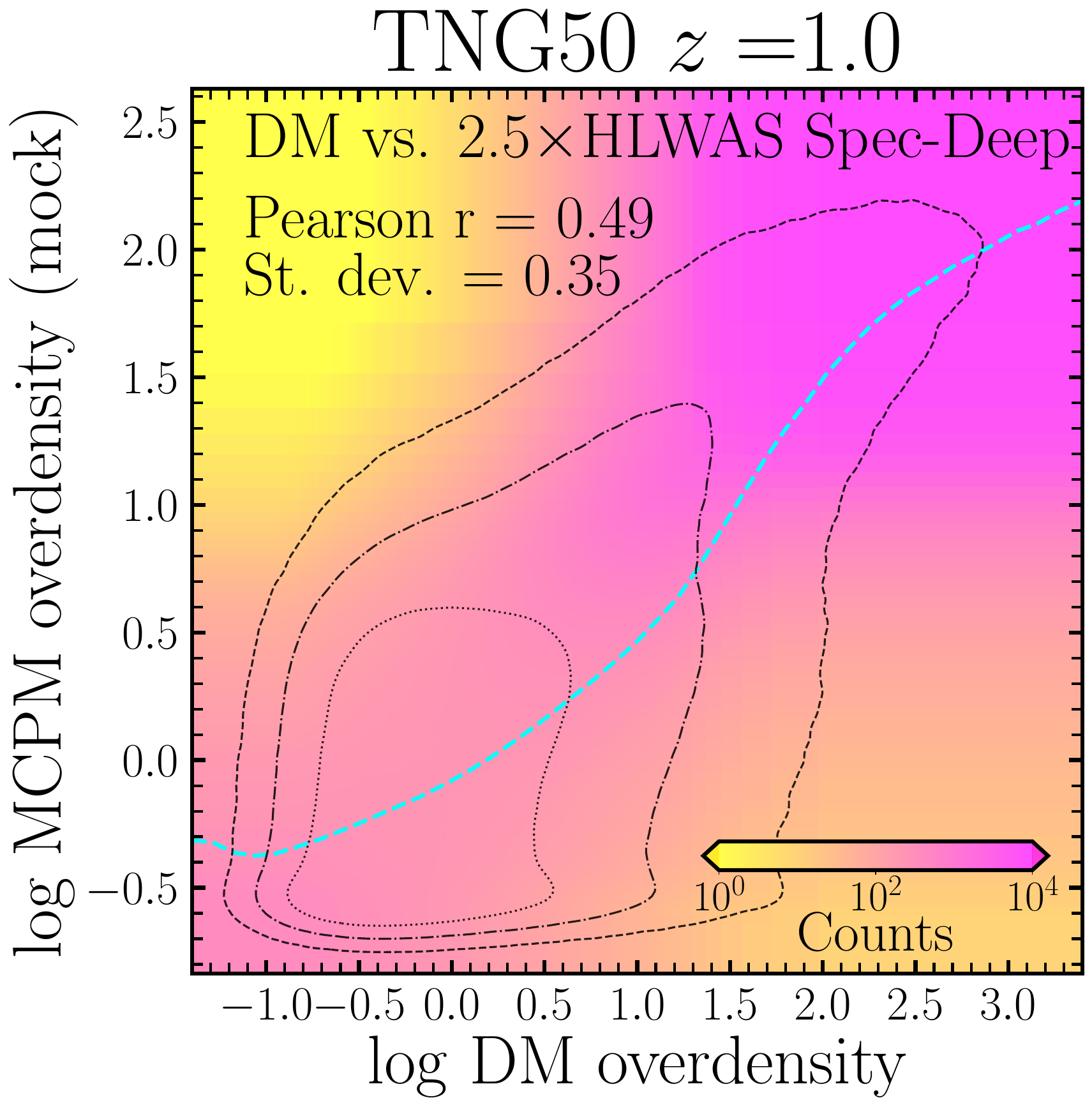}{0.3\textwidth}{}
\fig{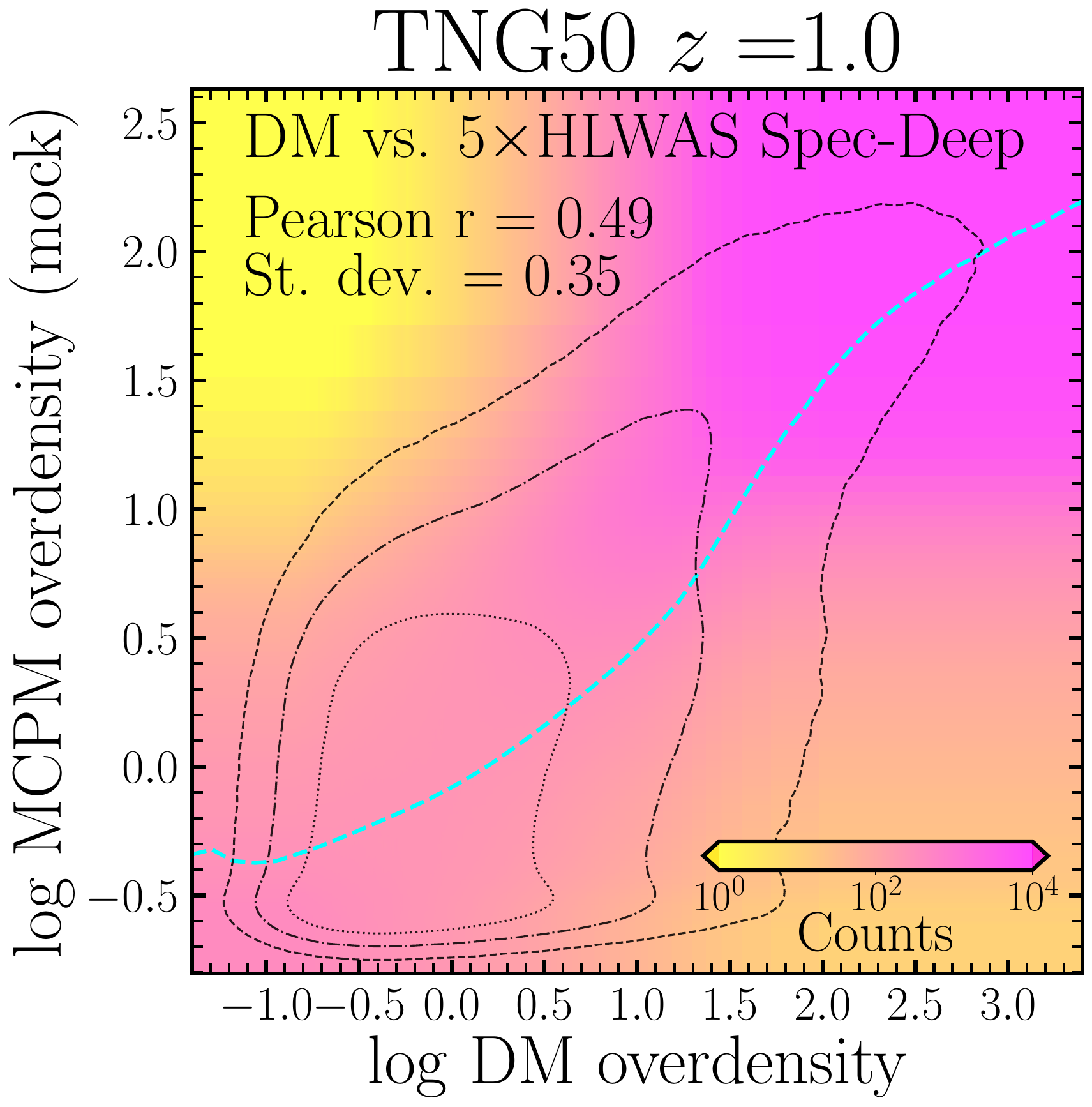}{0.3\textwidth}{}
}
\vspace{-25pt}
\gridline{
\fig{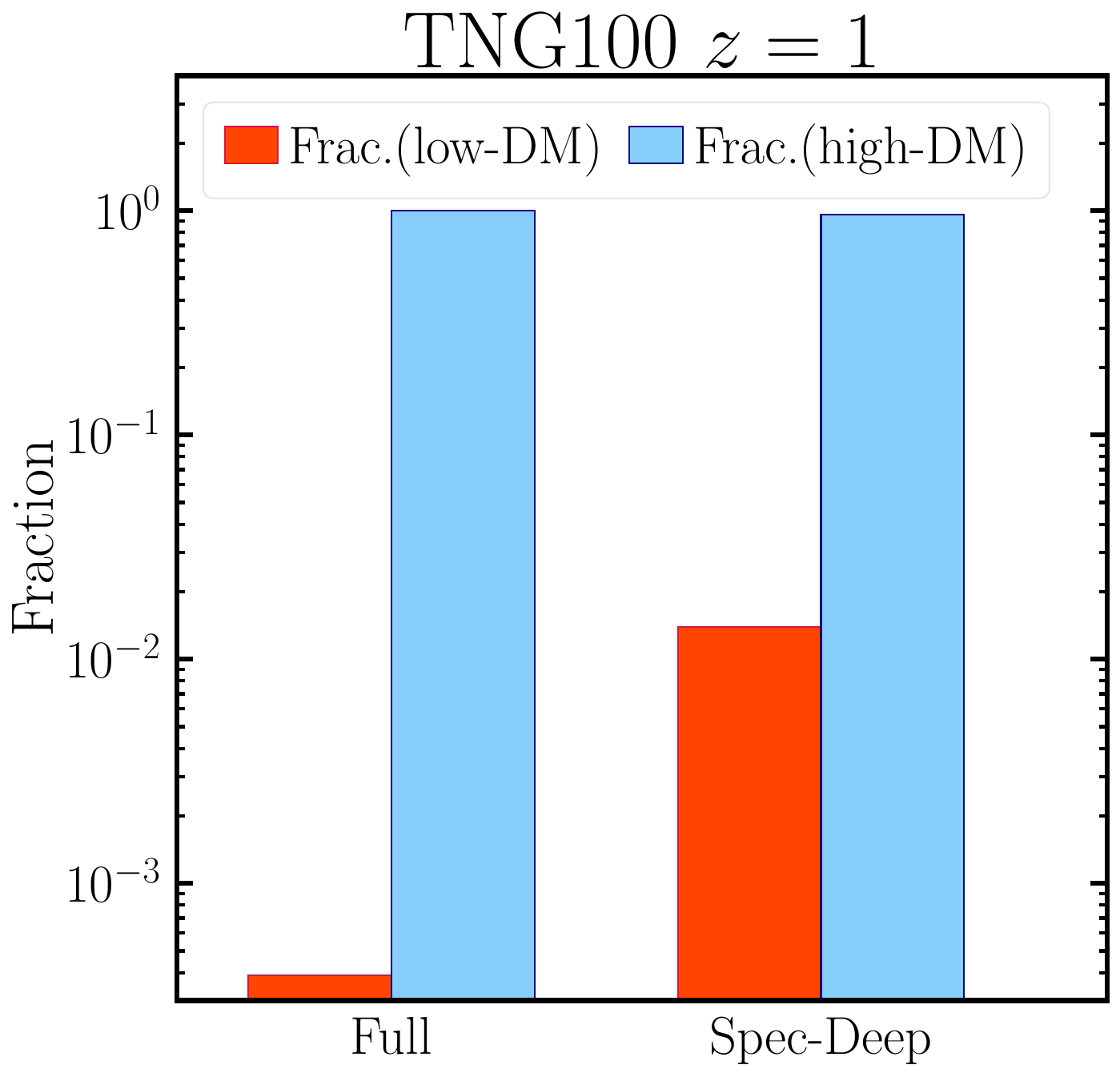}{0.285\textwidth}{}
\fig{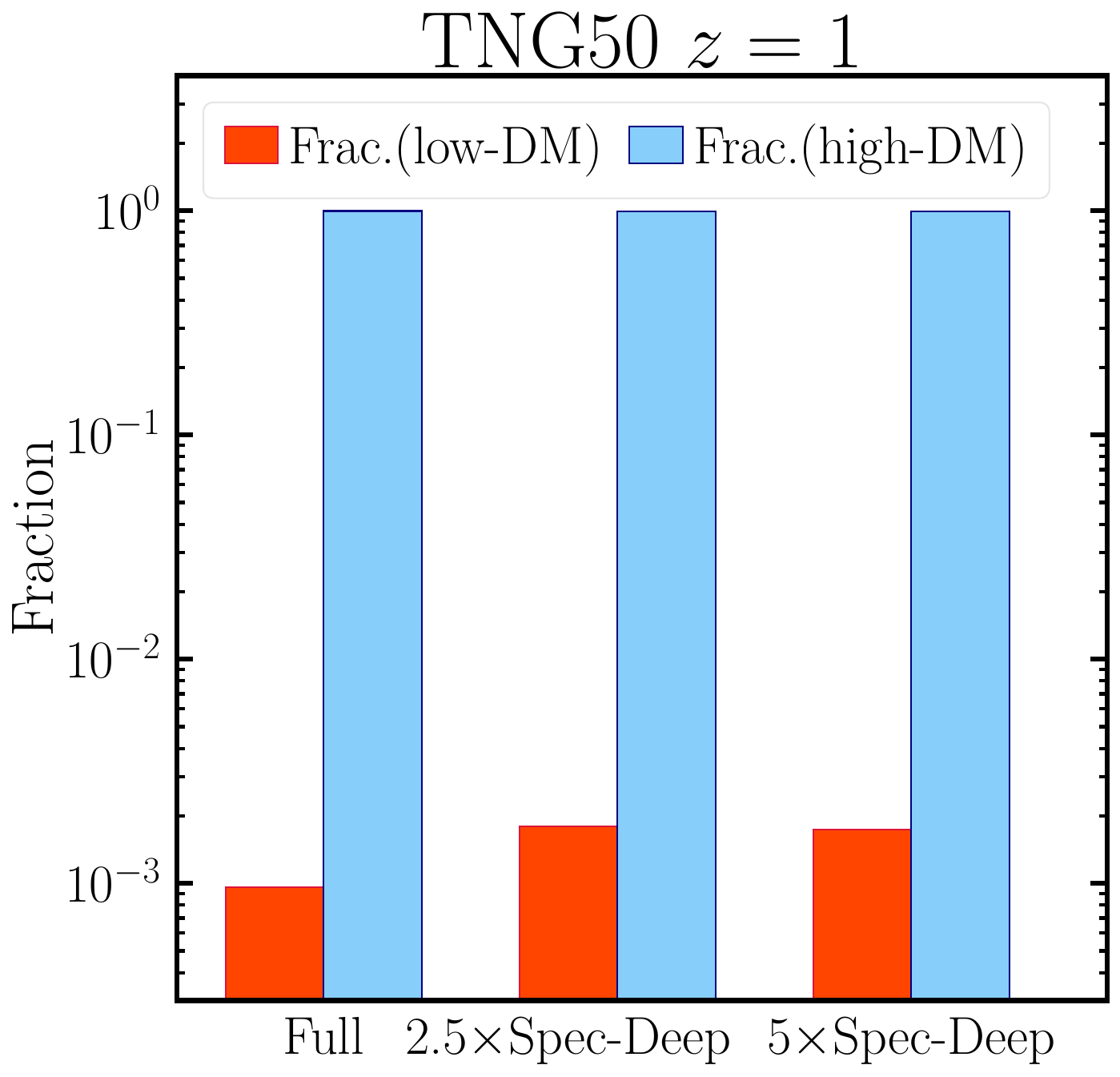}{0.285\textwidth}{}
}
\vspace{-25pt}
\caption{
Quantitative comparisons of the cosmic web from different galaxy samples at $z=1$.
\textit{Top row}: 2D histograms comparing DM matter overdensities ($x$-axis) with MCPM overdensities ($y$-axis) at the same physical location, for the ``full'' ${\logms}\geq8$ sample (left) and the mock HLWAS Spec-Deep sample (right) in TNG100. 
Dotted, dash-dotted, and dashed black contours enclose 50\%, 75\%, and 90\% of the distributions, respectively.
The cyan dashed curves represent the median relationship and Pearson correlation coefficients and median standard deviation are indicated as well.
\textit{Middle row}: the same comparison for three samples in TNG50: ${\logms}\geq8$ (left), HLWAS 2.5$\times$Spec-Deep (middle), and 5$\times$Spec-Deep (right). 
\textit{Bottom row}: the fraction of filament segments located in very low (bottom 25\%) DM overdensity regions (orange-red) and high (top 50\%) overdensity regions (light-blue) for the full and mock samples in TNG100 (left) and TNG50 (right).
}
\label{fig:denscomp}
\end{figure*}

We present visualizations of the cosmic web identified from different mock galaxy samples in Figure~\ref{fig:vis_mock}, representing filaments, galaxies, and the underlying matter density field similar to Figure~\ref{fig:vis}. The top row shows the TNG100 cosmic web at $z=1$ from the mass-complete ${\logms}\geq8$ sample \citepalias[as presented originally in][]{Hasan24} and the mock sample satisfying HLWAS-Deep photometric and spectroscopic depths (hereafter HLWAS Spec-Deep). It is immediately clear how incompletely and inaccurately the cosmic web is identified from HLWAS Spec-Deep, relative to the mass-complete sample. The cosmic web structure from the mass-complete sample traces the underlying mass distribution very well, owing to the excellent sampling of the galaxies used as tracers. The cosmic web structure from HLWAS Spec-Deep is very sparse, with far fewer filaments identified. Many regions of high matter overdensity (in dark-blue) do not have detected filaments passing through them. Several of the filaments that are identified do not trace matter overdensities and are therefore obviously spurious.   
These features are consequences of the lack of tracers (far fewer galaxies are seen in the right panel compared to the left) and the highly biased nature of the tracers (see Figure~\ref{fig:mock_gal}).

The bottom row of Figure~\ref{fig:vis_mock} visualizes three different mock cosmic web realizations for TNG50 at $z=1$. The left panel, similar to the top row, shows the cosmic web from the full ${\logms}\geq8$ sample in TNG50. The other two panels show the cosmic web from mock samples satisfying HLWAS-Ultra Deep photometric depths and 2.5 times the HLWAS-Deep spectroscopic depth (hereafter HLWAS 2.5$\times$Spec-Deep; middle) and 5 times the HLWAS-Deep spectroscopic depth (hereafter HLWAS 5$\times$Spec-Deep; right). 
The increase in spectroscopic depth from HLWAS Spec-Deep has a pronounced impact on the identified filamentary structure. Virtually all the highly overdense regions are traced by filaments identified from both the HLWAS 2.5$\times$Spec-Deep and 5$\times$Spec-Deep samples.
Furthermore, far fewer spurious filaments, i.e., not tracing matter overdensities, are detected in these realizations. Both the HLWAS 2.5$\times$Spec-Deep and HLWAS 5$\times$Spec-Deep samples mitigate the tracer-incompleteness issues that plague the HLWAS Spec-Deep sample. From this visual comparison, we do not find major differences between the 2.5$\times$Spec-Deep and 5$\times$Spec-Deep cosmic web maps.

In Figure~\ref{fig:denscomp}, we quantify the qualitative assessments above by (1) comparing the MCPM density field to the DM density field for the mass-complete samples and HLWAS mock samples, and (2) determining the fraction of filaments that are in underdense and overdense regions. 
For (1), we show smoothed 2D histograms of the DM vs. MCPM overdensity correlation in the top (for TNG100) and middle (for TNG50) rows, as well as the median relationship, contours enclosing 50\%, 75\%, and 90\% of the data, the Pearson correlation coefficient (where a higher value indicates a stronger linear correlation), and the median scatter or standard deviation of the relationship. For both TNG100 and TNG50, the correlation is strongest for the mass-complete sample, as expected. The HLWAS Spec-Deep sample yields a weaker correlation than the 2.5$\times$Spec-Deep and 5$\times$Spec-Deep samples.  Higher spectroscopic depth also produces less scatter in the relationship. 
However, the least dense regions are generally less well-traced by MCPM, with flatter relations and larger spreads in these relations.

The bottom row of Figure~\ref{fig:denscomp} shows bar plots representing the fraction of filament segments from these samples located in the lowest 25\% DM overdensities (orange-red), i.e., in underdense regions, and in the highest 50\% DM overdensities (light-blue), i.e., in overdense regions. The smaller the orange-red bars, the fewer spurious filaments are identified. From HLWAS Spec-Deep, the fraction of spurious filaments is about 1.5 dex higher than that from the full TNG100 sample. In contrast, the spurious fraction increases only slightly as we go to the mock samples from the full sample in TNG50. The fractions are almost identical between the 2.5$\times$Spec-Deep and 5$\times$Spec-Deep samples. Larger light-blue bars in this plot indicate that more filaments are in overdensities, which is the physical expectation. There is only a small difference in this fraction between the full TNG100 and HLWAS Spec-Deep samples, and no discernible change between the full and mock TNG50 samples. 
The two main takeaways from Figure~\ref{fig:denscomp} are therefore (1) $\geq2.5\times$ higher spectroscopic depth allows for a much more accurate cosmic web reconstruction, and (2) there is only a marginal improvement in the reconstruction from 2.5$\times$ to 5$\times$ deeper spectroscopy.

\begin{figure}[htbp]
\centering
\gridline{\fig{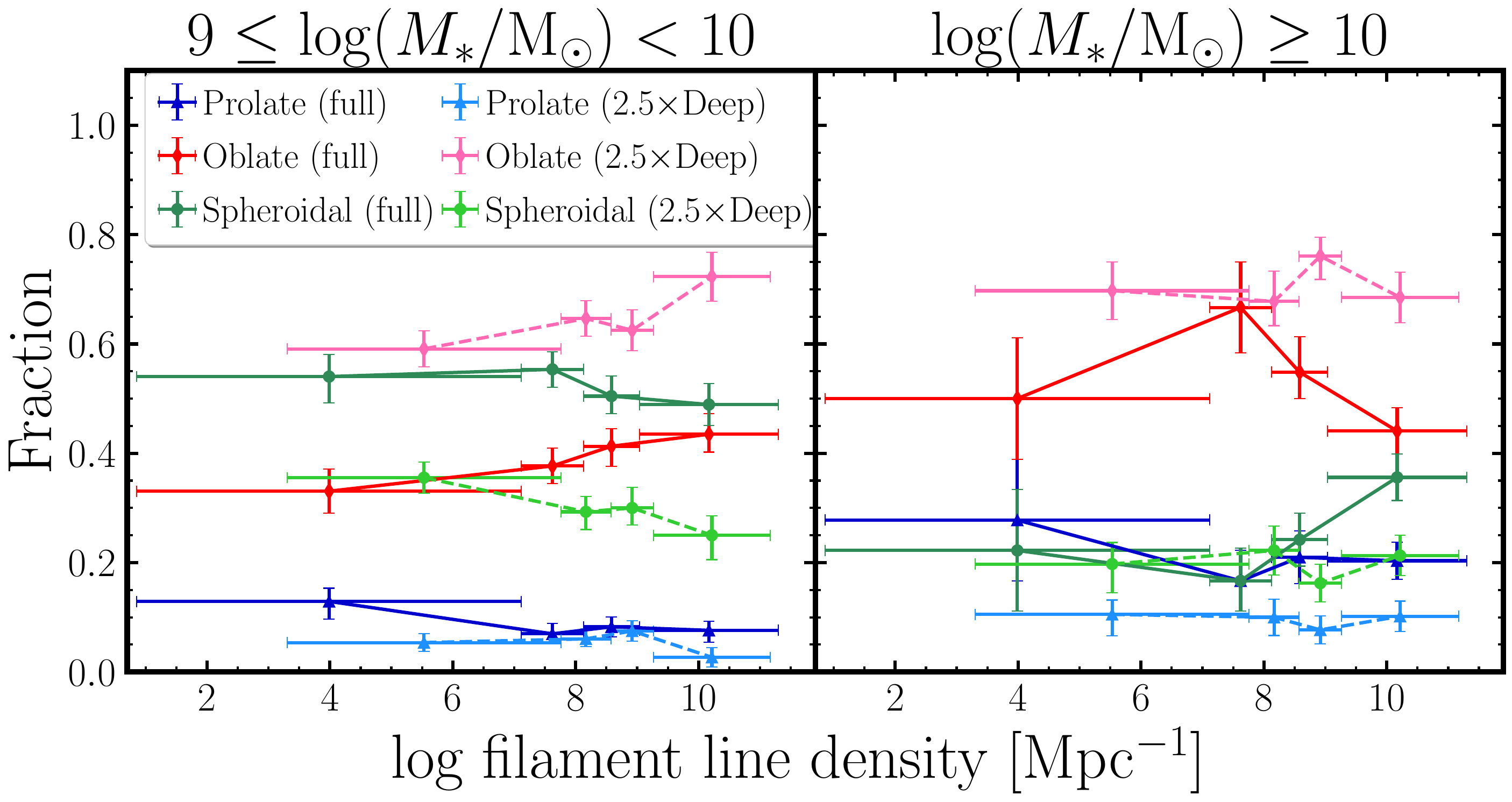}{0.465\textwidth}{}
}
\vspace{-25pt}
\gridline{
\fig{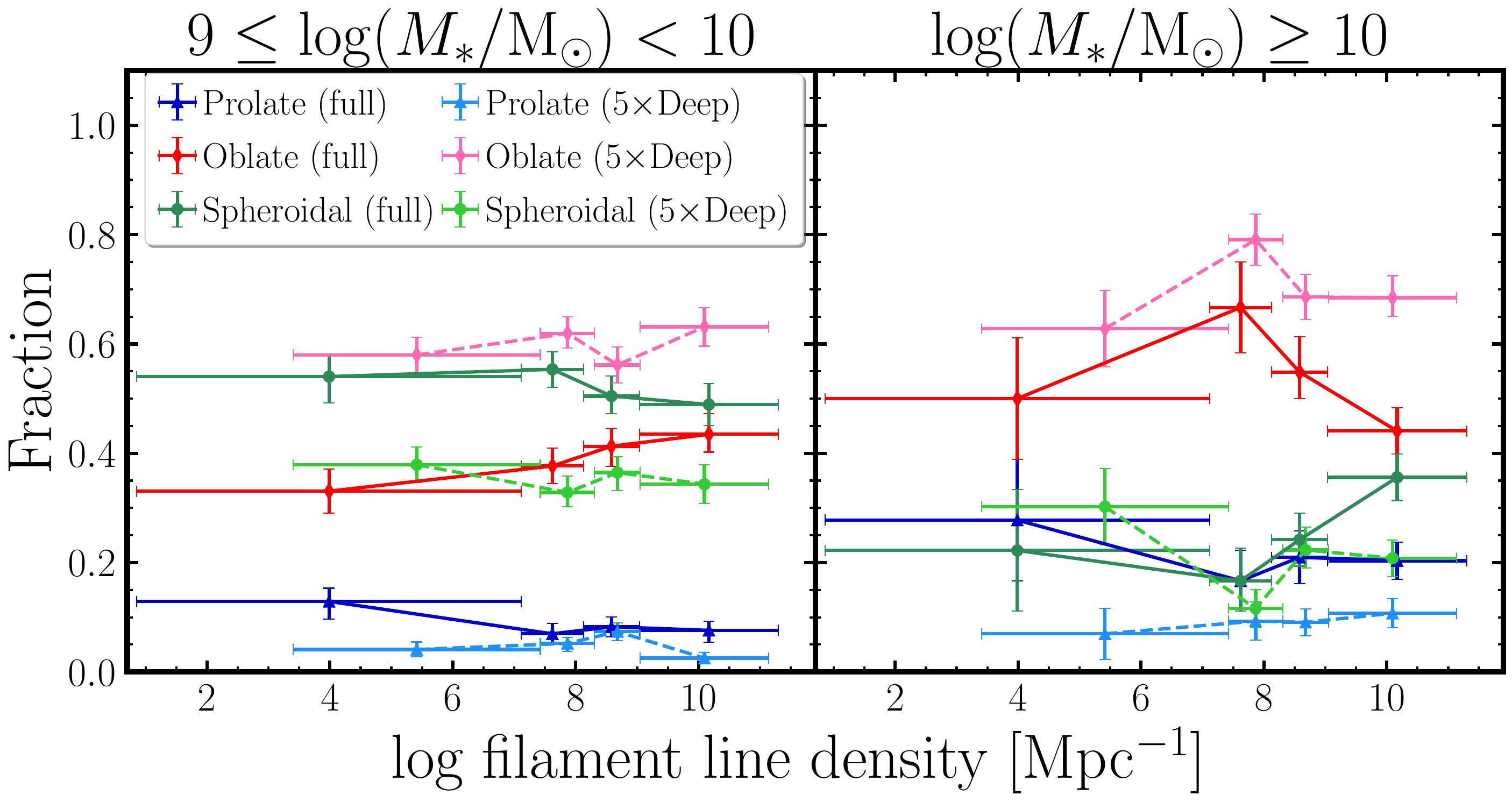}{0.465\textwidth}{}
}
\vspace{-25pt}
\caption{
The dependence of 3D morphological fractions (prolate, oblate, and spheroidal) on nearest filament density at $z=1$.
The mass-complete (``full'') sample is represented by darker colors and solid lines, while the mock samples are represented by lighter colors and dashed lines -- the top row is for the 2.5$\times$ HLWAS Spec.~Deep sample, while the bottom row is for the 5$\times$ HLWAS Spec.~Deep sample.
The left and right columns are for lower and higher stellar mass bins, respectively.
Deep spectroscopy can qualitatively, but not quantitatively, recover the true cosmic web-dependence of galaxy morphologies.
}
\label{fig:stats_comp}
\end{figure}

Finally, we return to the topic of testing the galaxy morphology-cosmic web connection and compare these statistics between the mass-complete and mock samples. We only conduct this experiment for TNG50. as the spatial resolution of TNG100 is fairly low and therefore not conducive to measuring morphologies as accurately as TNG50. In Figure~\ref{fig:stats_comp}, we present the morphological fraction as a function of nearest filament line density at $z=1$. The top row compares these statistics between the mass-complete and the HLWAS 2.5$\times$Spec-Deep cosmic web, while the bottom row compares the mass-complete and HLWAS 5$\times$Spec-Deep cosmic web. The statistics are divided into a lower-mass (${{\logms}=9-10}$) and higher-mass bin (${{\logms}\geq10}$), and $\pm1\sigma$ bootstrapped uncertainties are shown for all.
As in \S~\ref{sec:results}, we include only central galaxies and those galaxies within 2 Mpc of filament spines in this analysis.

We find that the deeper mock HLWAS samples are able to qualitatively capture much of the true filamentary dependence on galactic morphological properties, but they cannot quite unveil these trends quantitatively. Both the HLWAS 2.5$\times$Spec-Deep and 5$\times$Spec-Deep samples show a small increase in the low-mass oblate fraction and a small decrease in the low-mass prolate and spheroidal fractions with rising filament density. These trends are qualitatively consistent with the full sample. 
At higher masses, there is a somewhat greater discrepancy between the full and mock samples. Neither of the mock samples exhibits a rise in the high-mass spheroidal fraction in high-density filaments or the modest drop in the high-mass prolate fraction at these densities. The 5$\times$Spec-Deep sample does show a plateau and then drop with rising filament density for the high-mass oblate fraction. However, across the board, the actual morphological fractions are not quantitatively reproduced. Notably, increasing the spectroscopic depth from 2.5 to 5$\times$ HLWAS Spec-Deep at best only marginally improves the recovery of the true filamentary dependence of morphologies.

The experiment above implies that $\geq2.5$ times deeper spectroscopic flux limits on HLWAS can, in principle, allow us to qualitatively extract the true cosmic web-dependence of morphological fraction. Therefore, deeper grism spectroscopy than is currently planned for HLWAS is essential to reliably test the predictions herein. Deep grism spectroscopy is a requirement for a fairly complete sample of tracer galaxies, which helps reconstruct the cosmic web structure with high accuracy and completeness; this, in turn, helps paint a (more) complete picture of the impact of cosmic filaments on galaxy morphologies. High-quality photometry, combined with advanced techniques that exploit projected shapes and sizes, \citep[e.g.,][]{Pandya24} is necessary for accurate 3D shape measurements of real galaxies. Thus, there is reason to be optimistic about the first direct investigations of the impact of cosmic web accretion on the emergence of galaxy morphologies at $z\geq1$.
However, we stress that even a 5$\times$ deeper spectroscopic survey than HLWAS-Deep is unable to quantitatively reproduce the morphological fraction values that a mass-complete sample predicts. 
We discuss observational tests of our predictions further in \S~\ref{sec:forward}.

\subsection{Possible strategies for real observations}
\label{sec:strategy}

\begin{figure}[htbp]
\centering
\gridline{\fig{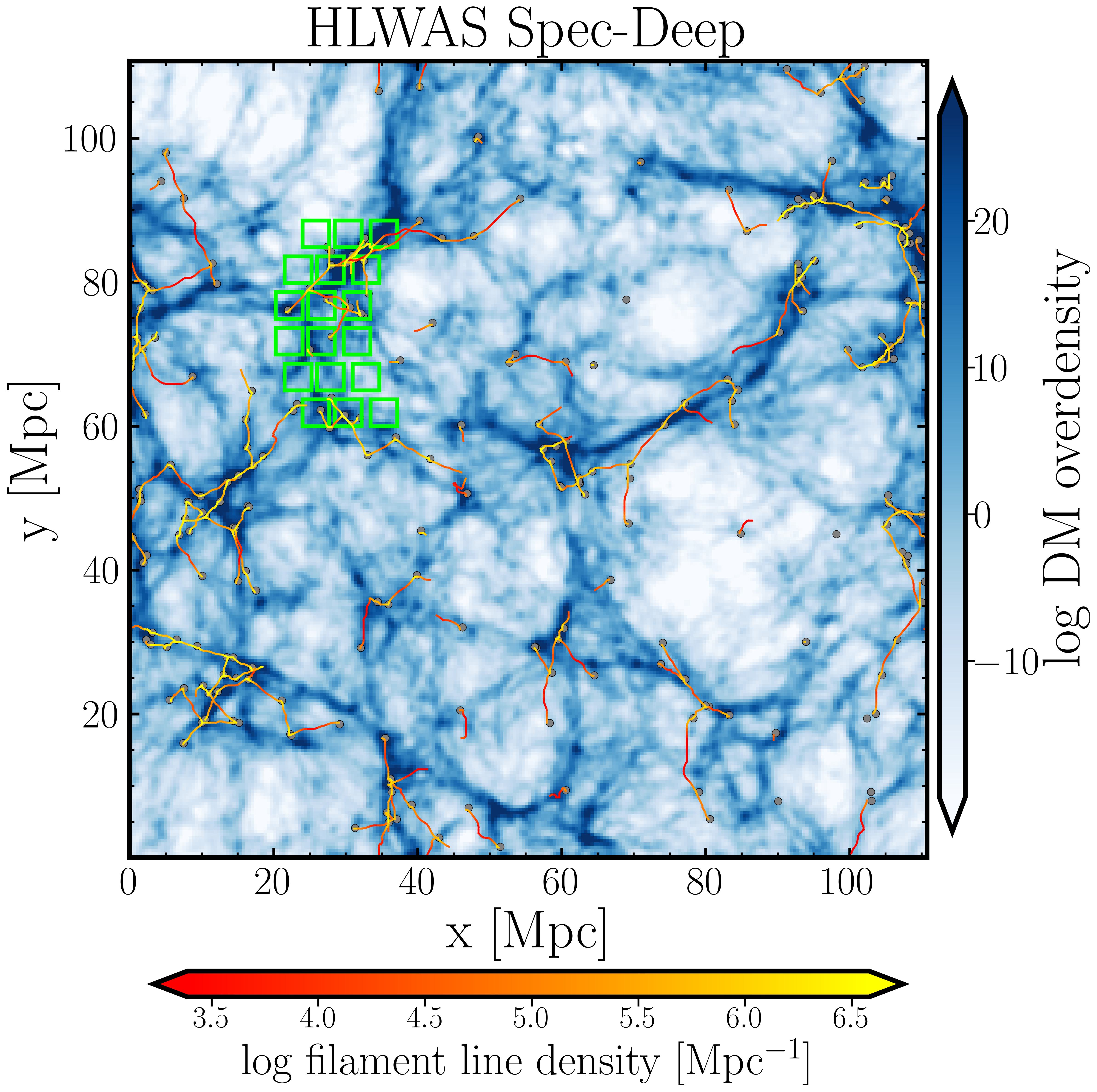}{0.42\textwidth}{}
}
\vspace{-25pt}
\gridline{
\fig{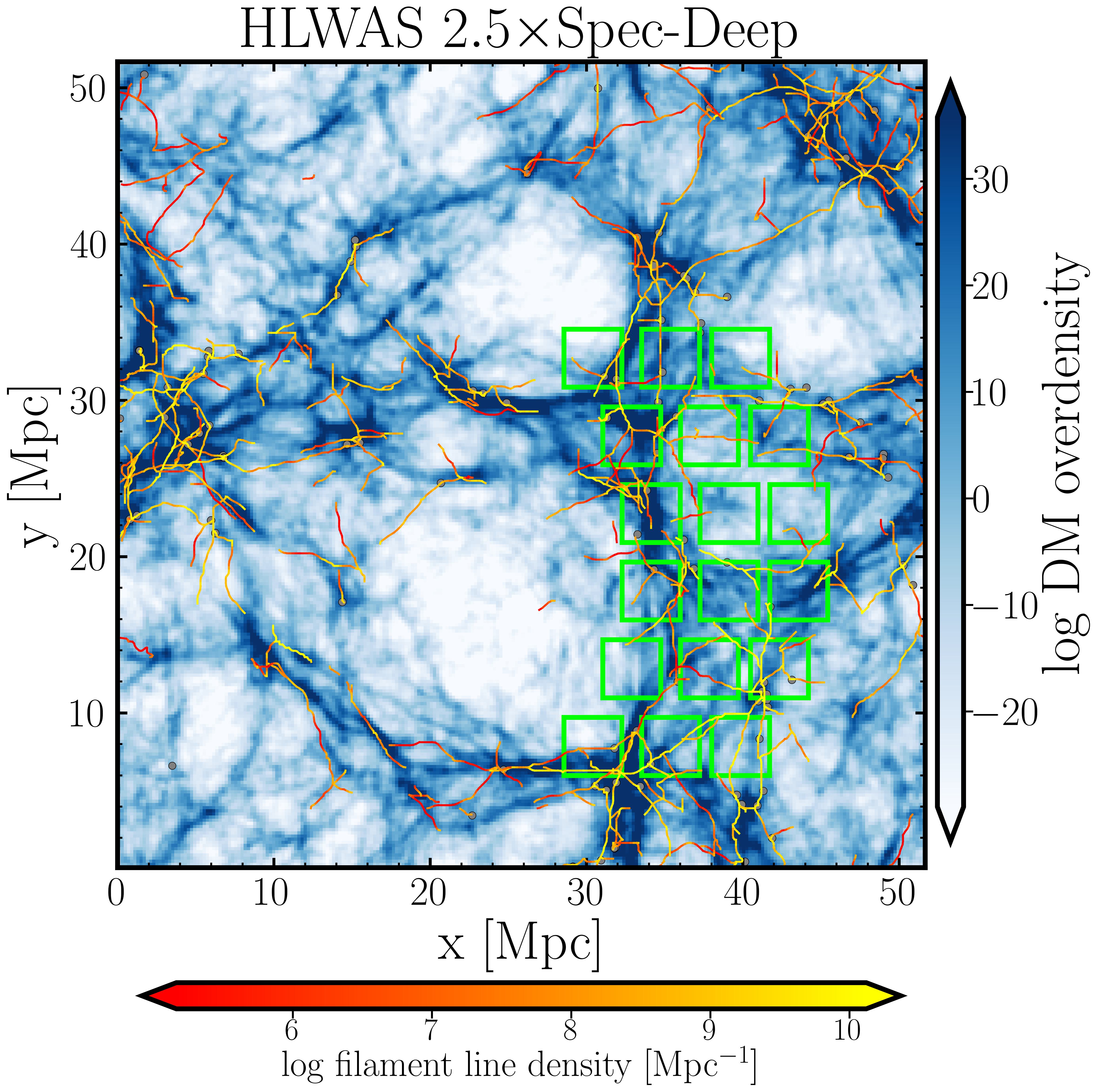}{0.42\textwidth}{}
}
\vspace{-25pt}
\caption{
Similar to Fig.~\ref{fig:vis_mock}; the cosmic web from HLWAS Spec.~Deep in TNG100 (top) and from HLWAS 2.5$\times$Spec-Deep in TNG50 (bottom) are shown, with the Roman WFI FOV overlaid in lime-green.
Targeted deep spectroscopy over selected overdense regions can allow better cosmic web reconstruction.
}
\label{fig:vis_mock_fov}
\end{figure}

In light of our results, we assert that a 2.5$\times$ lower line flux limit than that planned for the HLWAS Ultra-deep field is sufficient to trace the cosmic web with satisfactory accuracy and reveal its true effect on galactic morphological evolution. In practice, this would require a highly expensive observational effort on Roman. We estimate that it would require an appreciable fraction of the total general observing time of Roman's nominal 5-year mission (separate from the core community surveys) to achieve a 5$\times$ or even 2.5$\times$ flux limit over the entire HLWAS Ultra-deep field. 
Due to the practical limitations of undertaking this major task,
we consider alternative strategies to attain our goals.

From the top row of Figure~\ref{fig:vis_mock}, we note that while HLWAS Spec-Deep will not yield an accurate characterization of the cosmic web filaments, it will still identify most of the dominant galaxy overdensities (small gray circles embedded in the dark-blue regions). Accordingly, a viable strategy to optimize the identification of filamentary structure is to target the galaxy overdensities identified by spectroscopic redshifts from HLWAS Spec-Deep. These regions can then be targeted with much deeper spectroscopic observations to help reconstruct filaments in those regions with much higher fidelity. This would produce a much more efficient (and thus less expensive) mapping of cosmic filaments than performing very deep spectroscopy over the \textit{entire} volume of the HLWAS Deep or Ultra-deep field. We visually illustrate the size of the Roman WFI field-of-view (FOV) within the HLWAS Deep (top) and Ultra-deep (bottom) fields at ${z=1}$ in Figure~\ref{fig:vis_mock_fov}. The Roman FOV, which consists of 18 detectors covering a sky area of 0.281 {\sqdeg} (excluding gaps), is shown in green, following the specifications of the detector array arrangement outlined in \cite{Wang:2022_romanHLSS} and \cite{Schlieder24}. The same 2D slices are shown as Figure~\ref{fig:vis_mock}.  
Note that in practice, the gaps between the individual detectors can be filled by dithering \citep[see, e.g.,][]{Schlieder24}. 

Each individual pointing of Roman will easily cover virtually any identified galaxy overdensity. Several such pointings can be used to map filaments with much greater detail in most of the area covered by the HLWAS Deep field. For the much smaller Ultra-deep field, we would require $\sim$4 times fewer pointings to attain the same goals. 
This targeted strategy would ignore the voids, which contain very few galaxies but a significant portion of the volume of the cosmic web. However, one could tremendously reduce the observing time required to accurately map cosmic web filaments to just a fraction of the time it would take to observe an entire field. It is possible to optimize this observing strategy further by combining it with high-quality photometric redshifts, if available \citep[e.g.,][]{Yang21,Malavasi25_Euclid}. We reserve further discussion in \S~\ref{sec:forward}.


\section{Discussion}
\label{sec:discuss}

In this paper, we present one of the first statistical studies of the physical effect of the cosmic web on the morphological evolution of galaxies in a fully cosmological hydrodynamic simulation. We also forecast the detectability of the $z=1$ cosmic web from the observational limits of the planned HLWAS on Roman. 
Below, we discuss (1) our predictions of the TNG simulations in a physical context and (2) the prospects of observing the high-redshift cosmic web and testing our predictions.

\subsection{Physical Interpretations}
\label{sec:physical}

\subsubsection{Shape Alignments with Filaments}
\label{sec:discalign}

TNG50 predicts that the shape alignment of DM halos and stars with filaments decreases from $z=4$ to 0, but there are significant differences between the inner DM halo/galactic scale and the outer halo scale. 
While stars lose their preferential shape alignment by $z=3$, inner DM halos do so by $z=2$, and outer DM halos do so at $z=0$.
Shape alignments between halos and filaments that feed them are most prominent at large halo-centric scales. 
Our result of preferential shape alignment of outer halos for most of cosmic time is in agreement with many previous studies \citep[e.g.,][]{Hahn07,WK18,Lopez21}.
As discussed in \cite{GV18}, this is due to the outer regions of a halo being composed of matter that has been most recently accreted onto the halo, which flows in preferentially along the filament in which the halo resides.
Further into the halo, various internal mechanisms such as supernovae (SNe) and other star formation-induced phenomena dominate, likely helping to disrupt the shape alignment with filaments.
These effects also increase globally at later times \citep[e.g.,][]{FS20}.

In the $N$-body counterpart of TNG50, we see a smaller drop in the DM shape-filament alignment from the outer to the inner halo across all redshifts. 
Baryonic mechanisms evidently help disrupt the anisotropic accretion signal from filaments to halos as we go deeper into the halo. 
Several other studies also highlight the imprints of baryonic physics on DM halos in simulations \citep[e.g.,][]{Chan15,Butsky16,Anbajagane22}.
Additionally, as baryonic physics changes the small-scale matter distribution, this may have implications for the so-called S8 tension in cosmology \citep[e.g.,][]{McCarthy23}.
Nevertheless, even in TNG50-Dark, there is a significant decline in shape alignment at low redshifts, implying that halo mergers -- or possibly even filament mergers -- from gravitational forces alone can disrupt the early signature of shape alignment with filaments \citep[see, e.g.,][]{Cadiou20}. Increase in mergers at late times is known to increase misalignment with filaments \citep[e.g.,][]{Welker14}.

Prolate halos show significant shape alignment with filaments at almost all redshifts. This trend is strongest in outer halos in both simulations, while inner prolate halos show weaker alignment at high redshift ($z\gtrsim2$).
This implies that collimated (recent) mass accretion from filaments forms elongated halos along the direction of the filaments. 
Furthermore, we find that in prolate galaxies, the stars are strongly aligned with the outer DM halos down to $z=0$, indicating that torques from the halos align the stars in the same direction and could support the elongation of the stellar structure.
Our findings reinforce previous predictions from both $N$-body \citep[e.g.,][]{BS05} and zoom-in hydrodynamical simulations \citep[e.g.,][]{ceverino15,tomassetti16} in a fully cosmological context and lead us to a picture in which  the formation of prolate structures is linked to filamentary accretion.
While we acknowledge the prolate halo shape alignment as a statistical result, we highlight the potential to leverage the alignments of prolate structures to map the early cosmic web. \cite{Pandya19} predicted that early prolate galaxies in proximity should have their shapes intrinsically aligned with each other, as they form along the same filament. This alignment could be used to identify the direction of high-$z$ filaments.

DM shape alignments show significant mass-dependence in both TNG50 and TNG50-Dark. High-mass galaxies/subhalos show strong DM shape alignments with nearby filaments at the outer halo scale at all times. The shape alignments of inner halos are weaker, and in TNG50, a preferential alignment disappears at $z<2$.
Our results are consistent with those of \cite{GV18}, who studied an $N$-body simulation to find that at $z=0$, high-mass halos have the strongest shape-filament alignments and that the alignment increases further away from halo centers.
\cite{GV19} found qualitatively similar results in the EAGLE hydrodynamical simulations at $z=0$ \citep{Schaye15}.

Halo shapes are most strongly aligned with high-density filaments. In TNG50, this is only true in the outer halo, but in TNG50-Dark, this trend is also seen in the inner halo at $z\geq2$.
These findings are consistent with studies such as \cite{GV18}. 
We can physically explain our result by noting how mass accretes differently from filaments of different densities -- and equivalently thickness \citep[e.g.,][]{Cautun14}.
Halos in thin or low-density filaments grow by isotropic accretion of matter and do not grow aligned with the filament, whereas halos residing in thick or high-density filaments form earlier and only allow mass accretion along the axis of the filament \citep{Borzyszkowski17,WK18, Lopez19}. In this picture, early-forming halos are ``stalled'' and have their major axes aligned with their host high-density filament.

\subsubsection{Galaxy Morphologies and Filaments}

In addition to the internal mechanisms within galaxies, the external effect of mass accretion from the cosmic web also plays a sizable and complex role in determining the morphological properties of galaxies in TNG50. 
To first order, stellar mass and redshift are the primary determinants of galaxy morphology, but the nature of the host filament also has varying degrees of influence on several morphological properties.

Under the effect of gravity alone (i.e., in TNG50-Dark), more massive DM halos are more elongated. Inner halos are much more prolate than outer halos, which become mostly spherical at late times.
This long-established result is due to dissipationless DM falling anisotropically to halo centers and making them more triaxial, while outer halos accrete more isotropically to form spheroids \citep{BS05,Allgood06}. 
The inclusion of baryons results in both stars and halos becoming less prolate as they become more massive, with spheroidal or disky structures dominating the population at most times \citep[see, e.g.,][]{Butsky16}.

But are there common pathways by which galaxy and halo shapes evolve from early to late times?
The picture suggested by the {\sc VELA} simulation is that most high-redshift galaxies start out as prolate and undergo `wet compaction'' \citep{DB14,Zolotov15}, wherein gas accreting from the cosmic web or wet mergers loses angular momentum and forms stars near the halo core, transforming galaxies from prolates to compact spheroids and extended disks at masses ${\logms}\gtrsim9.5-10$ \citep{ceverino15,Lapiner23}.
It is not clear if the same physical mechanisms produce the diversity of galaxy morphologies in the TNG model. 
First, even as early as $z=4$, spheroidal galaxies are the dominant population in TNG50 and less than a third of ${\logms}=8-9$ galaxies are prolate.
\cite{Tacchella19} found that wet compaction events are not required to transform morphological structure; dry mergers could also do this. 
Furthermore, we do not find a transition mass in TNG50 above which galaxies are no longer prolate. In fact, there is a small population of high-mass (${\logms}>10$) prolate galaxies at $z<2$, which were plausibly formed by major dry mergers \citep[see the Illustris study of][]{Li18}.

TNG50 predicts a small but significant dependence of the 3D shapes of stars and DM halos on the nearest filament density after removing the stellar mass dependence. At fixed mass at $z\gtrsim1$, prolate halos are more likely to be found in lower density filaments, and spheroidal halos in higher density filaments. In terms of stellar structures, prolates are somewhat more likely to be found in low-density filaments at $z\gtrsim1$, while spheroidal (oblate) halos more strongly prefer high-density (low-density) filaments at later times.
Outer halo shapes are more sensitive than inner halos to neighboring filaments, as they experience more recent accretion than the inner halo. The complex gas physics associated with star formation explains why stellar shapes correlate more weakly with DM halos in the inner halo.

The dynamics of mass accretion from low to high density filaments are critical in understanding the connection between filament density and galaxy/halo shapes.
Since low-density filaments are thin, they are comparable in width to high-$z$ halos. 
Thus, matter can accrete smoothly along these filaments in a dipole fashion and form elongated structures. 
High-density filaments, on the other hand, are much wider than halos and form complex quadrupolar vortices \citep{Laigle15}. These can disrupt smooth, coherent accretion onto halos and help form fewer prolate structures. Instead, they can impart additional angular momentum to galaxies and help form disky or spheroidal structures.
It is somewhat surprising that the stellar spheroidal and oblate fractions are significantly correlated with filament density only at very late times. We postulate that at these times, a higher incidence of galaxy mergers in high-density filamentary environments could be responsible for this trend. In these environments, a large fraction of oblate galaxies could merge to form spheroids, thus explaining fewer oblates and more spheroids in high-density filaments at $z\lesssim0.5$. 

The stellar sAM shows the strongest residual dependence on filament density of any morphological property we study. Higher stellar sAM is associated with higher density filaments at fixed mass, an effect that lessens with time but persists to the present day. 
A key theoretical expectation of the cosmic web is that the large-scale tidal field generates high-vorticity flows in thick/higher density filaments \citep[e.g.,][]{Codis12,Laigle15}.
Thus, high-vorticity gas is supplied to galaxies residing in these filaments, which eventually forms stars which retain the high sAM of the accreted gas.
The vorticities are also larger at higher redshifts, explaining the stronger filamentary-dependence at earlier epochs.
We additionally found that at fixed mass, more rotationally-supported and less dispersion-dominated stellar structures reside in high-density filaments at $z\gtrsim1$. This trend also follows from high-vorticity gas accretion from high-density filaments, which enables stars to form in rotating disks.

Our findings on the filamentary dependence of kinematic morphological properties are consistent with and add to a growing body of work describing how the dynamics of mass accretion from the cosmic web are connected to halo and galaxy kinematics.
\cite{GV21} reported that massive halos in thick filaments have higher spin (sAM) than those in thin filaments at $z\lesssim2$.
\cite{Song21} showed that the high-vorticity gas flowing from filaments to halos at $z\sim2$ can lead to inefficient gas transfer to the galaxy scale and subsequent quenching \citep[see also][]{PR20}. 
At later times, the angular momentum of gas accreted from vorticity-rich filaments is lost more easily, as baryonic phenomena, such as disk instability, stellar feedback, and external effects such as merger activity, increasingly  dominate galaxies.
However, \cite{Woo25} recently found that higher-density filaments host relatively higher-sAM galaxies in TNG100 even at $z=0$.

Lastly, we find that more extended $z\geq1$ galaxies tend to live in somewhat higher density filaments than more compact galaxies. The latter also have much higher stellar sAM, consistent with high-density filaments transferring higher angular momentum gas to galaxies. 
Our results imply a connection in TNG50 between the sizes and spins of galaxies and the spin of matter accreting onto them from the cosmic web, which is consistent with the expectation of abundance matching models that propose that galaxy sizes are determined by the angular momentum of their parent halos, with very little dependence on stellar mass, morphology, or even redshift \citep[e.g.,][]{Kravtsov13,Somerville18}.
Based on nearly identical methods to this work. \cite{Woo25} concluded that galaxies in TNG100 have higher stellar sAM and size at fixed mass in high-density filaments, even at $z=0$.

\subsubsection{A combined narrative and further considerations}
\label{sec:disccomb}

\begin{figure*}[htbp]
\centering
\gridline{
\fig{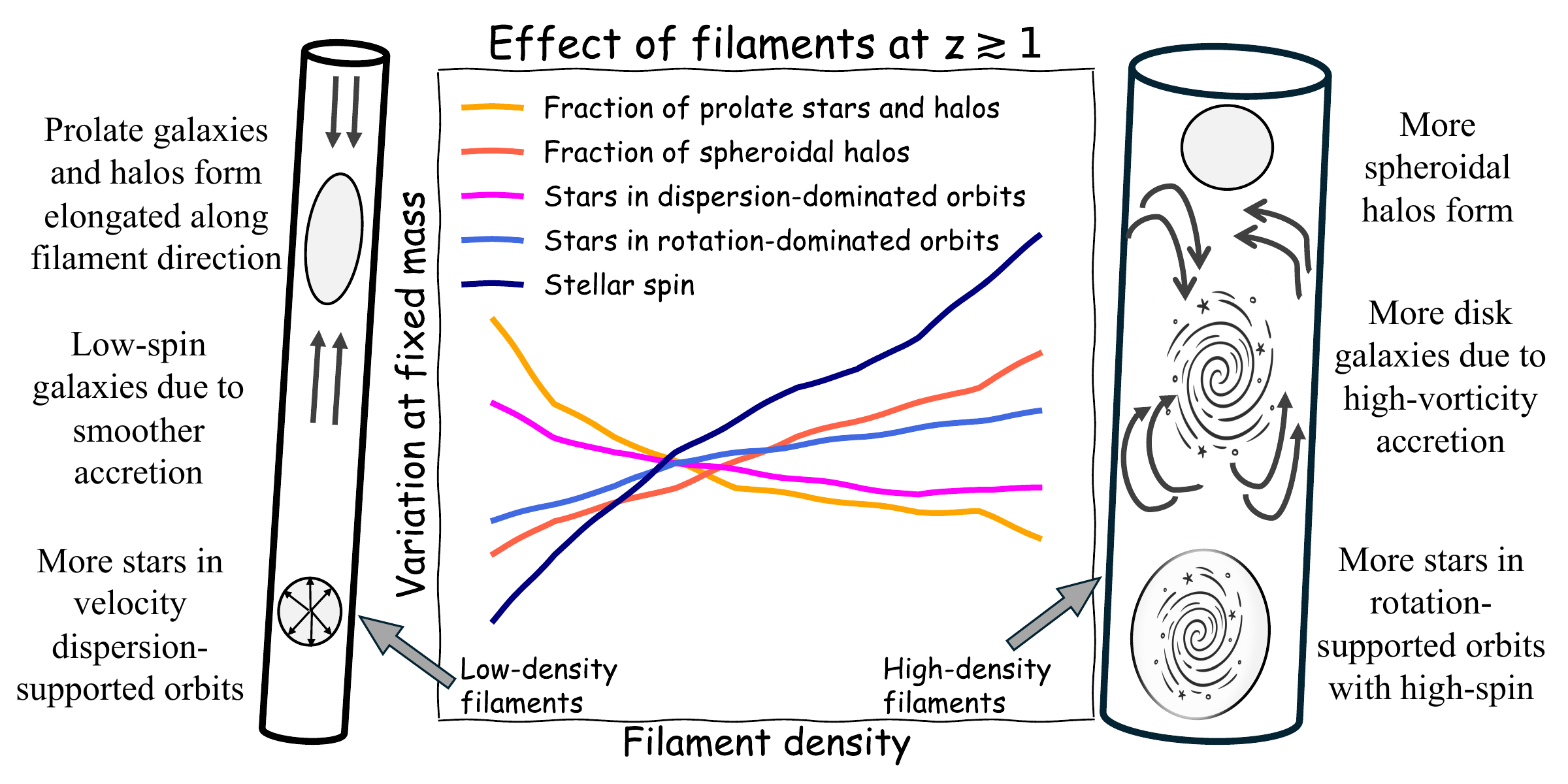}{0.8\textwidth}{}
}
\vspace{-25pt}
\caption{
A cartoon schematic visualizing the effect of filaments on the morphological evolution of galaxies at early times in TNG50.
The qualitative residual effect of filament density (after removing the mass-dependence) on various morphological characteristics of galaxies at $z\gtrsim1$, based on our results in \S~\ref{sec:results}, is shown by the sketch in the middle. 
To the left and right of this sketch, we illustrate the types of morphological phenomena associated with low- and high-density filaments, respectively.
While many of these relationships disappear at $z\lesssim1$, some remain and some even reverse.
}
\label{fig:cartoon}
\end{figure*}

Synthesizing our results together, we propose a simple cartoon schematic in Figure~\ref{fig:cartoon} to describe the effect of the cosmic web on the early ($z\gtrsim1$) morphological evolution of galaxies in TNG50. We sketch the rough qualitative dependence of various morphological properties on the density of the nearest filament in the middle and illustrate the types of morphological characteristics associated with lower- and higher-density filaments on the left and right, respectively. The arrows inside the low- and high-density filaments indicate the relative dynamics of accretion from different densities, as alluded to in the previous section.
We highlight a differential residual effect of filaments; low-density filaments are associated with the formation of prolate galaxies and halos, low-spin galaxies, and stars in velocity dispersion-dominated orbits, while high-density filaments are associated with the formation of spheroidal halos, oblate/high-spin galaxies, and stars in highly rotation-supported orbits.
In the last $\sim$8 Gyr, many of these trends disappear, but some remain -- such as the rise in stellar spin in high-density filaments -- and some new trends also appear -- such as the rise (fall) in spheroidal (oblate) stellar shapes with filament density.

While we do not study the shape of gas explicitly, this could be a key factor in the overall narrative. 
\citet{Pillepich19} found that the majority of ${\logms}=8-9.5$ galaxies in TNG50 form prolate star-forming gaseous structures at $z<5$. Yet, the fraction of prolate stellar structures is always subdominant.
In higher mass galaxies, the star-forming gas is dominated by oblate shapes at most times. Star-forming gas therefore typically forms disky or elongated shapes. This reveals an intrinsic difference in how stars and gas evolve. Namely, unlike gas, stars cannot dissipate energy and can therefore remain in high-velocity dispersion orbits for a long time, while gas can quickly settle into disky or boxy orbits with higher rotational support.
However, we note that the gas in a prolate galaxy, while being disky, can accrete at different angles and precess, affecting the orbits of the stars \citep[see][]{Meng21}.

It is vital to understand how the underlying physical model gives rise to the trends we uncover and how this relates to the cosmic evolution of real galaxies.
Virtually all cosmological hydrodynamical simulations predict that feedback is essential in forming a diversity of galactic morphologies \citep[see reviews by][and references therein]{Primack24,FB25}. \cite{Genel15} showed that feedback plays a key role in establishing  distinctive galactic structures in the Illustris simulation, and the same is true for its successor, TNG \citep{Pillepich19,Tacchella19}, and other simulations such as EAGLE \citep{Lagos17}. 
Unfortunately, the highly uncertain physics of feedback mechanisms such as SNe, stellar winds, and radiation pressure limits such simulations to only qualitatively reproducing broadly observed characteristics of galaxies, such as their morphological variety, but they often fall short of quantitatively reproducing many observations. Like \cite{Pillepich19}, we confirm that TNG50 produces too few low-mass prolate stellar systems at early times -- by factors of two or more relative to HST/JWST observations, depending on the redshift and mass \citep{vanderwel14,zhang19,Pandya24}.

One possibility to explain why TNG50 produces so few prolates and so many spheroids in the high-redshift low-mass regime is that the fiducial model requires some modification. 
\cite{tomassetti16} showed that the absence of feedback produces rounder galaxies, while higher-than-fiducial feedback forms more early prolates and late oblates.
This could hint at stellar feedback in TNG being weaker than necessary to reproduce the observed 3D shape fractions.
However, the TNG model was modified from its precursor, Illustris, to better match fundamental observations, such as the galaxy-color bimodality \citep{Nelson18}; thus, simply increasing or otherwise modifying the stellar feedback could very well cause tensions with other observed galaxy properties. 
The balance between uncertain parameters of the feedback models and the subgrid physics of hydrodynamical simulations is quite delicate.
Simplified star formation prescriptions in the subgrid model, calibration to lower-resolution simulations, and even the specific numerical scheme employed have sizable effects on the physical mechanisms that shape the baryonic structure of galaxies \citep[see, e.g., the discussion in][and references therein]{LH25}.

Regardless of the physical model, the numerical resolution of TNG50 itself may be insufficient for accurately modeling stellar shapes.
The gravitational force softening length of stars and DM is of the order of  $\sim$300~pc, while that of  gas is of the order of $\sim$70 pc, resulting in typical Voronoi sizes of $\approx$100 pc in the star-forming regions of galaxies \citep[see][]{Pillepich19}. While this is much higher resolution than TNG100 and most other cosmological simulations, it can still result in uncertain star particle orbits in most higher-redshift lower-mass galaxies with effective sizes of less than 1 kpc. As a result, the 3D shapes of many (low-mass) galaxies can be fairly uncertain.

It would therefore be useful to perform higher-resolution zoom-in re-simulations of such galaxies (and their halos) to understand the sensitivity of morphologies to numerical resolution alone, keeping the physical model fixed \citep[see][for an example of such zoom-ins]{Lu24}. This would enable us to assess how the statistics we find in this paper, such as overall morphological fractions and their variation with filamentary environment, change with resolution parameters such as gravitational force softening.
Additionally, repeating our analyses in other hydrodynamical models, particularly in high-resolution cosmological boxes such as FIREbox \citep{Feldmann23} and NewHorizon \citep{Dubois21}, would help gain further insight into the physical mechanisms that govern morphological evolution.

\subsection{Testing predictions: caveats and observational limitations}
\label{sec:forward}

In \S~\ref{sec:pred}, we presented some prospects for accurately mapping the early cosmic web with galaxy surveys with deep photometry and spectroscopy as would be enabled by Roman. The deepest emission line flux limits currently planned for the HLWAS would only be able to detect a very biased sample of massive and highly star-forming galaxies at $z\!=\!1$, which cannot accurately trace the filamentary structure. By boosting the spectroscopic survey sensitivity, more representative tracer galaxy samples can be obtained.
However, even the planned line flux limit of HLWAS-Deep is sufficient to identify most of the galaxy overdensities in the field. These specific regions can be spectroscopically followed up to reach a few times lower emission line flux limits, which represents a far more efficient and practically viable approach to map cosmic filaments than performing such observations over the entire HLWAS Deep or Ultra-deep fields. 
Generally, 2.5$\times$ deeper spectroscopy will enable a highly accurate and complete reconstruction of the $z=1$ cosmic web, but increasing this depth two-fold does not significantly improve the reconstruction.
Assuming that galaxy shapes and spins can be accurately measured, such deep surveys would be able to qualitatively recover the dependence of galaxy morphological properties on filament densities.

We present a fairly idealized analysis of the ability of high-redshift galaxy surveys to reconstruct the cosmic web. We made several assumptions in this process (see Appendix~\ref{app:mock}), and these will have different degrees of impact on the final reconstructed structure and its correlation with galaxy morphologies. Moreover, observational data is much more complex and uncertain to interpret than simulation snapshots. Accordingly, our predictions should be interpreted as the most optimistic scenarios for high-$z$ cosmic web reconstruction from wide-field surveys on Roman and other state-of-the-art observatories. We discuss some of the uncertainties in our results stemming from the assumptions we made.

One of the largest sources of uncertainty is the conversion from observational emission line flux limits to simulated SFRs. It is very challenging to generate detailed mock photometry of galaxies that self-consistently models the dust attenuation within each galaxy \citep[e.g.,][]{Nelson18,Donnari19}. This being beyond the scope of our paper, we simply assume that the dust attenuation in our simulated galaxies follows the population-averaged properties of observed $z\!\sim\!1$ galaxies \citep{Nedkova24} according to their stellar mass. If we instead allow the individual values to vary within the $\pm1\sigma$ confidence interval of the observations, the number of tracers in our mock samples does change. This change is within $\approx$10\% for the HLWAS 2.5$\times$ Spec-Deep and 5$\times$ Spec-Deep samples, but the mass and SFR completeness limits change by less than 10\%. Consequently, cosmic web reconstructions from these samples would not change significantly from our fiducial results. In contrast, the number and completeness of tracers in the HLWAS Spec-Deep sample are much more sensitive to the assumed dust attenuation. Despite this, our conclusion that the HLWAS Spec-Deep sample cannot accurately reconstruct the cosmic web remains unchanged.

Observations suggest that there is significant galaxy-to-galaxy variation in dust attenuation with not only stellar mass \citep[e.g.,][]{Battisti16}, but also SFR \citep[e.g.,][]{Kriek13}, metallicity \citep[e.g.,][]{Shivaei20}, and other properties.
By simply considering the mass-dependence, we sidestep the well-established complexities and uncertainties in dust extinction and scattering in galaxies \citep[][and references]{SN20} in favor of a statistical description of dust at $z\!\sim\!1$. Accurate modeling of dust attenuation in each individual galaxy is well beyond the scope of this paper, but we reserve such an exercise for future work.

There are also some uncertainties in our assumed mock Roman photometric magnitudes. We assume a single value for the average $J\!-\!H$ color observed at $z\!\sim\!1$ \citep{Skelton14}, instead of accounting for variation with mass or other properties. This carries a small uncertainty on the mock $H$-band magnitudes and would not significantly change the mock samples. There is an even smaller uncertainty in considering the Roman and HST $H$-bands to be equivalent. For TNG50, we also performed a more uncertain conversion from mock SDSS $z$-band magnitudes and $i\!-\!z$ colors to Roman $H$-band. However, this does not significantly alter the tracer samples we obtain either. A complete treatment of mock photometric magnitudes requires careful transformations between different filters \citep[see, e.g.,][]{Durbin23,Durbin25}, which is beyond the scope of this paper.

Thus far, we have only considered spec-$z$ for our cosmic web reconstruction, but not photometric redshifts (photo-$z$). 
Imaging is less expensive than spectroscopy; thus, significant effort has been expended in the last decade or so to explore photo-$z$ as a viable alternative to spec-$z$ for mapping the distribution of galaxies from large surveys \citep[e.g.,][]{Laigle16,MdO19,Yang21}.
However, the key underlying issue with this is that photo-$z$ typically has much lower precision than spec-$z$ -- typically 100-1000 times larger \citep[e.g.,][]{Salvato19} -- resulting in uncertainties of $\sim$100 Mpc or more in 3D physical location. This would make it very challenging to detect statistically significant filamentary structures on smaller scales, and thus the diversity of filamentary structures (including lower-density filaments) would be missed.

The higher redshift accuracy provided by spec-$z$ is needed to reconstruct more detailed filamentary structures over much smaller areas, such as the HLWAS Deep and Ultra-deep fields. 
We note that even grism-based redshift errors can sometimes be on the order of $\sim10$ Mpc \cite[e.g.,][]{Momcheva16}. This is enough to degrade the quality of the cosmic web reconstruction in 3D space, as shown by \citet{Pandya19}.
However, if multiple emission lines are confirmed, one can minimize the spatial error to a $\sim$few Mpc, even with much lower-resolution spectra than the Roman grism \citep[e.g.,][]{Bunker24,DEugenio25}.

Still, there have been some promising advances with regard to photo-$z$.
\citet{Laigle16} showed that by combining photometry from 30 bands from the NUV to the MIR in the well-studied COSMOS \citep{Scoville07} field, many photometric bands can increase photo-$z$ precision by $\sim$1 dex or more. One can also use medium-band photometry instead of the wide filters on Roman/WFI and most other space-based instruments \citep[see, e.g.,][]{Suess24}.
Techniques that simultaneously combine spec-$z$ and photo-$z$ to map galaxies across large scales are also gaining prominence.
\citet{Yang21} showed that such an approach could yield high redshift accuracy in massive halos ($\gtrsim10^{12.7}~{\Msun}$) with several ($\geq$10) galaxies.
We posit that photo-$z$ measurements can supplement the spec-$z$ identified in the HLWAS-Deep field, helping to better identify the galaxy overdensities that can be followed up with deeper spectroscopy. For instance, such photo-$z$ measurements could be obtained by the DESI Bright Galaxy Survey, which targets massive galaxies \citep{Hahn23}, plausibly anchoring the nodes of the cosmic web \citep{Cohn22}. 
Recently, \cite{Malavasi25_Euclid} demonstrated that the persistence (robustness) of the filaments identified by {\disperse} using photo-$z$ from Euclid is well-correlated with the persistence of filaments from spec-$z$.

We have performed the reconstructions in 3D Cartesian space, assuming very low uncertainty on spec-$z$ for all galaxies above observational limits, which presents a few practical challenges.
A secure spec-$z$ requires at least two emission lines with high-significance; while {\Ha} would be observable at $z=1$, other common lines such as {\Hb} or {\OIIIf} would not be in the WFI G150 grism pass-band ($\sim$1--1.9~{\micron}). We would therefore have to rely on less common lines such as {\SIIf}. Furthermore, the resolution of the grism may not be able to resolve some of the narrower features, which may be a concern, especially for lower-mass galaxies.

Redshift space distortions affect the distribution of observed structures in the Universe \citep[e.g.,][]{Kaiser84}, but the analyses in this paper have been limited to real space. While we expect redshift space distortions to be less of an issue at higher redshifts, this still needs to be taken into account and corrected for when applying our reconstruction methods to observed data. A simple correction was applied to MCPM tracers on low-$z$ SDSS data by \cite{Wilde23}, but in future work we plan to quantify the effect of these distortions in much more detail in our MCPM+{\disperse} methodology.

We assume negligible errors in the stellar masses, which are used as weights for the tracers 
in our structure identification approach. Precise estimates of {\ms} require SED-fitting on many bands of photometry, and the fewer the bands, the larger the errors on average. The HLWAS-Deep survey is slated to observe seven photometric bands from the optical to the NIR \citep{ROTAC25}, but further ancillary data may be needed to obtain more confident stellar masses. Some of this may be obtained from existing deep surveys with areal overlap, such as COSMOS. 
In follow-up work, we wish to explore and quantify the impact of both redshift and stellar mass uncertainties on the identified structure.

A major assumption in our statistical analyses of galaxy morphologies
using mock samples is that intrinsic shapes will be perfectly recovered in real data. In practice, we can only observe galaxies in 2D projection on the sky. The best available tools for inferring the intrinsic 3D shapes from these measurements provide a \textit{likelihood} that a galaxy has a given shape \citep[e.g.,][]{zhang19,Pandya24}, and therefore only allow statistical estimates of the fractions of 3D shapes. Thus, we cannot estimate the true 3D shape of any galaxy with 100\% confidence. The degree of uncertainty increases for fainter, higher-redshift, and more edge-on galaxies.

While there are several alternatives to using galaxies to trace the high-$z$ cosmic web, they all currently have clear limitations. 
One such approach is {\Lya} forest tomography, which uses the {\HI} {\Lya} absorption from the intergalactic medium (IGM) in the spectra of background sources such as quasars and galaxies to create a 3D map of the cosmic web \citep[e.g.,][]{Lee18,Horowitz19}. This method can reconstruct the cosmic web at $z\!\gtrsim\!2$ with only ground-based data, but it offers fairly coarse spatial resolution ($\sim$1--5 Mpc) and, in practice, requires a high density of background sources, thus limiting it to well-observed fields ($\approx$tens of Mpc in physical size at most).
Another approach is intensity mapping, where the integrated emission of common features such as {\HI} 21 cm, {\Lya}, or CO, can be used to map the cosmic web \citep[e.g.,][]{Fonseca17,Renard24}. While this technique can help trace the cosmic web out to very high redshifts ($z\sim7$), it provides quite poor resolution and usually suffers from significant foreground contamination.
Other promising tracers such as fast radio bursts \citep[e.g.,][]{Simha20,Khrykin24} and the thermal/kinetic Sunyaev-Zel'dovich effect \citep[e.g.,][]{Tanimura19,Boryyana25} are limited by the lack of statistically significant sample sizes.



\section{Conclusion}
\label{sec:conc}

We have analyzed the IllustrisTNG cosmological simulations to achieve two primary goals. (1) Understand the role played by the cosmic web in the morphological evolution of galaxies; and (2) predict the early cosmic web structure that can be identified from upcoming galaxy surveys with the Nancy Grace Roman Space Telescope. 
For (1), we reconstructed the cosmic web structure from ${\logms}\geq8$ galaxies at redshifts $z=4$, 3, 2, 1, 0.5, and 0 in the TNG50 simulation. We applied a well-calibrated method that combines density field estimation from the MCPM model \citep{Burchett20,Elek22} with the {\disperse} structure identification algorithm \citep{disperse2} to generate highly complete and accurate cosmic web filament catalogs. We first studied the stellar and DM halo shape alignment of galaxies with their nearest cosmic filaments across different redshifts. We then investigated the dependence of intrinsic 3D shapes and kinematic morphological properties of galaxies on their nearest filaments. 
For (2), we reconstructed the cosmic web at $z=1$ in both the TNG50 and TNG100 simulations, using both mass-complete and mock samples based on observational limits. We quantified how well different mock samples can uncover the true dependence of galactic morphological properties on the cosmic web. 
Our main findings are as follows.

\vspace{-2pt}

\begin{itemize}

  \setlength\itemsep{1pt}

    \item The shape-filament alignment of galaxies decreases with time, with stars, inner DM halos, and outer DM halos losing preferential shape alignment at $z=3$, 2, and 0, respectively. The strongest shape alignment is seen in halos that are prolate, high-mass (${\logms}\geq10$), and in high-density filaments. The outer halo shape alignment increases with mass at all redshifts, but for the inner halo, this only holds at $z\gtrsim2$ (\S~\ref{sec:align}).

    \item At $z\gtrsim1$,  there is a significant increase in the fraction of prolate (spheroidal) halos in higher (lower) density filaments at fixed stellar mass, as well as a small rise in prolate stellar structures in higher-density filaments. At $z\lesssim1$, higher stellar spheroidal (oblate) fractions are associated with high-density (low-density) filaments  (\S~\ref{sec:3dshape}).

    \item At fixed stellar mass, increasing stellar sAM of galaxies is strongly associated with increasing filament density at all times, albeit the relationship weakens considerably by $z=0$. At $z\gtrsim0.5$, a higher fraction of stars in rotation-supported/disk-like orbits and a lower fraction of stars in velocity dispersion-supported/bulge-like orbits are associated with high-density filaments (\S~\ref{sec:kin}).

    \item At early times, more extended galaxies have significantly higher stellar sAM and reside in somewhat higher density filaments than more compact galaxies (\S~\ref{sec:size}). These trends weaken at later times.

    \item Baryonic physics plays a sizable role in the evolution of galaxy morphologies, as (1) the DM-only TNG50-Dark simulation generates more prolate halos with increasing mass at any redshift, in direct contrast to the hydrodynamic TNG50 simulation, and (2) TNG50-Dark shows stronger shape-filament alignment in inner halos than TNG50.

    \item The deepest grism spectroscopic survey planned on the HLWAS on Roman will \textit{not} enable an accurate and complete reconstruction of the cosmic web at $z\!=\!1$ (\S~\ref{sec:mockres}). This is because the proposed spectroscopic line flux limit of HLWAS-Deep leads to incomplete galaxy samples consisting of rarer galaxies with high mass and SFR (\S~\ref{sec:mocksamp}), thus making them poor tracers of the large-scale matter distribution. Increasing the spectroscopic depth by a factor of $\geq$2.5 times HLWAS-Deep ($\geq$2.5$\times$ fainter line flux limit) yields much more complete galaxy samples that enable much improved cosmic web reconstruction at $z\!=\!1$. Such mock galaxy samples will be able to qualitatively uncover the true cosmic web-dependence of galaxy morphological fractions, but fall short quantitatively.

    \item HLWAS-Deep will still be able to identify most of the galaxy overdensities, which can be followed up with deeper spectroscopy in those specific regions (\S~\ref{sec:strategy}). Thus, it is possible to efficiently map the early cosmic web with Roman.
    
\end{itemize}

\vspace{-3pt}

We have presented a picture wherein the morphological evolution of galaxies, including their 3D shapes, kinematics, sizes, and shape alignment with nearby filaments, is determined by a complex interplay between the accretion of matter from the large-scale cosmic web and secular processes occurring near galactic centers. 
The formation of prolate and low-spin galaxies is linked to mass accretion from predominantly low-density filaments, whereas the emergence of disks and highly-ordered systems is strongly associated with internal galactic processes such as star formation-driven feedback but also with accretion from high-density filaments.
The relative effect of the cosmic web on internal mechanisms declines with time, but even at later times, the effect of the cosmic web on galaxy morphology does not disappear completely.

Many of our predictions are likely dependent on the technical details of the TNG model, including the implementation of baryonic physics and the numerical resolution of TNG50. 
Similar analyses in separate high-resolution cosmological simulations or zoom-in re-simulations of halos in TNG50 can help constrain the effects of such uncertainties on the evolution of galactic structure.
Understanding the source of tensions with observations (including, notably, the high-redshift prolate fraction) is also crucial for understanding the causal effects of phenomena such as outflows and stellar winds.

We also determined that properly mapping the early cosmic web and testing our predictions in the real Universe will be challenging, but possible. With its combination of high-sensitivity grism spectroscopy and HST-like imaging resolution over a $\sim$100 times wider field of view, Roman can help reconstruct the $z\geq1$ cosmic web in unprecedented detail and unveil its role in the emergence of galactic morphologies. While pushing to lower line flux limits results in much higher exposure times, we showed that a practically effective strategy is to focus on patches of overdensities that can be followed up to identify most of the filamentary structures. Such targeted deep spectroscopy may be achieved not only with Roman but also with much more sensitive spectrographs on JWST.
In addition, supplementing spectroscopic redshifts with high-quality photometric redshifts -- especially over well-studied patches of the sky -- will go a long way in accurately mapping the 3D positions of distant galaxies (and therefore the large-scale structure).

This work is our first attempt at predicting the early cosmic web with large surveys, and we defer detailed investigations to follow-up studies. We deliberately limit our predictions to the deep but small fields of the HLWAS, as it is infeasible to survey low-mass galaxies over larger areas. In the future, we will apply our reconstruction method to much larger volumes comparable to the medium and wide tiers of the HLWAS, which will provide spectroscopic redshifts of millions of galaxies \citep[e.g.,][]{Zhai21,Wang:2022_romanHLSS,Madar24}. We aim to employ larger volume simulation boxes such as the $\sim$Gpc-sized MilleniumTNG \citep{Pakmor23} and mock lightcones based on semi-analytic models \citep{Yung23}. This will help us develop a more complete understanding of the large-scale structure of the Universe and its role in regulating galaxy evolution.

\vspace{-5pt}


\appendix
\label{sec:app}

\restartappendixnumbering 

\section{Mock galaxy sample selection based on photometry and spectroscopy}
\label{app:mock}

We first estimate mock magnitudes of TNG galaxies in the Roman $H$-band, i.e., F158W filter.
We utilize the dust-corrected mock magnitudes calculated according to the procedure presented in \citet{Nelson18}. 
Stellar magnitudes are modeled by taking into account dust attenuation from both (1) extinction due to dense star-forming clouds surrounding young stars \citep{CF00}, and (2) extinction and scattering from the spatial distribution of neutral gas and metals in the galaxy \citep{Calzetti94}. Each stellar population spectrum is convolved with this dust attenuation and by an observed passband, and the magnitudes are computed within 30 pkpc radius from the galaxy center.
For TNG100, we use mock 2MASS $J$-band \citep{Skrutskie06} magnitudes as also utilized by \citet{Donnari19}. This passband is almost equivalent to the F125W filter of HST/WFC3 as well as the F129W filter of Roman/WFI, albeit somewhat narrower than these space-based filters, and prone to atmospheric telluric absorption features 
\citep[see][]{Durbin23}. 
More detailed filter transformations between a variety of NIR filters, including those on HST and Roman, are now also available \citep{Durbin25}.
However, we do not adopt these, as we only require approximate magnitude values, and small uncertainties are unlikely to change our results on a population basis.

We then convert the $J$-band magnitudes to $H$-band using observed $J\!-\!H$ colors from HST. We use observed F125W-F160W colors of galaxies in the five HST frontier fields observed by the CANDELS \citep{Grogin11,Koekemoer11} and 3D-HST \citep{Brammer12} surveys, as presented in \citet{Skelton14}. Based on their reported (roughly field-invariant) median F125W-F160W colors for extended sources, we adopt an average $J\!-\!H$ color of 0.3 for each galaxy at $z\!=\!1$. While this is not as precise as implementing a magnitude-dependent $J\!-\!H$ color, the variations are small enough that this would not strongly change the mock magnitudes and therefore the selection of our galaxy samples.

For TNG50, $J$-band magnitudes are not publicly available, so instead we use mock SDSS \citep{York00} $z$-band magnitudes and $i\!-\!z$ colors. We derive an empirical transformation from SDSS $z$-magnitudes and $i\!-\!z$ colors to Roman F158W magnitudes. We do this using synthetic photometry based on Flexible Stellar Population Synthesis \citep[FSPS;][]{Conroy09} models. In these models, the stellar populations span a range of ages (0.1--5~Gyr) and redshifts ($0.5\!\leq\!z \!\leq\!3$), assuming a \citet{Chabrier03} initial mass function (IMF), solar metallicity, and the \citet{Reddy15} dust attenuation curve (see more below). Each model spectrum was redshifted and convolved with the SDSS $i$, $z$ and Roman F158W filter transmission curves using the {\sc speclite} package \citep{speclite}. A linear relation was then fit between the mock SDSS $i\!-\!z$ color and the offset between mock Roman F158W and SDSS $z$-band magnitudes as a function of redshift, enabling us to estimate F158W magnitudes for TNG50 galaxies.

We also convert spectroscopic line flux limits to SFR limits, making a few assumptions, and adopt these limits to make additional sample selections for our galaxies. Note that below we adopt SFRs averaged over 10 Myr timescales, using measurements presented in \citet{Donnari19}, as this is appropriate for the short timescales of star formation traced by nebular emission lines such as {\Ha}. Regardless, these values are not severely different from the SFR averaged over shorter/longer timescales or even the instantaneous SFRs across most of the sample.
In order to convert observed emission line flux, $F_{\mathrm{obs}}$, to SFR, we adopt the following common scaling relation
\begin{equation}
\label{eq:sfr}
    SFR = C~4 \pi D_{L}^2~F_{\mathrm{obs}}~10^{0.4A_{\lambda}},
\end{equation}
where $D_{L}$ is the luminosity distance, 
$A_{\lambda}$ is the wavelength-dependent dust attenuation, and $C$ is a constant factor to scale the observed luminosity to SFR. 
According to \citet{KE12}, the constant value for the {\Ha} emission line luminosity in the equation above is $C\!=\!1.86\!\times\!10^{41}$. Following \citet{MD14}, we also divide this constant by 2.369, to convert from the \citet{Kroupa01} to the \citet{Chabrier03} IMF, calculated using FSPS models.

$A_{\lambda}$, in magnitudes, is defined as
\begin{equation}
\label{eq:dust}
    A_{\lambda} = \frac{k_{\lambda} A_V}{R_V},
\end{equation}
where $A_V$ is the V-band dust attenuation in magnitudes and the factor $k_{\lambda}/R_V$ is derived from the specific dust attenuation law assumed.
We assume the \citet{Reddy15} dust law, which is based on measurements of $z\!\sim\!1.4\!-\!2.6$ galaxies from the Keck/MOSDEF survey \citep{Kriek15} and is more representative of higher-redshift galaxy SEDs than more common local dust models \citep[e.g.,][]{Calzetti00,Gordon03}. The \citet{Reddy15} dust law, while close in shape and normalization to these more widely-used dust laws, yields $\approx$20\% lower SFRs.

Once the dust model is chosen, $A_V$ is the only variable in Eq.~\ref{eq:dust}. 
We adopt a stellar mass-dependent $A_V$ from the UVCANDELS survey \citep{Nedkova24}. We calculate the median $A_V$ for a given {\ms} at the redshift bin $0.5\!\leq\! z \!\leq\! 1.5$, which is appropriate for our $z\!=\!1$ analysis. We use a smoothed median function to assign $A_V$ to each galaxy in TNG, given its {\ms}. The median $A_V$ is less than 0.5 at ${\logms}\!<\!10$, and rises sharply at higher masses, in agreement with previous observations at $z\!\sim\!1$ \citep[e.g.,][]{Reddy15,Battisti20,SN20}.



\begin{acknowledgements}

We dedicate this paper to the memory of our co-author Joel Primack, who was a pioneer of the modern theory of galaxy formation.
We thank the anonymous referee for their through reading and valuable suggestions to improve this paper.
This project was born out of fruitful discussions at the 2023 and 2024 Santa Cruz Galaxy Workshops. We also acknowledge insightful and interesting discussions with S. Faber, D. Koo, A. Yung, F. van den Bosch, C. Welker, J. Blue Bird, L. Sales, S. Kassin, E. Sukay, F. Jiang, P. Xu, I. Medlock, and M. Huertas-Company.
F.H. was supported by the STScI Director's Discretionary Research Award D0101.90361. 
F.H. and J.N.B. were supported by the National Science Foundation (NSF) LEAPS-MPS award No. 2137452. 
V.P. was supported by: (1) NASA Hubble Fellowship grant HST-HF2-51489 awarded by the Space Telescope Science Institute, which is operated by the Association of Universities for Research in Astronomy, Inc., for NASA, under contract NAS5-26555, and (2) NSF Astronomy \& Astrophysics grant No. 2307419.
D.N. was supported by the NSF grant AST-2307280. This research used resources of the National Energy Research Scientic Computing Center (NERSC), a DOE Office of Science User Facility supported by the Office of Science of the US Department of Energy under contract No. DEAC02-05CH11231 using the NERSC award HEP-ERCAP0024028.

\end{acknowledgements}


\end{CJK*}


\bibliographystyle{aasjournal_nikki}
{\tiny \bibliography{Refs}}

\end{document}